\documentclass[letter,11pt]{article}
\usepackage{jheppub}
\usepackage{bbold}
\usepackage{booktabs}
\usepackage{multirow}

\def\as{\alpha_s}

\newcommand{\muF}{\mu_F}
\def\t{{\bar t}}
\def\b{{\bar b}}
\def\nf{{n_f}}
\def\nl{{n_l}}
\def\MSbar{\overline{\mathrm{MS}}}
\def\mub{\mu_b}
\def\k{\kappa}

\title{Heavy--flavor parton distributions without heavy--flavor matching prescriptions}

\author[a,b]{Valerio Bertone,}
\author[c]{Alexandre Glazov,}
\author[d]{Alexander Mitov,}
\author[d]{Andrew Papanastasiou,}
\author[e]{Maria Ubiali}

\affiliation[a]{Department of Physics and Astronomy, VU University Amsterdam, De Boelelaan 1081, NL-1081, HV Amsterdam, The Netherlands}
\affiliation[b]{Nikhef, Science Park 105, NL-1098 XG Amsterdam, The Netherlands}
\affiliation[c]{DESY Hamburg, Notkestrasse 85 D-22609, Hamburg, Germany}
\affiliation[d]{Cavendish Laboratory, University of Cambridge, Cambridge CB3 0HE, UK}
\affiliation[e]{DAMTP, University of Cambridge, Wilberforce Road, Cambridge, CB3 0WA, UK}

\note{Preprint numbers: Cavendish-HEP-17/12, DAMTP-2017-40, DESY-17\_168, NIKHEF/2017-052, NSF-ITP-17-139}

\abstract{We show that the well-known obstacle for working with the zero-mass variable flavor number scheme, namely, the omission of ${\cal O}(1)$ mass power corrections close to the conventional heavy flavor matching point (HFMP) $\mub=m$, can be easily overcome. For this it is sufficient to take advantage of the freedom in choosing the position of the HFMP. We demonstrate that by choosing a sufficiently large HFMP, which could be as large as 10 times the mass of the heavy quark, one can achieve the following improvements: 1) above the HFMP the size of missing power corrections ${\cal O}(m)$ is restricted by the value of $\mub$ and, therefore, the error associated with their omission can be made negligible; 2) additional prescriptions for the definition of cross-sections are not required; 3) the resummation accuracy is maintained and 4) contrary to the common lore we find that the discontinuity of $\as$ and pdfs across thresholds leads to improved continuity in predictions for observables. We have considered a large set of proton-proton and electron-proton collider processes, many through NNLO QCD, that demonstrate the broad applicability of our proposal.}

\begin{document} 
\maketitle
\flushbottom

\section{Introduction}\label{sec:intro}

It is well established that a proton $p$ undergoing an inelastic collision at a scale $Q>m_p$ will exhibit non-trivial heavy quark content. Heavy quarks are the ones whose mass $m$ is large enough that the strong coupling at that scale is perturbative, i.e. $\as(m)\ll1$. In practice, only the charm and bottom quarks are considered massive while the top quark is too massive to be relevant for currently accessible collider energies (which might change at future high energy colliders). In this work we consider the simplified situation of a single massive flavor which for convenience we choose to be the bottom. We will return to the issue of including charm and top in the Conclusions.

This work is based on the collinear factorization approach \cite{Collins:1989gx,Collins:1985gm,Collins:1998rz} within which one identifies two types of contributions to the heavy quark parton distribution function (pdf): a perturbative and an intrinsic one. In this study we will not consider the intrinsic component since it is very small and, in fact, all recent attempts \cite{Ball:2016neh,Ball:2015dpa,Ball:2015tna,Ball:2017nwa,Hou:2017khm} to derive the intrinsic charm component of the proton from experimental data have been consistent with vanishing intrinsic charm (intrinsic bottom would be even smaller). We note that not including intrinsic heavy flavor component is mostly for convenience, however, and this is a problem which is orthogonal to the scope of the current work. Therefore, for the rest of this work we will simply speak of heavy flavor pdfs and will always have in mind the perturbative component.

The heavy flavor decoupling theorem \cite{Symanzik:1973vg,Appelquist:1974tg,Collins:1978wz} provides the natural framework for discussing heavy flavor pdfs. When a quark of mass $m$ is probed at a scale $Q$ one has two unambiguous limiting behaviors: if $Q\gg m$ then the quark is effectively massless; all ${\cal O}(m)$ contributions can be neglected, large quasi-collinear logs $\sim\ln^n(Q/m)$ are resummed by the DGLAP evolution and the quark behaves exactly as the light quarks. In the opposite limit, when $Q\ll m$ the quark is very heavy and can be integrated out from the theory. Corrections behaving like ${\cal O}(1/m)$ are neglected and, effectively, the heavy quark disappears from the theory. 

There exists a substantial interval of scales $Q$ not very different from the mass $m$, $Q\sim m$, which is not covered by the two above-mentioned asymptotic regimes and where power corrections of $m$ are important. The theory provides us with only a minimal guidance about this intermediate region and it is this inherent ambiguity which is the subject of this work as well as the vast majority of past works on including heavy flavors in pdfs. 

In the interval of energies from $Q\sim m$ down to $Q\ll m$ the dependence on the mass $m$ can be reconstructed, at least at the conceptual level, in the following way. In any observable, there are three basic ingredients where the dependence on the mass $m$ appears: the pdfs, the strong coupling $\as$ and the partonic cross-section $d\hat\sigma$. In the limit $Q\ll m$ the pdfs and $\as$ evolve with $\nf=\nl$ active flavors. As emphasized extensively in ref.~\cite{Collins:1998rz}, since they both are renormalized in the $\MSbar$ scheme -- which is mass independent -- one can extend their evolution all the way up to scales $Q\sim m$. On the other hand the mass dependence in $d\hat\sigma$ can also be included by computing all diagrams that include the heavy quark in the final state. This way, one can have a prediction with full dependence on the mass $m$ from scales $Q\sim m$ down to $Q\ll m$. This is the usual fixed-flavor (FF) picture for describing charm, bottom and top production.

The difficulties arise when one tries to extend this result towards high energies $Q\gg m$. Following the decoupling approach discussed above, at some scale $Q\equiv\mub\sim m$ one switches from the $\nf=\nl$ description to a $\nf=\nl+1$ one, where the massive flavor is now considered as massless. In the following we will refer to the point $\mub$ as heavy flavor matching point (HFMP).
\footnote{Our notation is such that the subscript labels the specific heavy flavor, i.e. $\mu_c$ for charm and $\mu_t$ for top. As mentioned above, depending on the context $\mub$ indicates either the $b$ flavor or a generic heavy flavor.}
We will avoid calling it threshold to avoid confusion with physical thresholds. For all scales $Q>\mub$ then one evolves the pdfs and $\as$ with $\nl+1$ flavors and introduces pdfs for the heavy flavor and anti-flavor. As usual, matching relations for the coupling and pdfs at the scale $\mub$ have to be imposed. Those are fully known through NNLO \cite{Collins:1986mp,Buza:1995ie,Buza:1996wv} and partly known beyond \cite{Ablinger:2010ty,Blumlein:2012vq,Ablinger:2014lka,Ablinger:2014uka} for the pdfs, and through four loops \cite{Schroder:2005hy,Chetyrkin:2005ia} for $\as$. The above construction represents what is known as a variable flavor number scheme (VFNS) \cite{Aivazis:1993pi} or, specifically, a zero-mass one (ZM-VFNS). 

The ZM-VFNS described above is consistent. However, it has an important shortcoming: the intermediate region $Q\sim m$ in the $\nl+1$ scheme misses power corrections ${\cal O}(m)$ that are important numerically (indeed, possibly dominant) in that region. A number of approaches have been proposed in the past, starting with the ACOT proposal \cite{Aivazis:1993pi}, which allows for the power corrections present in the FF result to be incorporated in the ZM-VFNS prediction. Many other such proposals exist \cite{Thorne:1997ga,Thorne:1997uu,Kramer:2000hn,Thorne:2006qt,Tung:2006tb,Nadolsky:2009ge,Forte:2010ta,Guzzi:2011ew,Han:2014nja} and approaches of this type are known as general mass VFNS (GM-VFNS). These proposals have been refined in a number of phenomenological applications \cite{Cacciari:1998it,Forte:2015hba,Forte:2016sja,Olness:1987ep,Olness:1994zn,Bonvini:2015pxa,Bonvini:2016fgf} and in studies on the rationale behind the choice of either scheme \cite{Maltoni:2007tc,Maltoni:2012pa,Degrande:2015vpa,Lim:2016wjo}. 

Schematically, the GM-VFN schemes work in the following way (see ref.~\cite{Forte:2010ta} for details): 
\begin{equation}
d\sigma^{\rm (GM-VFNS)} = d\sigma^{(\nl+1)}(Q,m) + d\sigma^{(\nl)}(Q,m) - d\sigma^{(\nl,0)}(Q,m)\,,
\label{eq:GMVFNS}
\end{equation}
where $d\sigma^{(\nl+1)}(Q,m)$ is the ZM-VFNS result, $d\sigma^{(\nl)}(Q,m)$ is the one in the $\nf=\nl$ scheme with full mass dependence in the perturbative coefficient function and $d\sigma^{(\nl,0)}(Q,m)$ is its massless limit (which contains $\ln(m)$ and $m$-independent terms but no ${\cal O}(m)$ ones). Such a construction naturally converges to the ZM-VFNS one for $Q\gg m$ and has the added benefit that at scales $Q\sim m$ also ${\cal O}(m)$ power corrections that are part of the massive $\nf=\nl$ calculation are included. Note that since in most GM-VFNS pdfs and $\as$ are transformed into the same scheme, either $\nf=\nl$ or $\nf=\nl+1$, the notation in eq.~(\ref{eq:GMVFNS}) emphasizes the number of flavors in the partonic cross-sections and not the ones in $\as$ and pdfs.

While the above construction is undoubtedly an improvement over the ZM-VFNS for scales $Q\sim m$, GM-VFN schemes suffer from certain ambiguities. First, ${\cal O}(m)$ terms in reactions initiated by heavy flavors are usually not introduced. Presumably those are suppressed numerically by the smallness of the heavy-flavor pdf; we revisit these power corrections in sec.~\ref{sec:mubchoice}. Second, $d\sigma^{\rm (GM-VFNS)}$, as defined in eq.~(\ref{eq:GMVFNS}), does not behave as desired for $Q\sim m$ because the difference $d\sigma^{(\nl+1)}(Q,m) - d\sigma^{(\nl,0)}(Q,m)$ does not automatically vanish there. To restore the expected $d\sigma^{\rm (GM-VFNS)}\to d\sigma^{(\nl)}(Q,m)$ behavior in this limit, one typically suppresses this difference by hand
\footnote{While in the calculation of DIS structure functions all schemes implement a suppression, such a suppression is not necessarily introduced in calculations of Tevatron or LHC processes where the typical scales are much larger than the heavy quark mass $m$. See for example refs.~\cite{Campbell:2003dd,Forte:2015hba,Bonvini:2015pxa,Forte:2016sja,Bonvini:2016fgf}.}.
In the existing literature this suppression is implemented in two ways: either by multiplying that difference by an arbitrary function which vanishes below $Q=m$ or by introducing a specially designed DIS-inspired rescaling (known as $\chi$-rescaling) of the partonic variable $x$. See ref.~\cite{Forte:2010ta} for a detailed discussion on this point.

In this work we demonstrate that a numerically accurate prediction can be achieved within the simplest possible variable flavor scheme: ZM-VFNS. To that end only one modification needs to be made compared to the conventional formulation: move the matching point $\mub$ that separates the $\nf=\nl$ and $\nf=\nl+1$ schemes towards higher values, i.e. from the conventional value $\mub=m$ to $\mub=\kappa \cdot m$ with the parameter $\kappa$ as high as 10. The detailed justification for, and implications of, such a choice are given in the following section. The remainder of the paper is then devoted to the phenomenological study of our proposal.

\section{The proposed idea}\label{sec:idea}

The single most relevant piece of information about our work is the position of the matching point $\mub$. In order to explain why and how we change it we need to first revisit its status. 

All existing pdf sets and virtually all papers on the subject (with handful of exceptions we discuss below) set $\mub=m$. We have been unable to trace this choice to any specific proposal in the literature. It seems to us that this choice stems from a combination of two results. First, as discussed in sec.~\ref{sec:intro}, decoupling implies that the matching point $\mub$ has to be of the order of the mass $m$. This, of course, does not imply that $\mub$ has to be set equal to $m$ and, in fact, there is a considerable range of values around $m$ that satisfy this requirement. It is precisely this range that we will be exploring in this work. 

Second, the requirement for continuity of the pdfs across the HFMP was historically very influential. The matching conditions for $\as$ and the full set of pdfs $f_i$ read
\begin{eqnarray}
\as^{(\nl+1)}(\mub^2)&=&\as^{(\nl)}(\mub^2)\left(1+\sum_{n=1}^{\infty}c^{(n)}(\mub^2/m^2)\left[\alpha_s^{(\nl)}(\mub^2)\right]^n \right)\,, \label{eq:match}\\
f^{(\nl+1)}_i(x,\mub^2)&=& \sum_j\left(\delta(1-x)\delta_{ij} + \sum_{n=1}^{\infty} K^{(n)}_{ij}(x,\mub^2/m^2)\left[\alpha_s^{(\nl)}(\mub^2)\right]^n \right)\otimes f_j^{(\nl)}(x,\mub^2)\,,\nonumber
\end{eqnarray}
where the sum over $j$ goes over all flavors but the heavy one and $\otimes$ stands for the usual integral convolution in the variable $x$.

It is obvious from eqs.~(\ref{eq:match}) that at the leading order (LO) both the coupling and pdfs are continuous for any value of $\mub$. At higher orders the coefficients $c^{(n)}$ and $K^{(n)}_{ij}$ are polynomials in $\log(\mub^2/m^2)$. A peculiar feature of the NLO result is that the ``constant" (i.e. $\log(\mub/m)$-independent) terms in both $c^{(1)}$ and $K^{(1)}_{ij}$ are zero. This implies that the coupling and pdfs are continuous also at NLO if the matching scale is chosen to be the mass, i.e. if $\mub=m$. Naively, the requirement for continuity seems desirable and much attention to it has been devoted in the past \cite{Thorne:1997ga,Thorne:1997uu}. However, this is not so as we demonstrate in the following.

First, it turns out that the continuity of $\as$ and pdfs across the matching point $\mub$ for $\mub=m$ is {\it accidental}. It does not persist at higher perturbative orders in the space-like region \cite{Buza:1995ie,Buza:1996wv} or for related quantities in the time-like case \cite{Cacciari:2005ry,Mele:1990cw,Melnikov:2004bm,Mitov:2004du}. Therefore, one has to accept that the argument for continuity by a special choice of the matching point $\mub$ is {\it invalid}. 

Second, pdfs are not observables and their continuity is not a formal theoretical requirement. Perhaps somewhat surprisingly, in this work we demonstrate precisely the opposite: the presence of discontinuities across $\mub$ in $\as$ and pdfs is, in fact, beneficial for the continuity of predictions for observables at higher perturbative orders. Technically this happens because the discontinuities in $\as$ and various pdfs conspire in such a way that they tend to largely compensate each other. This happens in a variety of processes and observables (both inclusive and differential). Unsurprisingly, this observation reminds us that the theory we work with is incredibly self-consistent! At this point we can introduce our proposal in the following way: 

\centerline{\bf \underline{Proposal}:}
\begin{itemize}
\item For the observable of interest (which could be inclusive or fully differential) choose the value (or functional form) of the factorization scale;
\item Decide on the value of $\mub$. In this work we explore values for $\mub$ as large as $10 m$;
\item Events with kinematics for which $\muF\leq\mub$ are computed in the $\nf=\nl$ scheme and full mass dependence is to be retained;
\item Events with kinematics for which $\muF > \mub$ are to be evaluated in a scheme with $\nf=\nl+1$ active flavors where the mass $m$ is set to zero.
\end{itemize}
In effect this is a ZM-VFNS scheme with a HFMP that is set to a (much) higher value than in all existing pdf sets. For short, we will sometimes call it ``variable $\mub$ approach".

With the exception of appendix~\ref{app:BGMPUfits}, in the rest of this work we focus on the bottom pdf. For added clarity, we make our proposal explicit with the following $b$-pdf specific equation:
\begin{equation}
\hskip -2mm
d\sigma =
\begin{cases}
f^{(n_l=4)}_i(\mu_F^2,\, \mub^2) f^{(n_l=4)}_j(\mu_F^2,\, \mub^2) d\hat\sigma_{ij}^{\rm 4F}(\mu_{F,R}^2,\,\as^{(\nl=4)},\,m_b\neq0), &\text{for $\mu_F < \mub$}
\\[10pt]
f^{(n_l=5)}_k(\mu_F^2,\, \mub^2) f^{(n_l=5)}_l(\mu_F^2,\, \mub^2) d\hat\sigma_{kl}^{\rm 5F}(\mu_{F,R}^2,\,\as^{(n_l=5)},\,m_b=0), &\text{for $\mu_F > \mub$}\,,
\end{cases}
\end{equation}
where the indices run over the following values: $i,j \in \{g,u,\bar{u},d,\bar{d},s,\bar{s},c,\bar{c}\}$ and $k,l \in \{g,u,\bar{u},d,\bar{d},s,\bar{s},c,\bar{c},b,\bar{b}\}$.

\subsection{Addressing some obvious questions}

In the following we address a number of questions about our construction.

\begin{enumerate}
\item What about the mass power corrections ${\cal O}(m)$ neglected above the HFMP? Our proposal is not exact and indeed for scales above the point $\mub$ it misses power corrections ${\cal O}(m)$. However, unlike existing VFNS realizations, we have a parameter that controls the size of the error we make by neglecting these power corrections. For a given threshold the actual error that is made is ${\cal O}(m^2/\mub^2)$ 
\footnote{Assuming that all other kinematic scales are bounded from below by the factorization scale $\muF$.}
and could be as small as $1\%$ for $\mub=10 m$. This is a negligible effect at the current level of experimental and theoretical precision. Even for $\mub=5 m$ the error is around $4\%$, i.e. very small. We would like to stress that none of the existing schemes have a way of parametrically estimating the error they make.
\item Can this proposal be complemented with existing, or future, GM-VFNS schemes like ACOT \cite{Aivazis:1993pi} or FONLL \cite{Forte:2010ta}? Yes. Our proposal does not preclude the inclusion of power corrections ${\cal O}(m)$ above the HFMP. However, by construction, such corrections will be small and therefore typically there is no need for them at the current accuracy level (we verify this explicitly in appendix~\ref{app:BGMPUfits}). One of the main goals behind our proposal is to have a scheme which is both accurate and very straightforward to implement.
\item How high should the value of $\mub$ be? Clearly it should not be very high because if $\mub\gg m$ then collinear resummation will be spoiled. On the other hand, the power corrections missed 
in the massless calculation above the HFMP are suppressed by ${\cal O}(m^2/\mub^2)$ and the motivation for choosing large values for $\mub$ is that they can be made tiny. In fact, as we demonstrate at length in the rest of this paper, a value of $\mub$ that is as high as $\mub=10 m$ is allowed and, therefore, for such a choice any power corrections are about $1\%$ effect and thus rendered phenomenologically irrelevant. 
\item Is resummation accuracy maintained? Yes, if the HFMP is not chosen too high. We have checked that even for $\mub=10 m$ the asymptotic behavior of cross-sections for $Q\gg m$ is maintained.
\item What is the distinction between choosing $\mub$ and the factorization scale $\muF$? There is a fundamental difference between these two scales. While $\mub$ is totally process independent - in effect it only knows about the proton and its choice should reflect that - the factorization scale $\muF$ is process and observable dependent. This scale implements the distinction between short- and long-distance physics specific to the way the proton is probed in a given measurement. For this reason the choice for $\muF$ should be made prior to, and independently of, that for $\mub$.
\item What about intrinsic heavy flavor? In the context of our work a possible intrinsic heavy flavor contribution will only affect the boundary condition at the HFMP used subsequently by the DGLAP evolution. If needed it could be implemented on top of our considerations. 
\item What is the relation to fragmentation functions? Although we have not explored the implications of our proposal for the case of fragmentation functions we see no reason why it would not apply directly there, too. One should note that our situation corresponds to the case of heavy flavor contribution to the fragmentation of light hadrons. The fragmentation of heavy flavored mesons is different and may have to be analyzed with more care.
\end{enumerate}

\subsection{Is the value of $\mub$ a matter of choice or part of the theoretical uncertainty?}\label{sec:mubchoice}

At this point one may wonder if the value of $\mub$ is a matter of making a suitable choice or if it should be considered apriori undetermined and the ambiguity due to its variation around the point $\mub=m$ considered as part of the theoretical systematics. 
The latter possibility has been explored in three recent papers \cite{Bonvini:2015pxa,Bonvini:2016fgf,Bertone:2017ehk} which, to the best of our knowledge, are the only ones where values for $\mub$ different from the standard one $\mub=m$ have been considered. For physical predictions, refs.~\cite{Bonvini:2015pxa,Bonvini:2016fgf} pick the canonical value $\mub = m$ as the central choice and use variations about this point to estimate the associated theoretical uncertainties.

Our motivation for moving away from the canonical choice $\mub=m$ is different from the ones in the above-mentioned references. We explain it next.

As far as $\as$ and pdfs are concerned any choice $\mub\sim m$ is equally correct and therefore the choice of $\mub$ is unimportant
\footnote{Only in the sense of working to a given perturbative order.}.
This ``translational invariance" of $\mub$ is, however, not respected by the perturbative differential cross-sections (i.e. coefficient functions). While in the $\nf=\nl$ scheme full mass dependence is retained, this is not the case for the $\nf=\nl+1$ scheme where power corrections ${\cal O}(m)$ are missed. It is these missing power corrections that make the theory predictions depend on the position of the HFMP.

In this work we take the viewpoint that the correct position of the HFMP is a matter of choice: it is the position where these corrections are minimized. Such a position may not be unique; it may belong to a finite range whose size is determined by the uncertainty tolerance level. Values of $\kappa$ in the interval $\kappa=5-10$ seem to satisfy this requirement for $b$-production. Once $\kappa$ is chosen within such a range, its variation within that range would be indicative of missing higher-order corrections related to perturbative matching. Indeed, in this work we see that the inclusion of higher-order corrections in the pdf matching conditions (and to a lesser degree in the pdf evolution) leads to a systematic reduction of the sensitivity of observables to the position of the HFMP $\mub$ for sufficiently high values of $\mub$.

Finally, we would like to stress that there are both technical and conceptual reasons why ${\cal O}(m)$ corrections are always missing. In existing GM-VFNS implementations this is due to the introduction by hand of rescaling or damping functions. These are checked in known cases but their applicability for any process or to higher-orders is not obvious or automatic. ${\cal O}(m)$ corrections are also neglected in higher-order calculations with incoming massive quarks. The reason for this is twofold: on the technical side such calculations are uncommon and technically challenging \cite{Kretzer:1998ju}. On the conceptual side, even if one makes the effort to compute a cross-section with full mass dependence for initial-state massive quarks and then subtracts all quasi-collinear singularities as appropriate for a massless $\MSbar$ subtraction, it is known that collinear factorization is violated starting at NNLO with two initial state massive fermions \cite{Doria:1980ak,DiLieto:1980nkq,Catani:1985xt}. More about this topic can be found in the textbook by Collins \cite{Collins:2011zzd}.

\section{The effect of changing HFMP on $\as$ and pdfs}

In this work, we produce 6 sets with varying $b-$HFMP's, in which the value of $\mub$ is set to $\k\cdot m$ with $\k=1,2,4,6,8,10$. For reasons of technical convenience we base our study on the NNPDF3.0 family of pdf sets, although it could be performed with any existing pdf set. The pdf set with $\k=1$ coincides with the standard NNPDF3.0 pdf set with $\as(m_Z)=0.118$. The strong coupling constant is fixed at the conventional reference scale, i.e. the $Z$-pole mass, with $m_Z=91.2$ GeV and for $\nf=5$. All pdfs are parametrized below the charm threshold ($m_c=1.275$ GeV) at the scale $Q_0=1.0$ GeV, with $\nf=3$. 

\begin{figure}[t]
\centering
\includegraphics[width=0.33\textwidth]{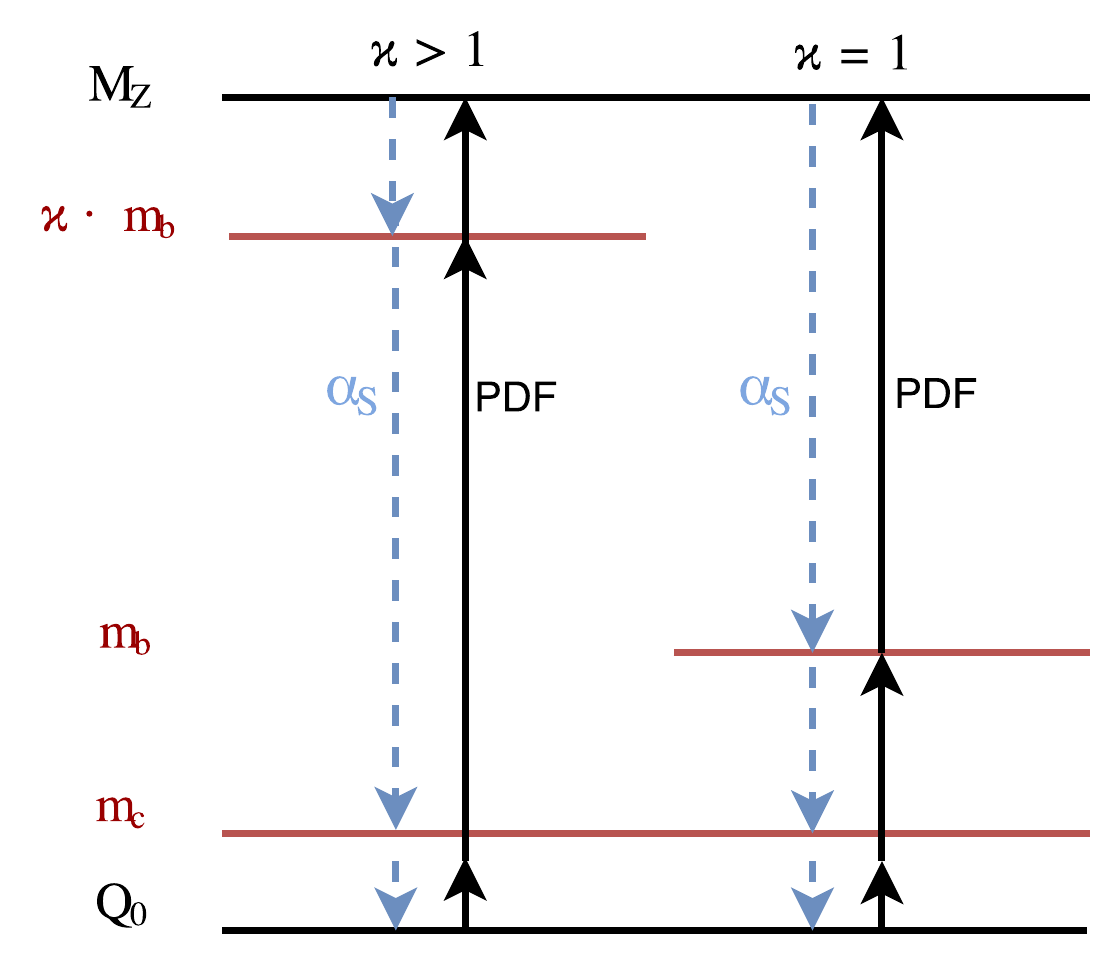}
\caption{\label{fig:diagram} Schematic illustration for the evolution of $\as$ and pdfs in our approach contrasting the canonical case $\k=1$ with the case $\k>1$ as proposed in this work. The dashed blue lines correspond to the backward evolution of the strong coupling constant (starting form an initial scale $m_Z$) while the black solid lines correspond to the forward evolution of pdfs starting form an initial scale $Q_0$.}
\end{figure}

Following the NNPDF setup, in our analysis the strong coupling constant is evolved from the initial scale $m_Z$ to the scale $Q$ of the data included in the fit, starting with $\nf=5$ and crossing all HFMP between $m_Z$ and $Q$. In turn, the pdfs are evolved from the initial scale $Q_0$ crossing at least the charm HFMP and up to the scale at which data are available.
\footnote{This is because only data with $Q^2>3.5$ GeV$^2$ are included in the NNPDF3.0 analysis.}
This is illustrated schematically in fig.~\ref{fig:diagram}.

Notice that in the NNPDF3.0 default set, the top mass is set to infinity, thus the maximum number of active flavors in both pdfs and $\as$ evolution is $\nf=5$. The hadronic observables included in the NNPDF3.0 analysis are computed in the ZM-VFN scheme, i.e. no charm or bottom quark mass effects are included in the computation of Drell-Yan, jets or vector boson production cross sections. Instead, in the computation of DIS observables, the FONLL scheme is adopted, in which the massive calculation in the vicinity of the heavy quark HFMP is matched to the massless computation far above it. 

In order to fully define the sets with $\mub=\k \cdot m$, with $\k>1$, one needs to specify their values at the initial scale $Q_0=1$ GeV. This could be done in two ways. One may wish to obtain the boundary condition for each set with $\k>1$ by refitting the data as described above and evolving consistently with HFMP set at $\mub=\k\cdot m$, as appropriate for that set. This way, in general, one will obtain initial conditions which are different for pdf sets with different $\k$. Alternatively, one may simply require that the initial condition for any value of $\k$ be the same (and in particular be the same as the canonical case $\k=1$). The rationale for the latter possibility is that one may imagine a lattice QCD-based prediction (see e.g. refs.~\cite{Alexandrou:2017huk,Nocera:2017war} for latest developments) which presumably will be insensitive to the heavy quark content at high scales. 
\footnote{We thank Gavin Salam for a discussion that led us to this argument. We also thank Jianwei Qiu for an illuminating discussion.}

One may wonder if in practice choosing one or the other approach leads to significant differences in the pdf boundary conditions. We have checked this explicitly by producing pdf sets based on both approaches. We observe very small differences, with PDFs at the initial scale deviating from each other by less than a tenth of the pdf uncertainty bands for all values of $x$. As a result we have decided to use in this work only pdfs whose initial value is the same as the one of the standard NNPDF3.0 set.

We next study the effect from changing the HFMP on the scale evolution of $\as$ and all pdfs. We plot them as functions of the factorization scale $Q$ at NLO and NNLO. The pdfs are shown for four values of the partonic momentum fraction $x$.
\begin{figure}[t]
\centering
\includegraphics[width=0.33\textwidth,angle=-90]{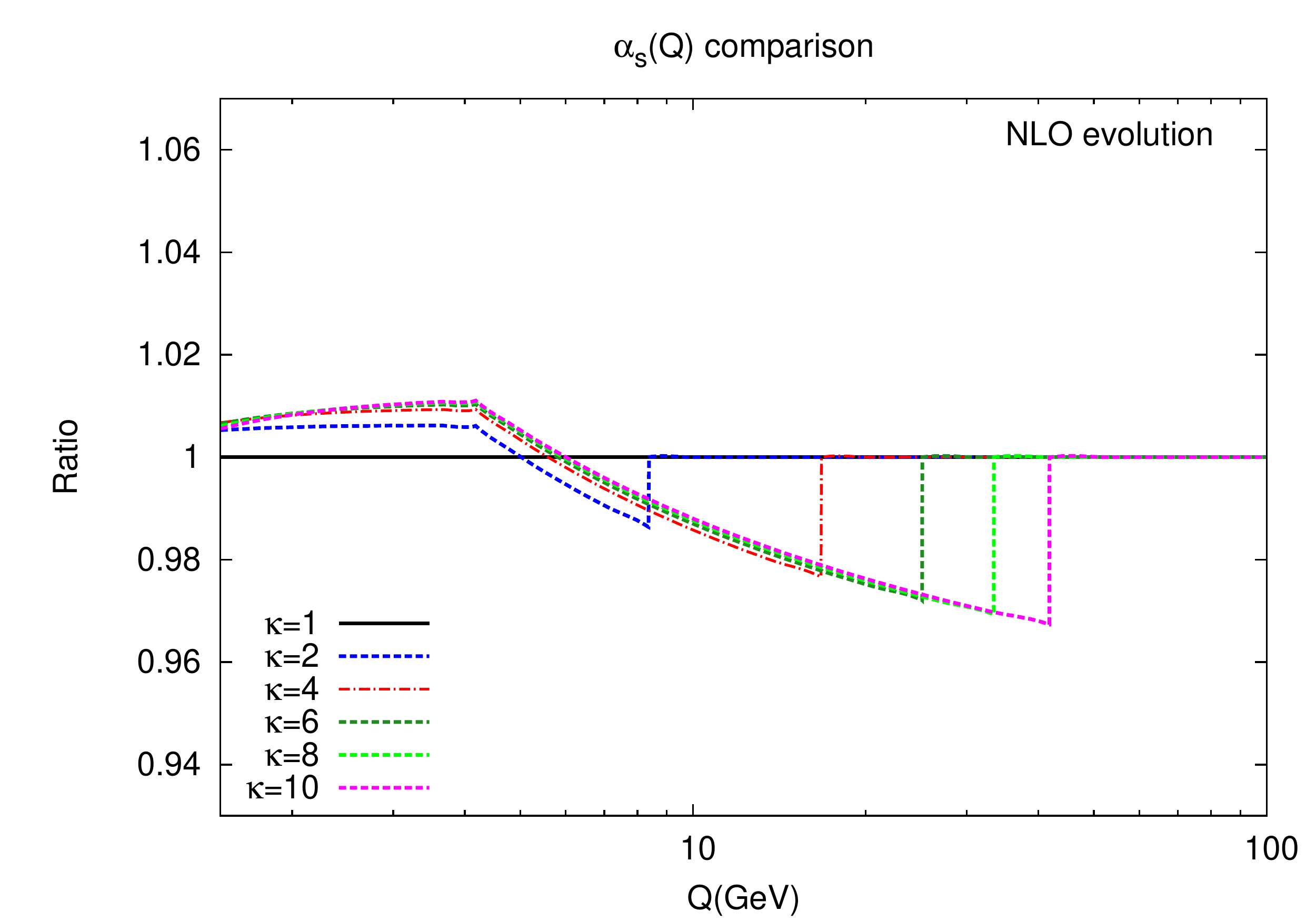}
\includegraphics[width=0.33\textwidth,angle=-90]{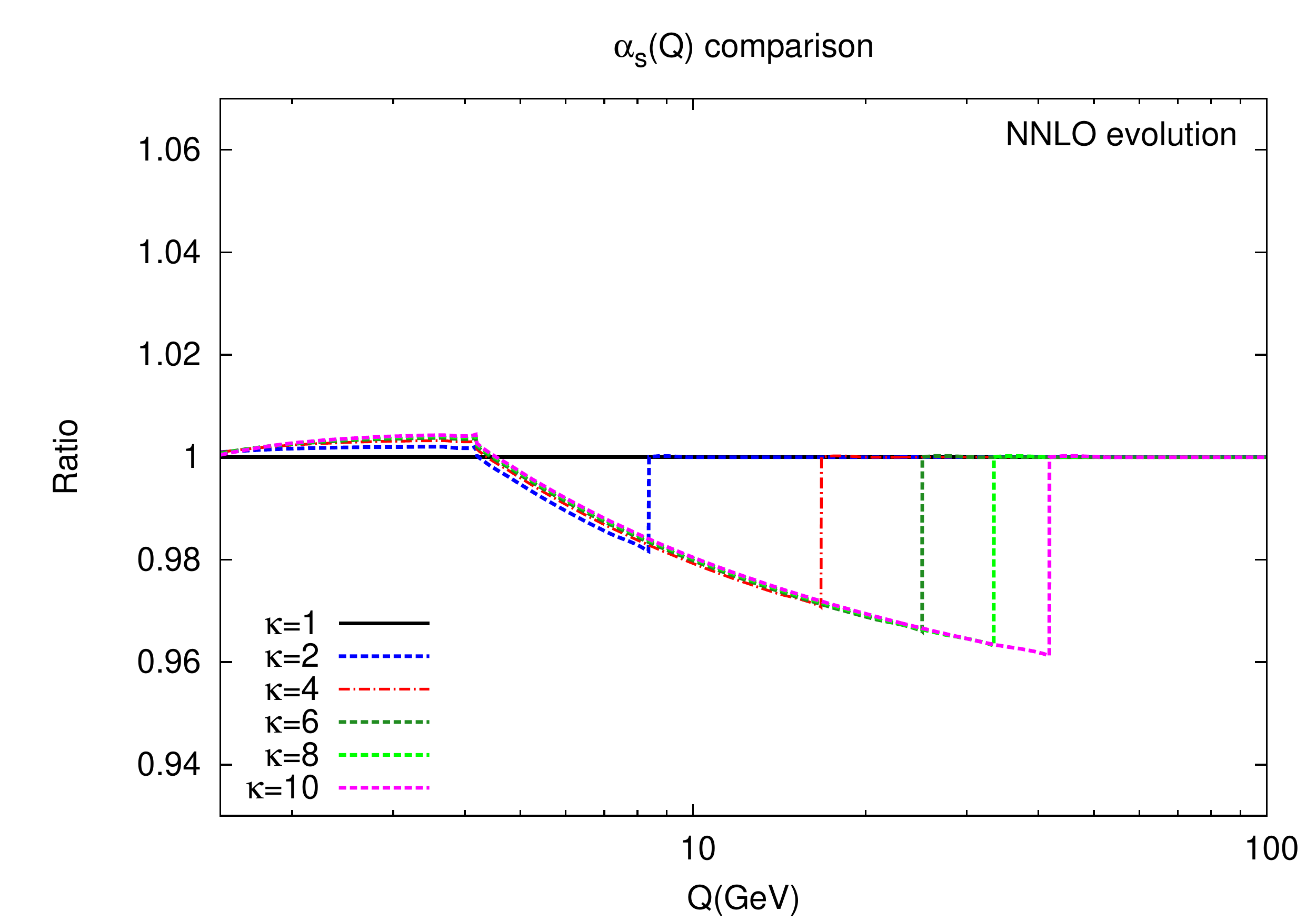}
\caption{$\alpha_s(Q)$ evolution at NLO and NNLO for various $b$-HFMP values. Shown is the ratio between $\k=$2,4,6,8,10 and the standard $\k=$1.}
\label{fig:alphas}
\end{figure}

The evolution of $\as$ is shown in fig.~\ref{fig:alphas}. In fig.~\ref{fig:gluonpdf} we show the gluon pdf while the bottom pdf is shown in fig.~\ref{fig:bottompdf}. The evolution of the quark singlet including (or not) the bottom flavor is shown in fig.~\ref{fig:singletb} (fig.~\ref{fig:singlet}), respectively. In all cases we observe significant and progressively increasing discontinuities as the value of $\k$ increases. As we demonstrate in the following sections, however, these discontinuities cancel each other in observables. 

Although this may appear somewhat counterintuitive, we would like to point out that the discontinuity in the matching conditions has a beneficial effect on the $b$-quark pdf. As can be seen from fig.~\ref{fig:bottompdf} going from LO towards NNLO the discontinuity in the matching condition increases and the bottom pdf at large scales becomes {\it less} sensitive to the position of the threshold. This happens because the large discontinuity at NNLO allows the $b$-quark pdf to DGLAP-evolve starting from a non-zero value immediately after the HFMP. As an example for how pdf continuity negatively affects pdfs one can take the LO case, not shown here, where the matching condition is continuous for any position of the HFMP. For this reason the $b$-quark pdf has to start from zero for any position of the HFMP, which means any two pdfs derived with different HFMPs will only very slow converge towards each other at large $Q$.

Finally, we also notice that in all cases, for very large $Q$ all pdf sets with different HFMPs converge towards each other and this convergence is well within the pdf uncertainty. This is an important validation of our approach which needs to preserve the asymptotic behavior for large $Q$ (i.e. to preserve the DGLAP evolution and the resummation of terms $\sim\ln^n(Q/m)$).

\section{Testing the approach with LHC processes}

Our idea -- to work with pdfs with high heavy flavor matching points -- is motivated formally, not phenomenologically. Therefore, in this section we want to verify that it works in practice by quantifying to what extent the position of the HFMP affects a broad range of LHC processes and observables. We will be paying particular attention to the discontinuities in inclusive and differential cross-sections across the HFMP and the effect of changing HFMP on precision LHC observables.

\subsection{Effect of varying HFMP on processes sensitive to $b$-quark pdf}\label{sec:threshold-switch}

We consider the following three processes that are sensitive to the $b$-content of the proton: the total $t$-channel single top cross-section at LHC 13 TeV with $\mu_F=\mu_R=m_t/2$, the total cross-section for the process $b\bar{b}Z$ at LHC 13 TeV and the differential cross-section of $Z+b$-jet at LHC 13 TeV 
\footnote{After this paper was completed a new measurement \cite{Aaboud:2017skj} of $\gamma+b$-jet appeared which would be an interesting process to study within our approach.}
as a function of $p_T(J_b)$ with
\begin{equation}
\mu_F = {1\over 4}\sqrt{m_Z^2 + \sum_j p_{T,j}^2}\,.
\label{eq:muFbjet}
\end{equation}
The coefficient $1/4$ in the equation above is chosen so that for small $p_{T,j}$ the scale $\mu_F$ is small enough and crosses as many values of $\mub$ as possible (see fig.~\ref{fig:disc3procs}). At leading order, the process $b\bar{b}Z$ contains $b{\b}\to Z$ in the 5 flavor scheme (FS) while in the 4FS it is given by $gg\to Zb\b$. At higher orders it is defined in such a way that it involves all diagrams containing a $b\b Z$ vertex dressed with QCD radiation, as appropriate. 

All calculations through NLO QCD are derived with the {\sc\small MadGraph5\_aMC@NLO} code \cite{Alwall:2014hca}. All public and private pdf sets we use are incorporated through the LHAPDF library \cite{Buckley:2014ana}. The 5FS NNLO corrections to the $t$-channel single top cross-section are not computed from first principles but are obtained by rescaling the NLO 5FS results with an NNLO $K$-factor derived from the results of refs.~\cite{Brucherseifer:2014ama,Berger:2016oht}. Specifically, the proxy to the NNLO $t$-channel single-top cross-section in the 5FS is
\begin{equation}
\sigma^{\rm NNLO~proxy}(\k) = \left( \frac{\sigma^{\rm NNLO}}{\sigma^{\rm NLO}} \right) \sigma^{\rm NLO}({\rm NNLO~pdf},\k)\,,
\end{equation}
where the ratio $\sigma^{\rm NNLO}/\sigma^{\rm NLO}$ is constructed from the 13 TeV inclusive cross-section numbers in table 1 of ref.~\cite{Berger:2016oht} (i.e. for $\k=1$) and $\sigma^{\rm NLO}({\rm NNLO~pdf},\k)$ is calculated exactly, using the relevant pdf set with $\k\ge 1$. Given the extremely flat scale dependence and small NNLO correction such an approach is more than adequate for our purpose. For this reason we have not extended the NNLO curves for scales outside the $\mu_F$ range considered in refs.~\cite{Brucherseifer:2014ama,Berger:2016oht}, which is sufficient for our goals. 

The 5FS cross-section for $b\bar{b}Z$ has been calculated at NNLO by using a private code~\cite{Maltoni:2005wd}, which has been cross-checked at LO and NLO against {\sc\small MadGraph5\_aMC@NLO}. The 4FS cross-section has been computed with {\sc\small MadGraph5\_aMC@NLO}. In both calculations the coupling of the $Z$ boson to light quarks is set to zero, so that only bottom-initiated diagrams are considered.

\begin{figure}[t]
\centering
\includegraphics[trim=0.4cm 0.0cm 0.2cm 0.4cm,clip,width=0.32\textwidth]{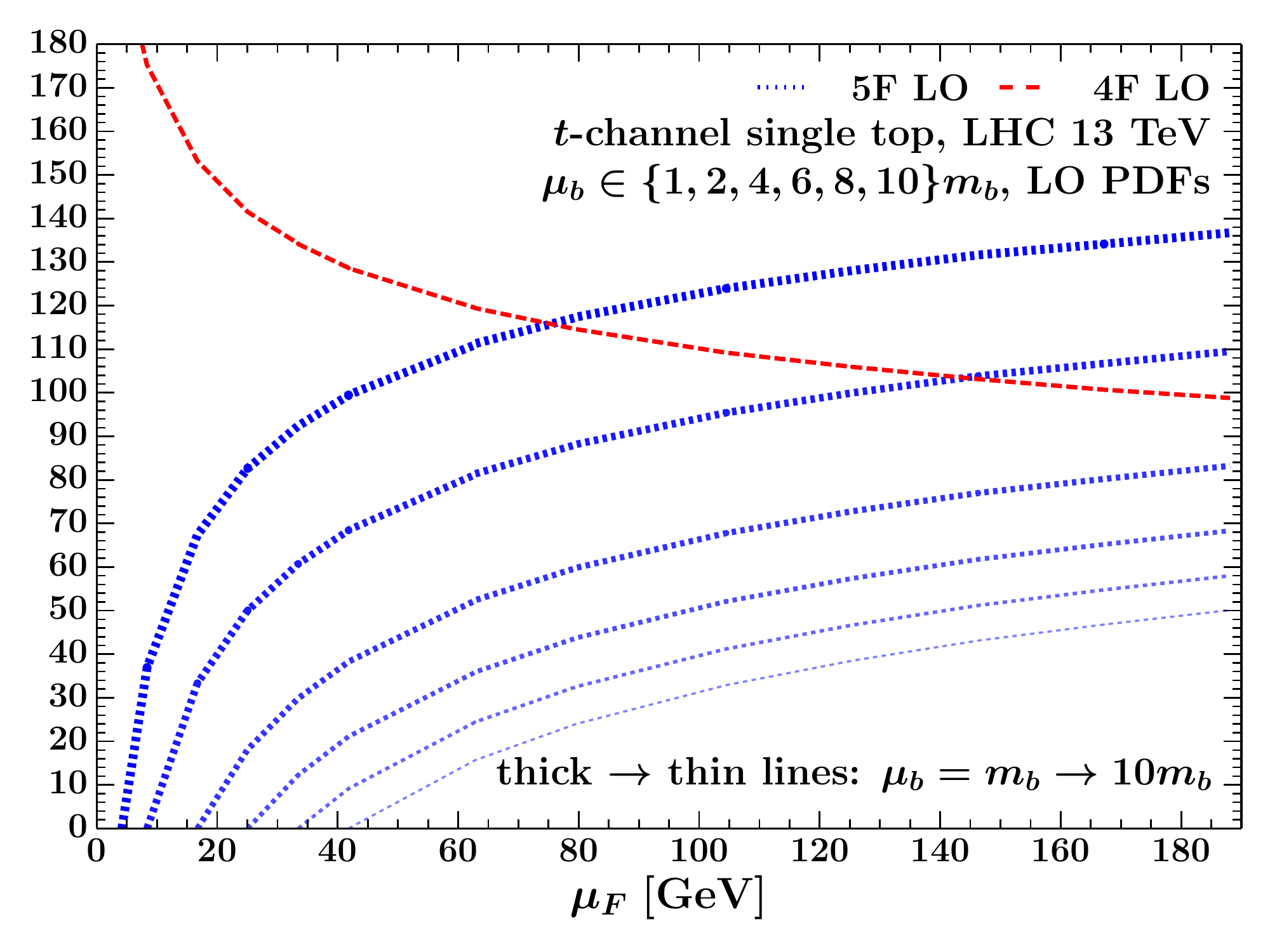}
\includegraphics[trim=0.4cm 0.0cm 0.2cm 0.4cm,clip,width=0.32\textwidth]{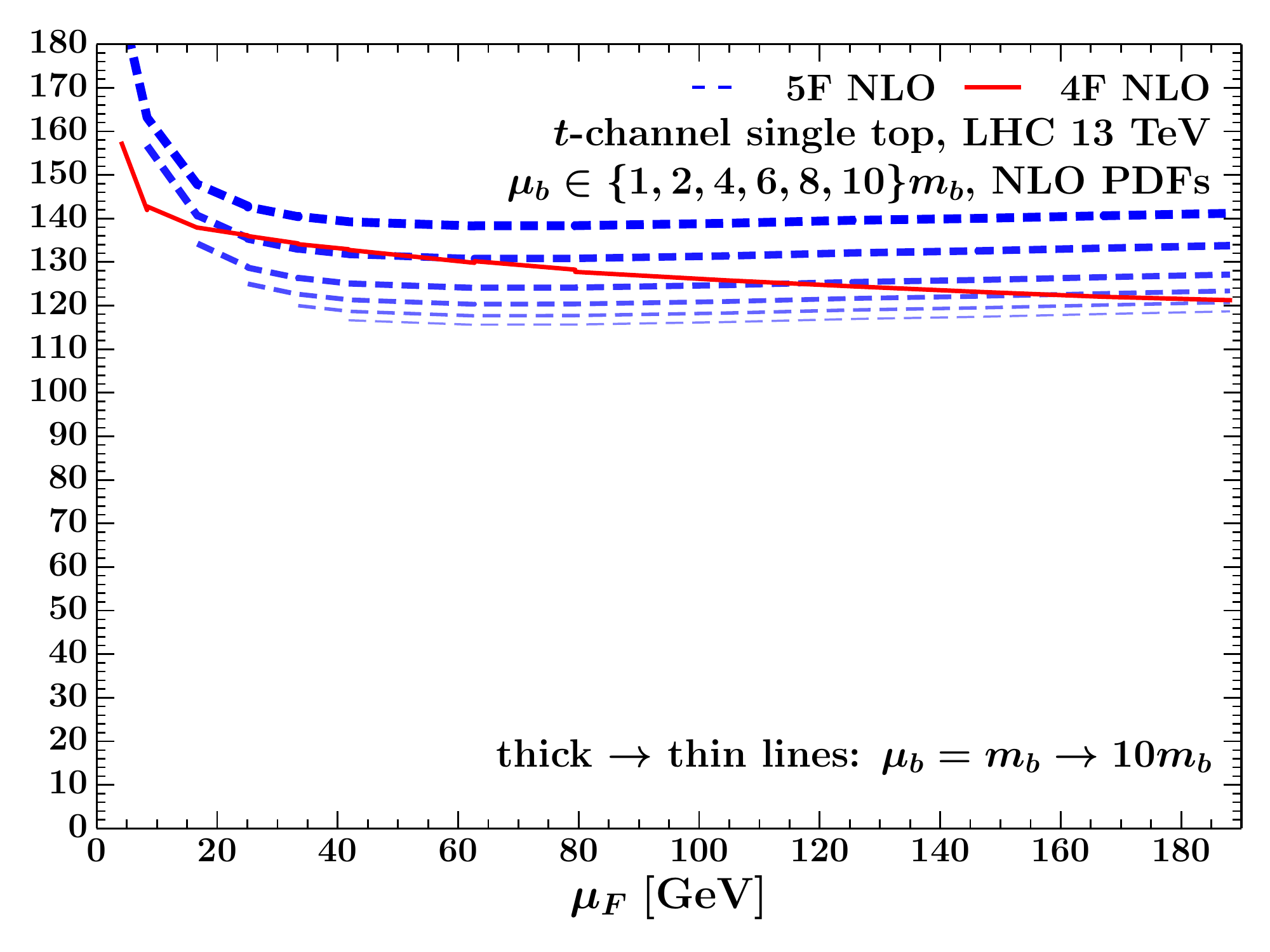}
\includegraphics[trim=0.4cm 0.0cm 0.2cm 0.4cm,clip,width=0.32\textwidth]{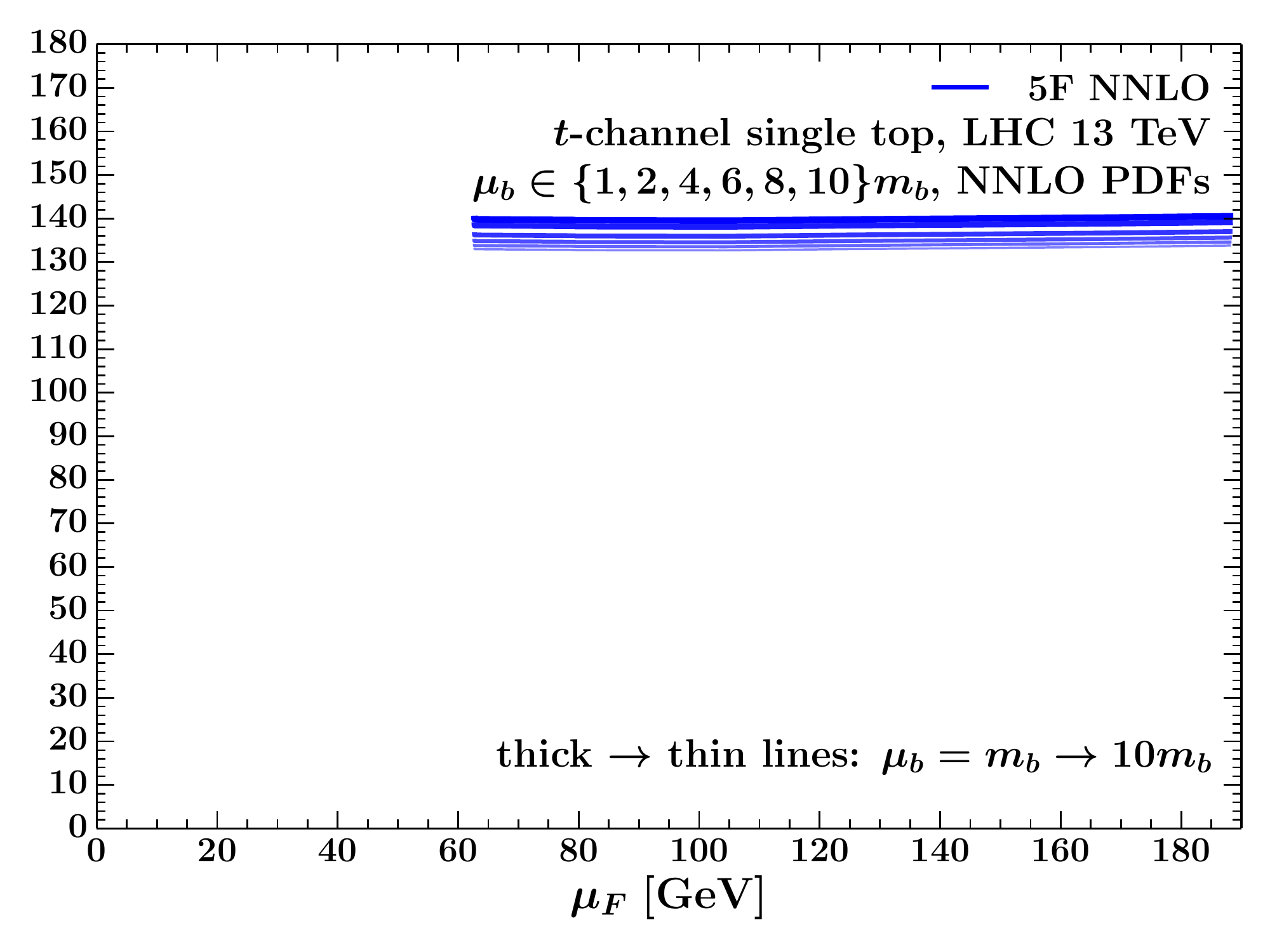}
\caption{Single top $t$-channel cross-section at LHC 13 TeV at LO (left), NLO (center) and NNLO (right) as a function of $\mu_F$ for several values of $\mub$.}
\label{fig:tj-disc}
\end{figure}

\begin{figure}[t]
\centering
\hskip-3pt
\includegraphics[trim=0.5cm 0.0cm 0.5cm 0.1cm,clip,width=0.33\textwidth]{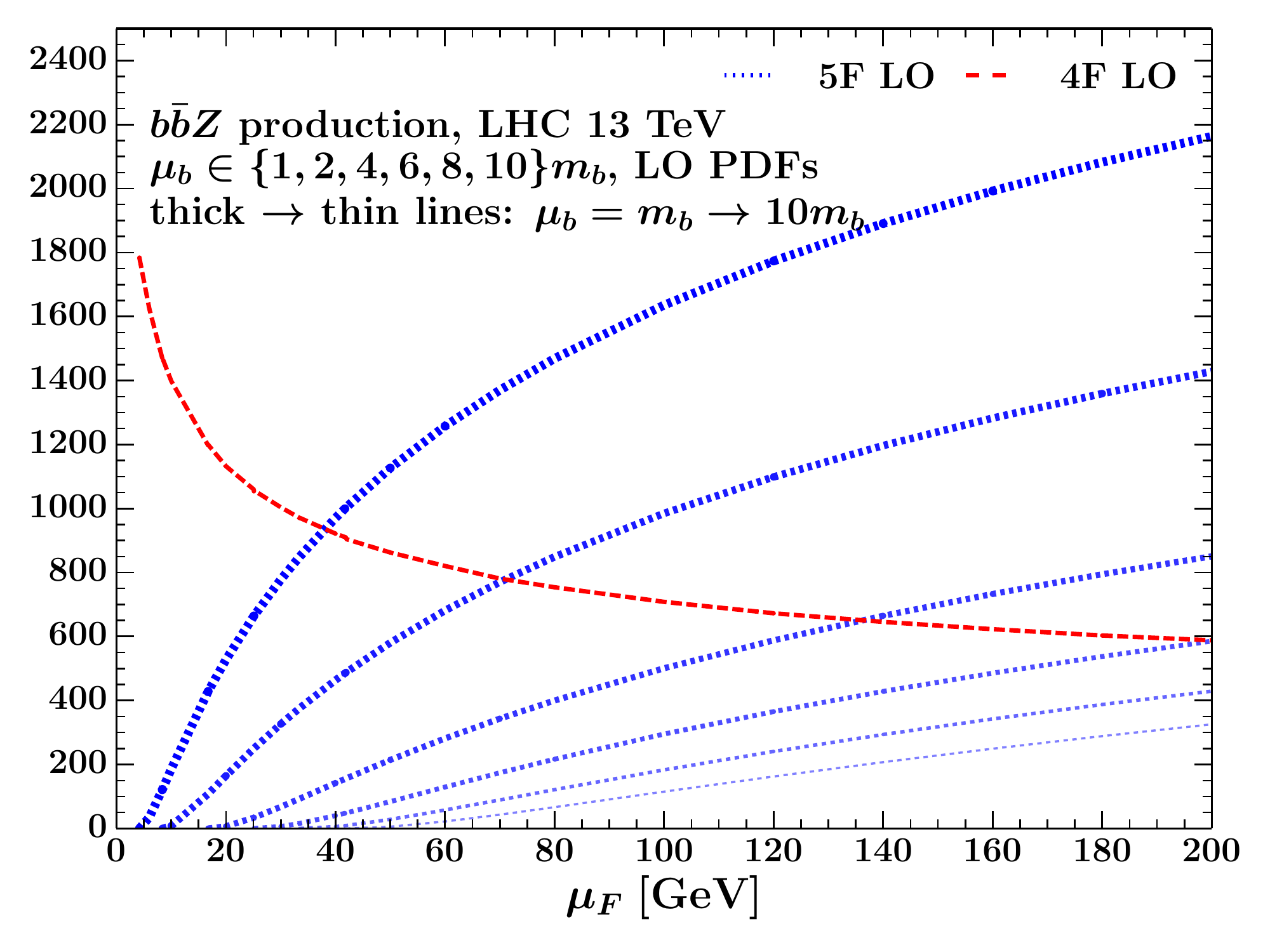} \hskip-3pt
\includegraphics[trim=0.5cm 0.0cm 0.5cm 0.1cm,clip,width=0.33\textwidth]{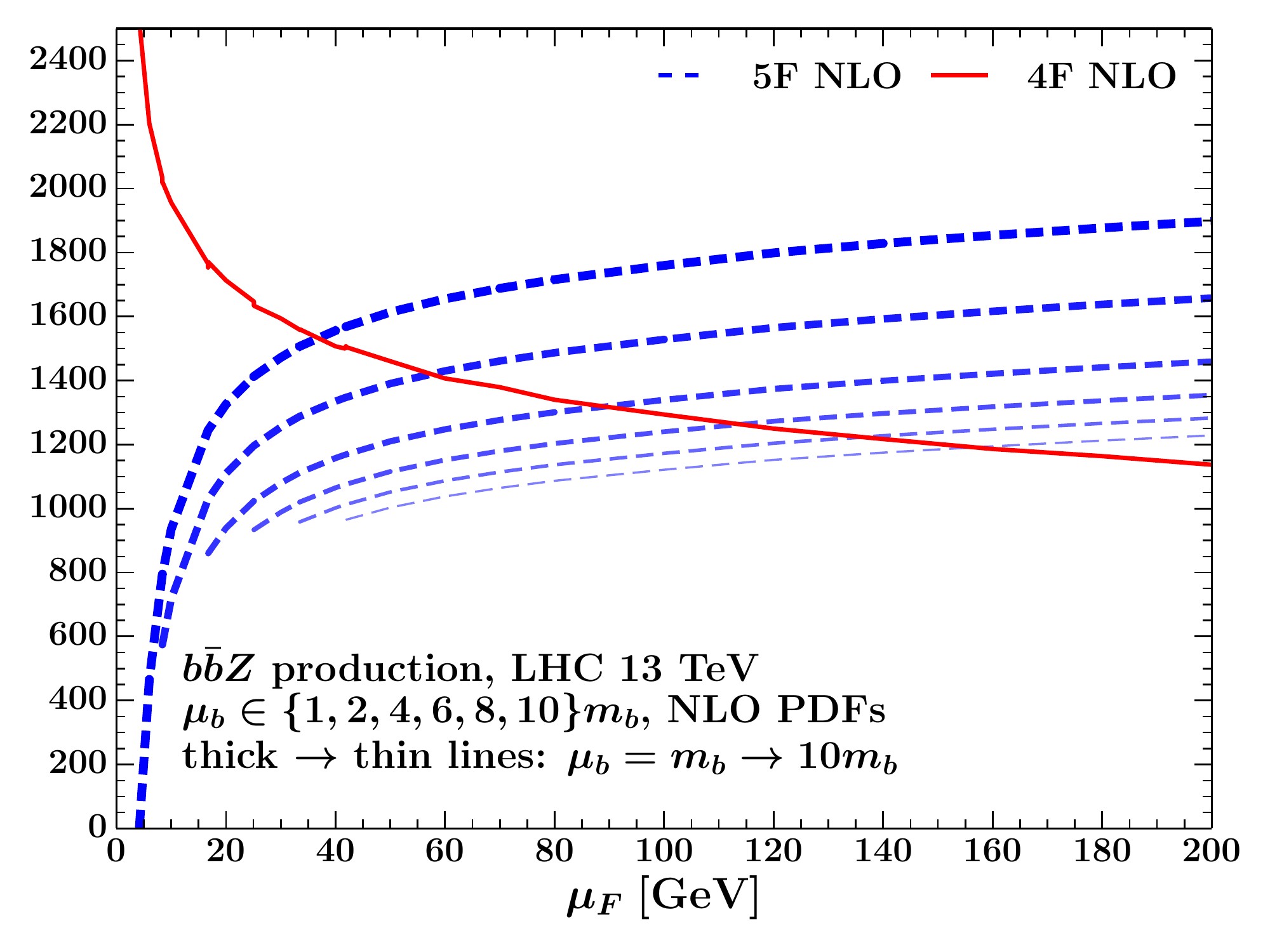} \hskip-3pt
\includegraphics[trim=0.5cm 0.0cm 0.5cm 0.1cm,clip,width=0.33\textwidth]{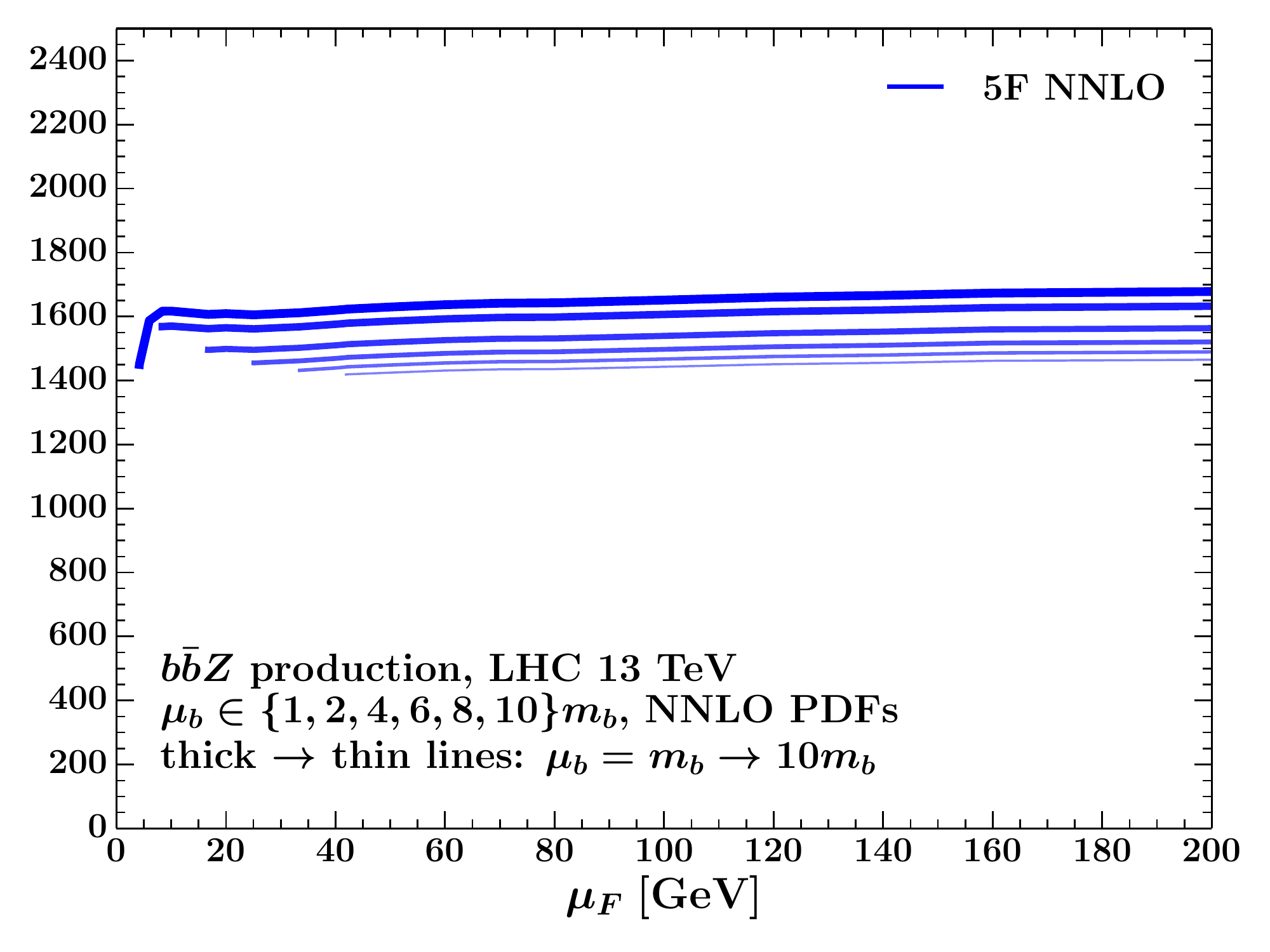}
\caption{As in fig.~\ref{fig:tj-disc} but for the total $b\bar{b}Z$ production cross-section.}
\label{fig:zbb-xs-mub-dep}
\end{figure}

\begin{figure}[t]
\centering
\includegraphics[trim=0.2cm 0.0cm 0.4cm 0.1cm,clip,width=0.49\textwidth]{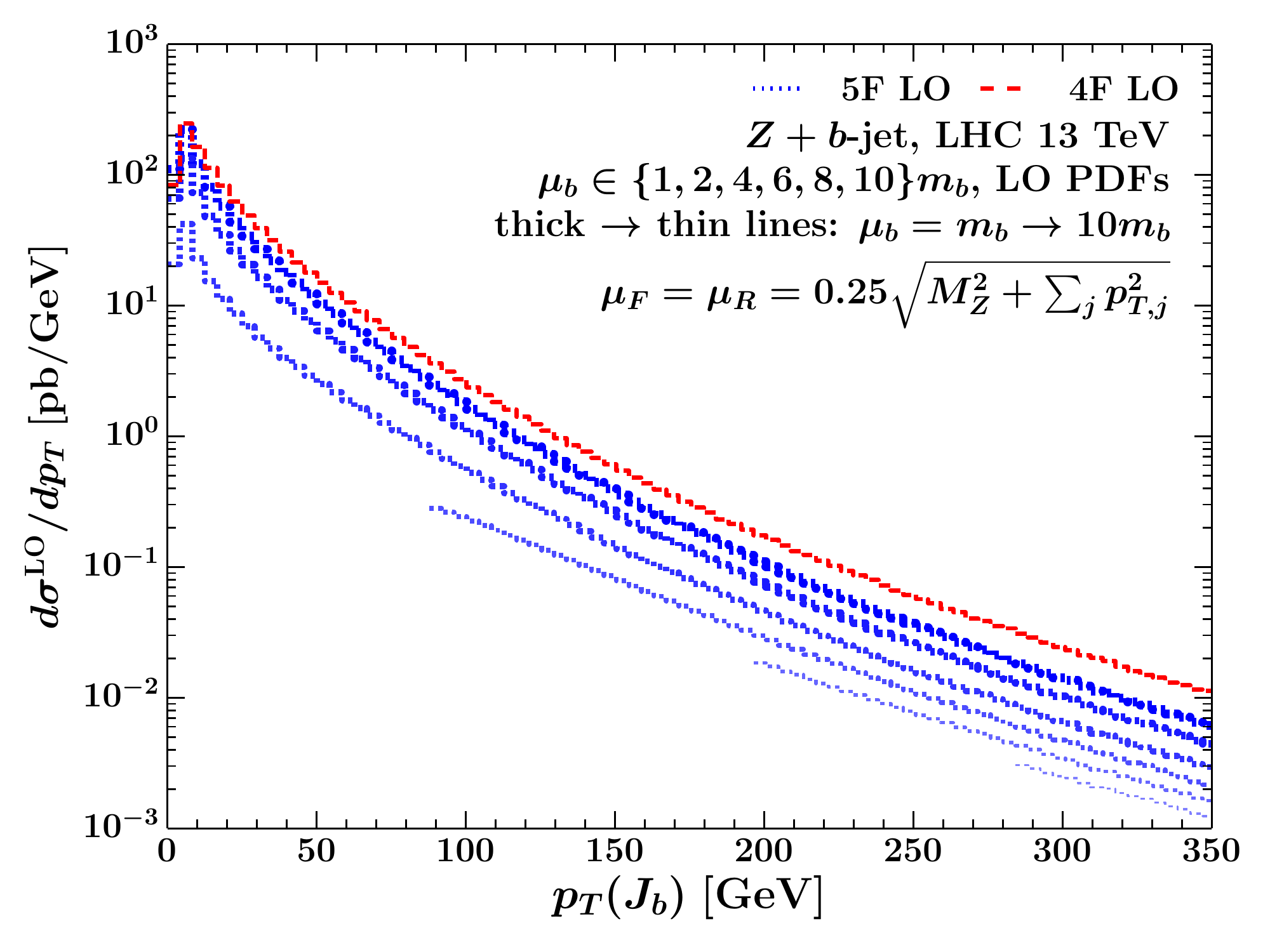} 
\includegraphics[trim=0.2cm 0.0cm 0.4cm 0.1cm,clip,width=0.49\textwidth]{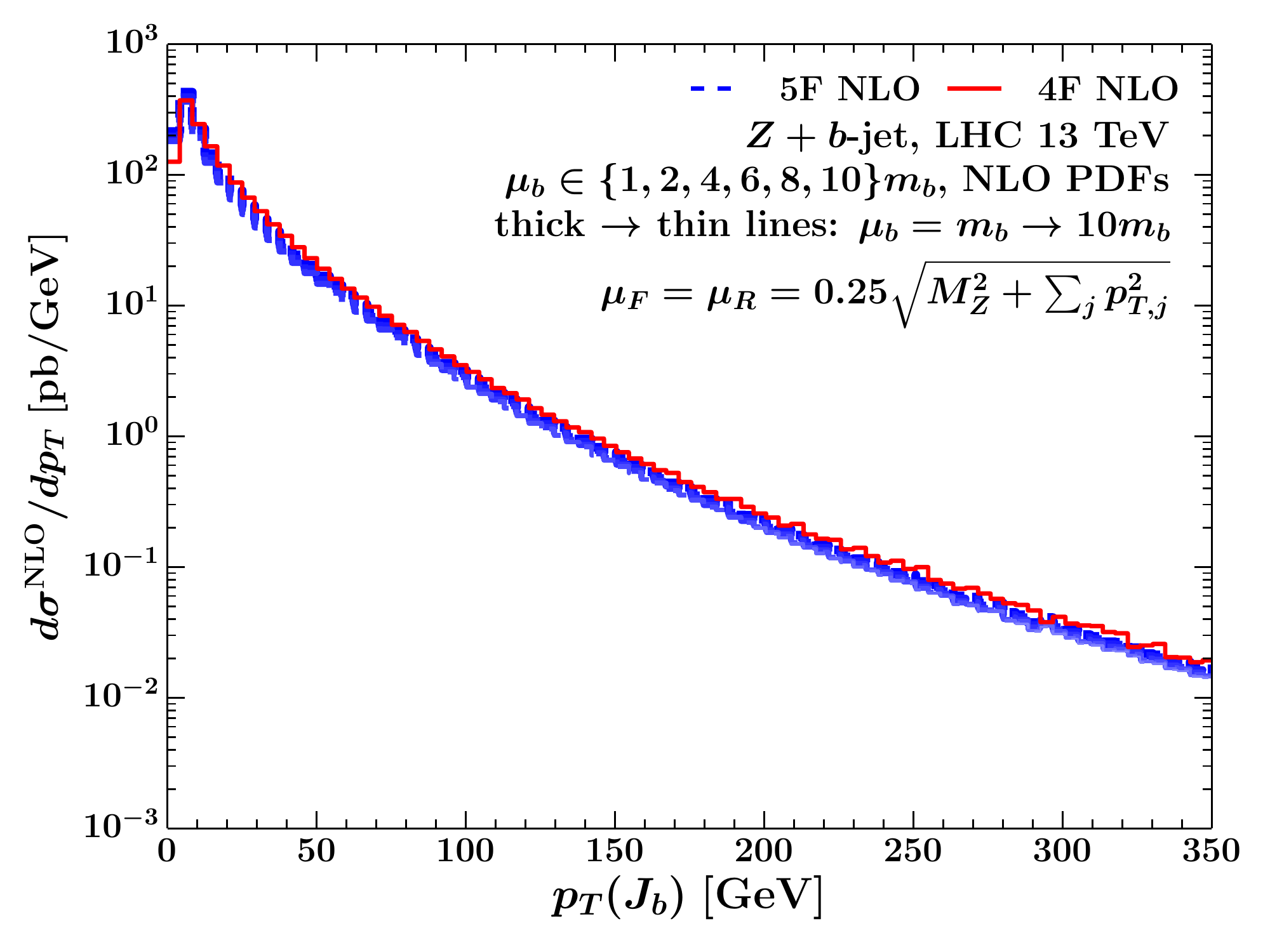}
\caption{The differential cross-section of $Z+b$-jet at LHC 13 TeV as a function of $p_T(J_b)$ at LO (left) and NLO (right) for several values of $\mub$.}
\label{fig:zbj-xs-mub-dep}
\end{figure}

We first consider the cross-sections with absolute normalization. They are plotted in figs.~\ref{fig:tj-disc}, \ref{fig:zbb-xs-mub-dep}, \ref{fig:zbj-xs-mub-dep} in both 4FS (red line) and 5FS, the latter computed for a range of $\mub$ values (blue lines). For all processes we present the predictions of orders (LO,LO) and (NLO,NLO) while for $t$-channel single top and $b\bar{b}Z$ we show the (NNLO,NNLO) prediction for the 5FS only (the 4FS cross-sections are not known at NNLO). With $({\rm N}^n{\rm LO},{\rm N}^m{\rm LO})$ we denote the order of a hadronic cross-section computed with perturbative cross-section accurate at order ${\rm N}^n{\rm LO}$ and pdf of order ${\rm N}^m{\rm LO}$. Unless specified otherwise, when we say a cross-section is of order ${\rm N}^n{\rm LO}$ we mean $({\rm N}^n{\rm LO},{\rm N}^n{\rm LO})$, i.e. the perturbative part and pdfs are of the same order.
%
%
%

A number of generic features can be observed in figs.~\ref{fig:tj-disc}, \ref{fig:zbb-xs-mub-dep}, \ref{fig:zbj-xs-mub-dep}. First we notice that by going to higher orders the scale dependence decreases uniformly for all values of $\mub$. While this result should be expected, it is a nice check that the approach works as anticipated. Second and more important, all 5FS curves that correspond to different values of $\mub$ become more and more tightly packed together as we go to higher perturbative orders. This means that at higher orders the predicted cross-sections become less sensitive to the position of $\mub$. These two observations apply universally for all three different processes.

\begin{figure}[t]
\centering
\includegraphics[trim=0.3cm 0.0cm 0.4cm 0.5cm,clip,width=0.32\textwidth]{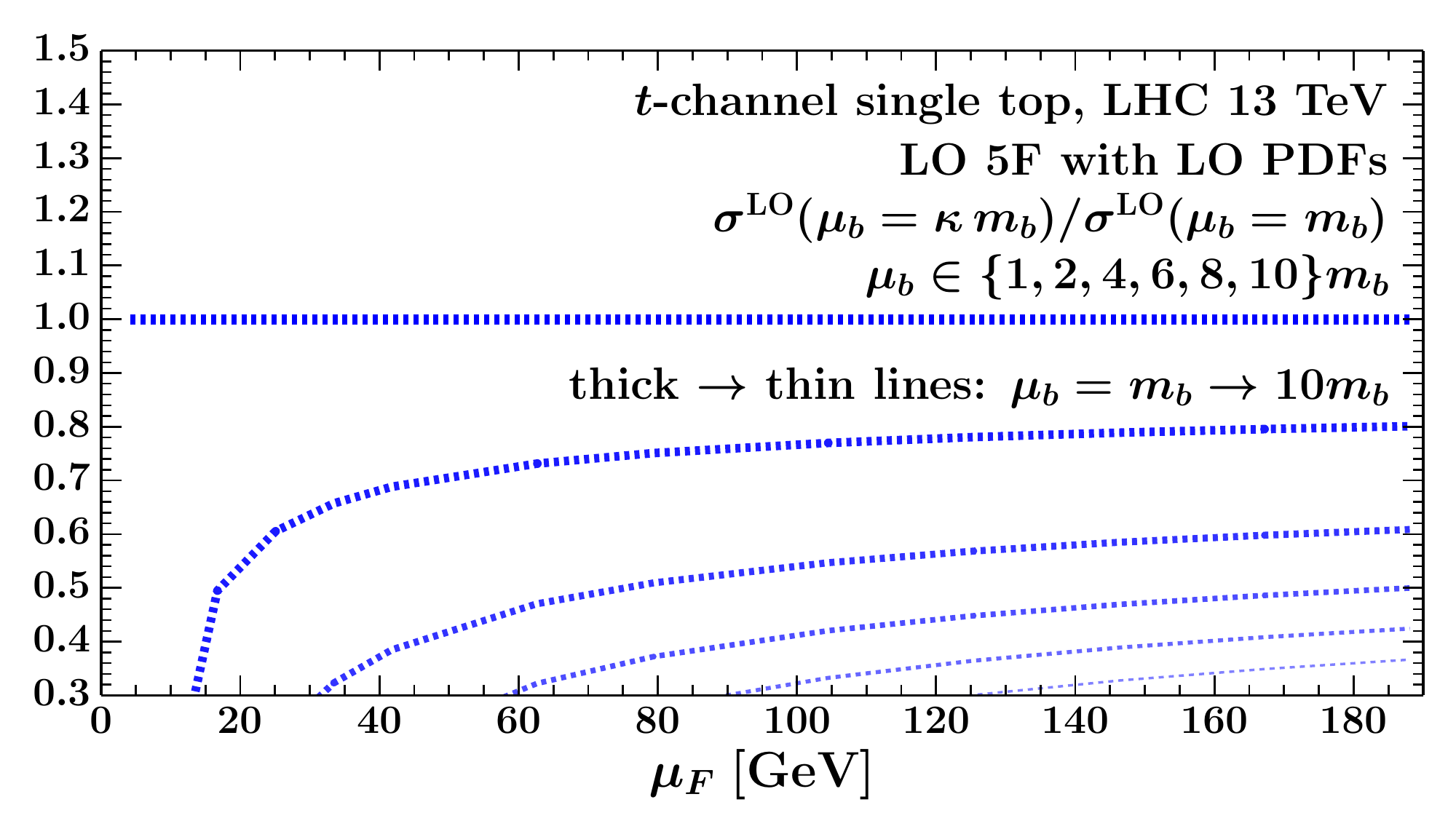}
\includegraphics[trim=0.3cm 0.0cm 0.4cm 0.5cm,clip,width=0.32\textwidth]{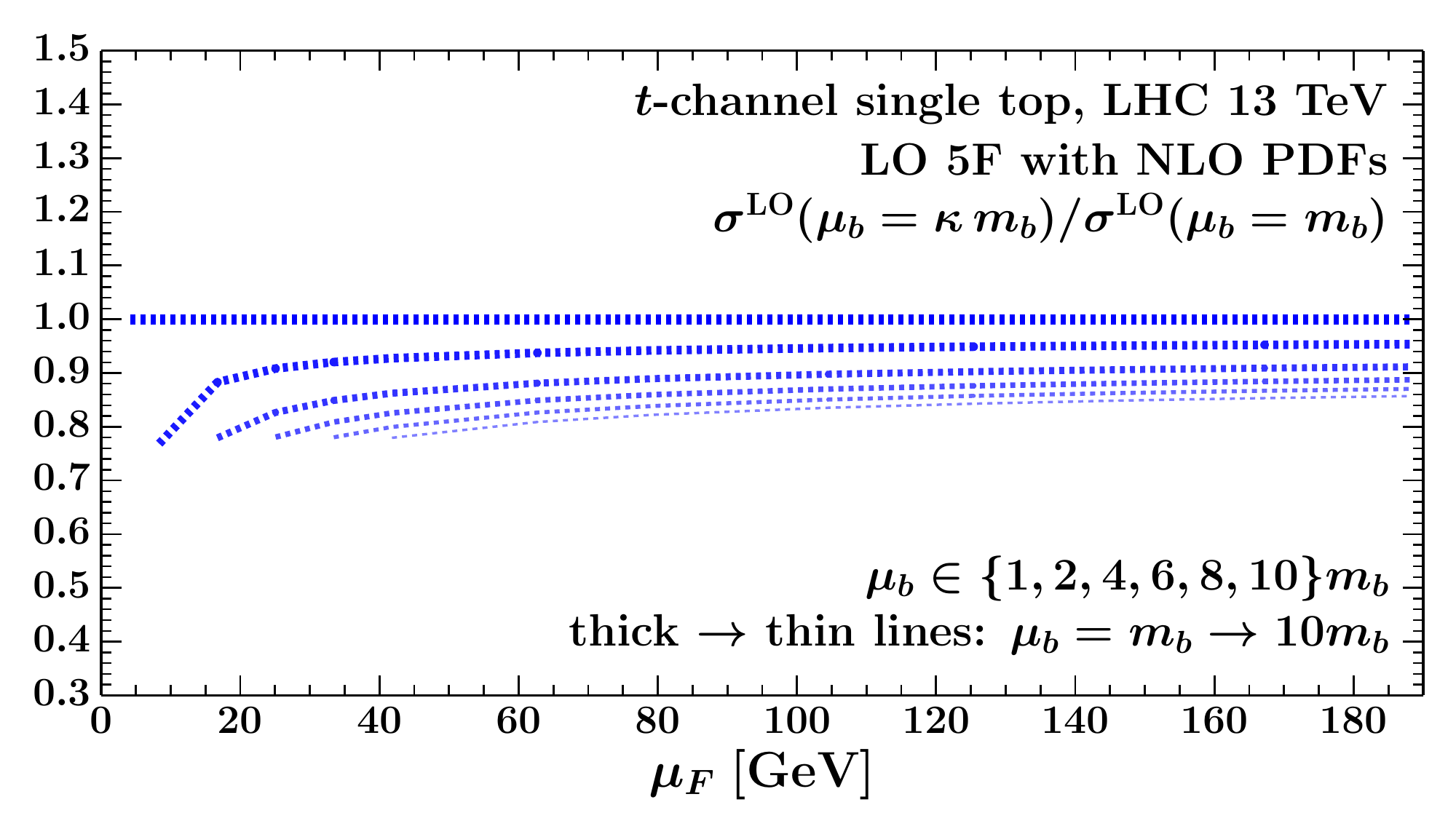}
\includegraphics[trim=0.3cm 0.0cm 0.4cm 0.5cm,clip,width=0.32\textwidth]{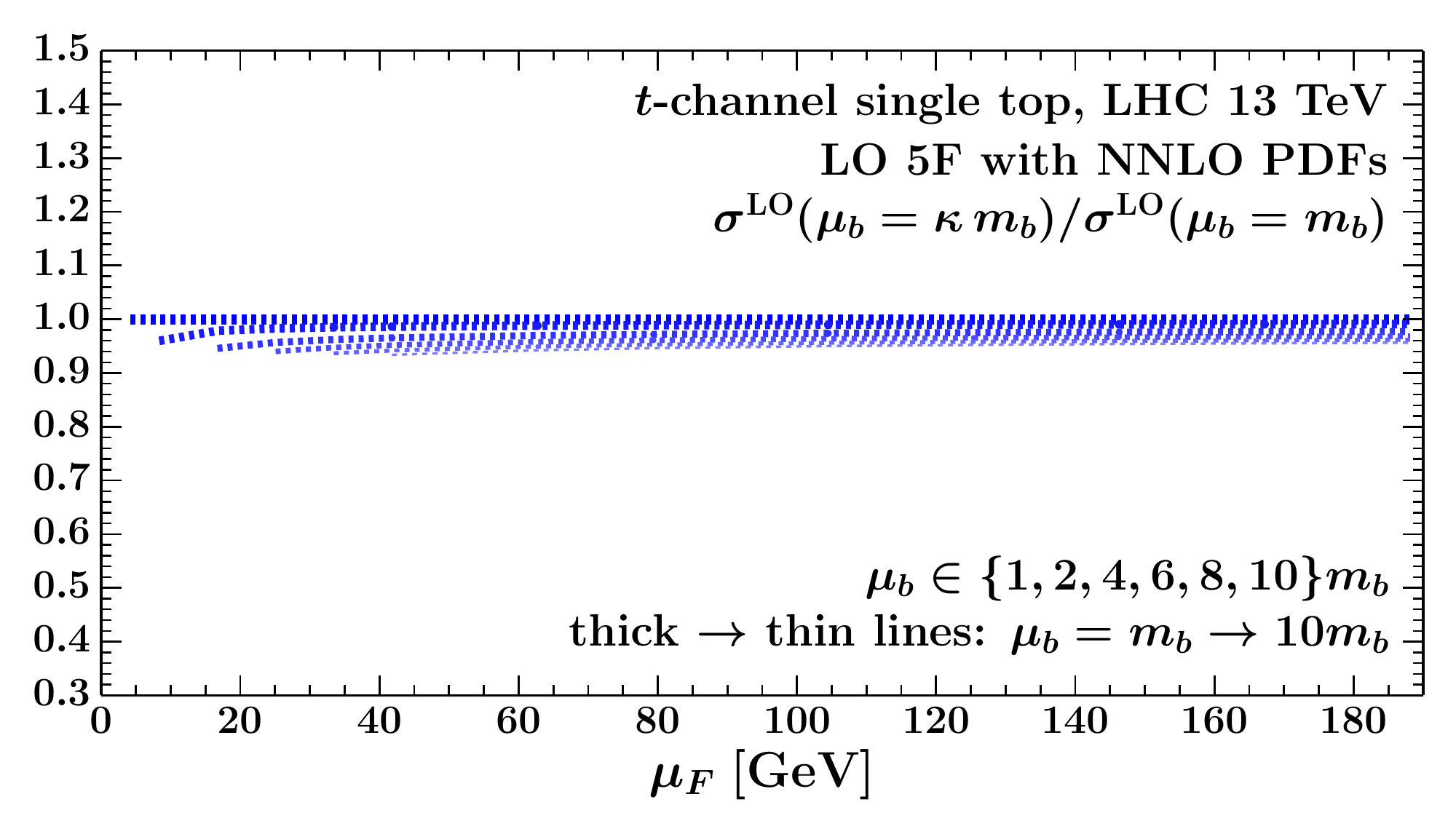} \\
\vskip -3.5mm
\hskip 51mm
\includegraphics[trim=0.3cm 0.0cm 0.4cm 0.5cm,clip,width=0.32\textwidth]{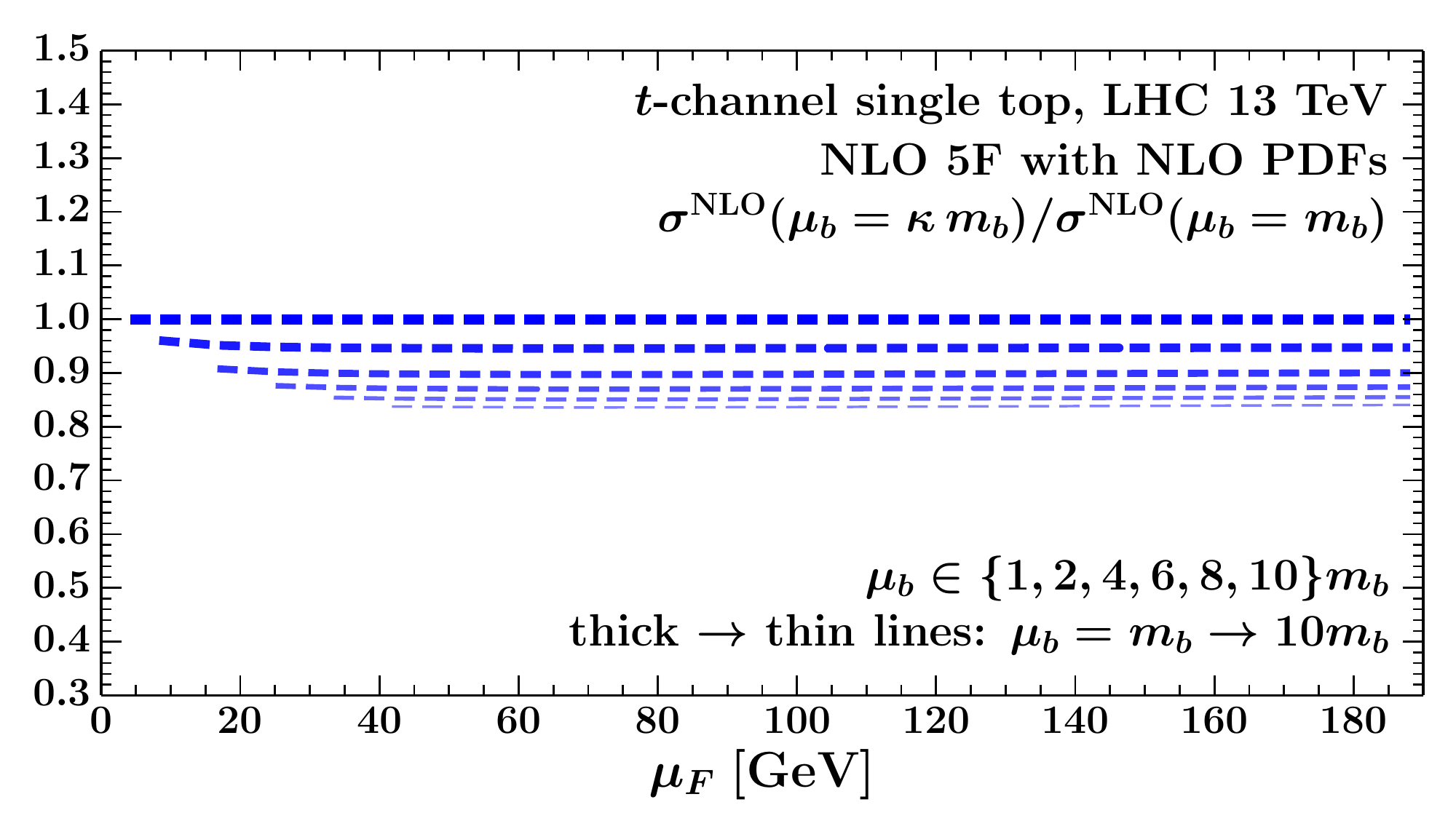}
\includegraphics[trim=0.3cm 0.0cm 0.4cm 0.5cm,clip,width=0.32\textwidth]{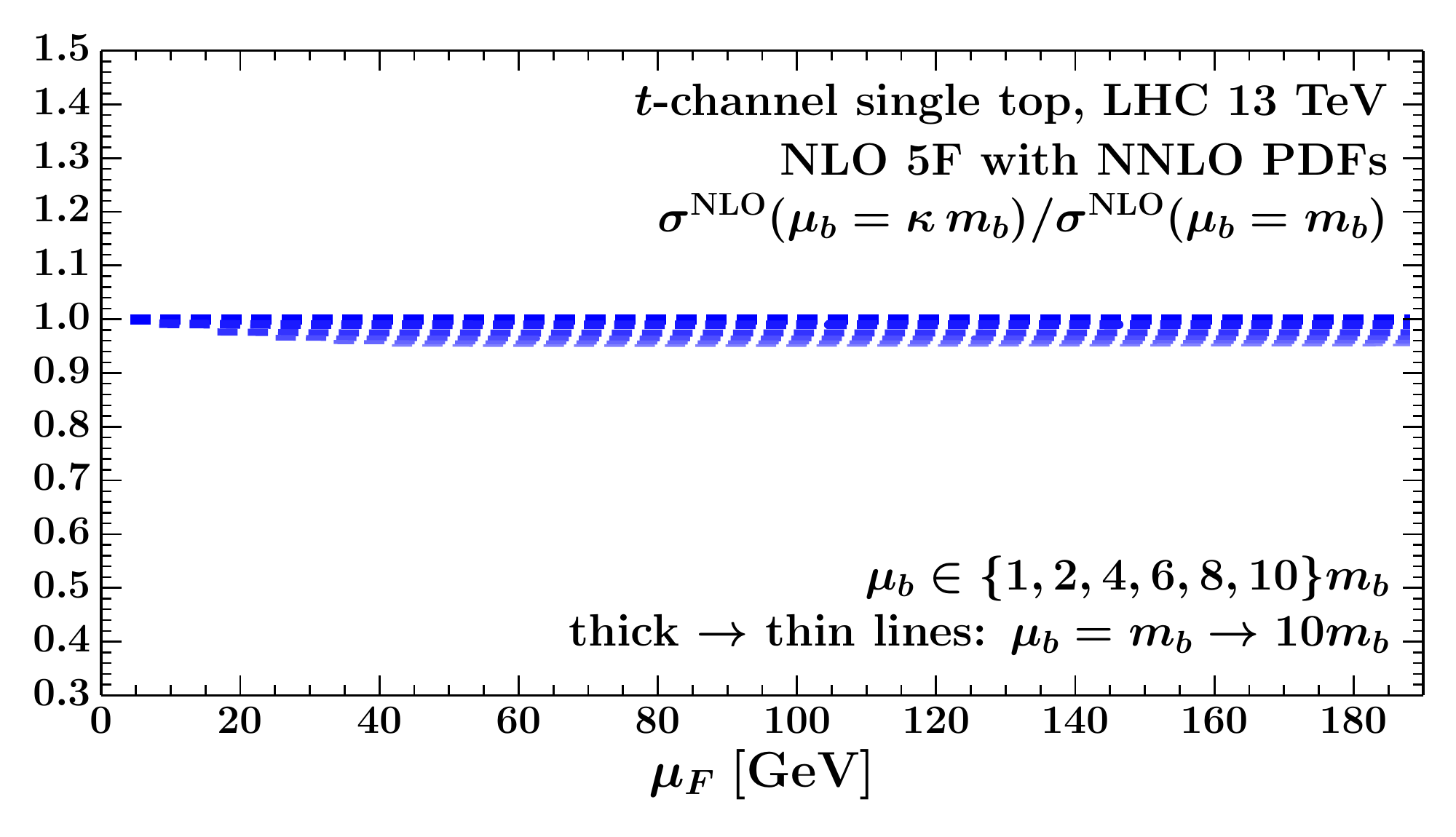} \\
\vskip -2mm
\caption{The ratio eq.~(\ref{eq:ratio}) as a function of $\mu_F$ for $t$-channel single top at LHC 13 TeV.}
\label{fig:tj-mub-dep}
\end{figure}

\begin{figure}[t]
\centering
\includegraphics[trim=0.3cm 0.0cm 0.5cm 0.5cm,clip,width=0.32\textwidth]{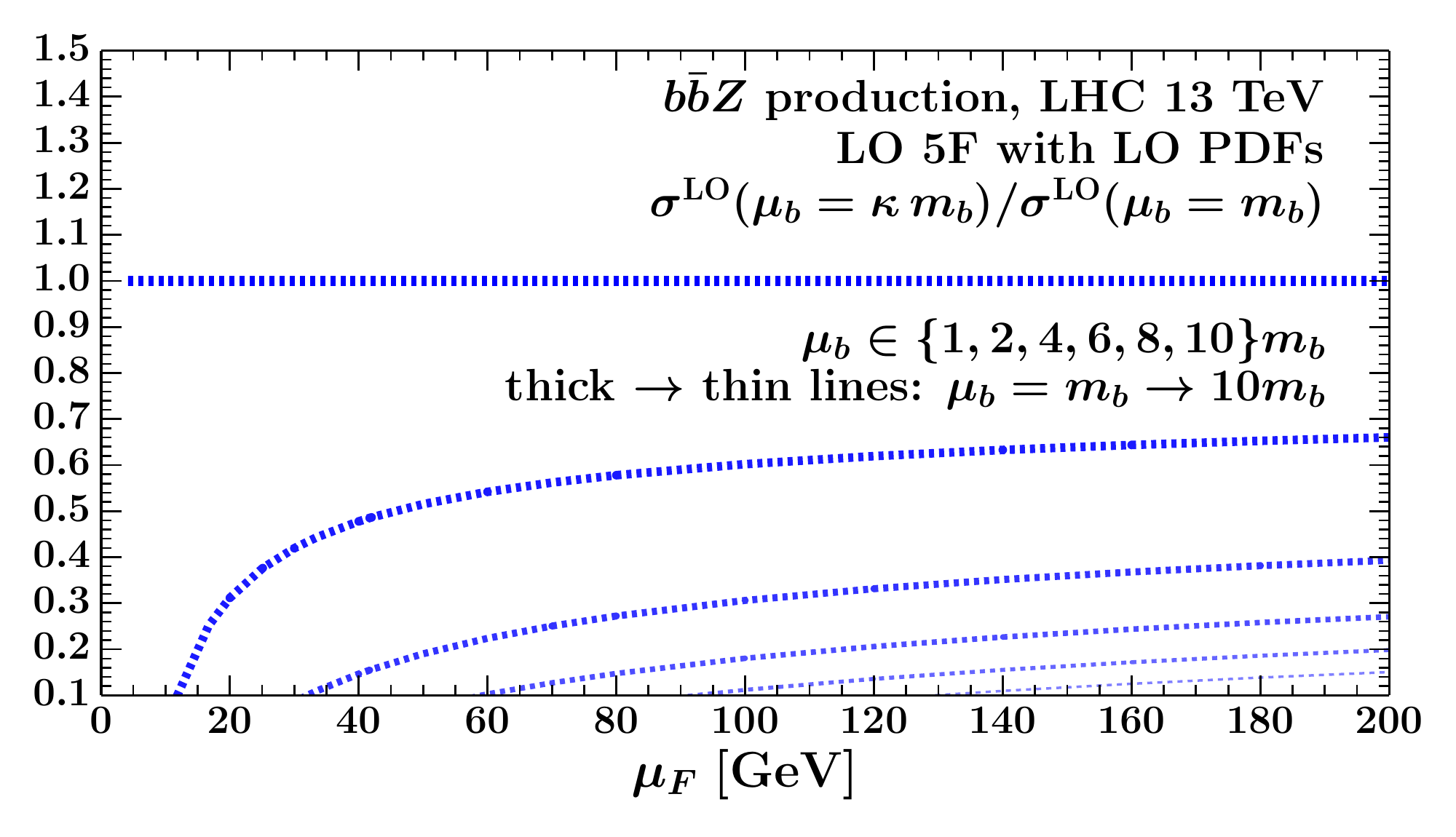}
\includegraphics[trim=0.3cm 0.0cm 0.5cm 0.5cm,clip,width=0.32\textwidth]{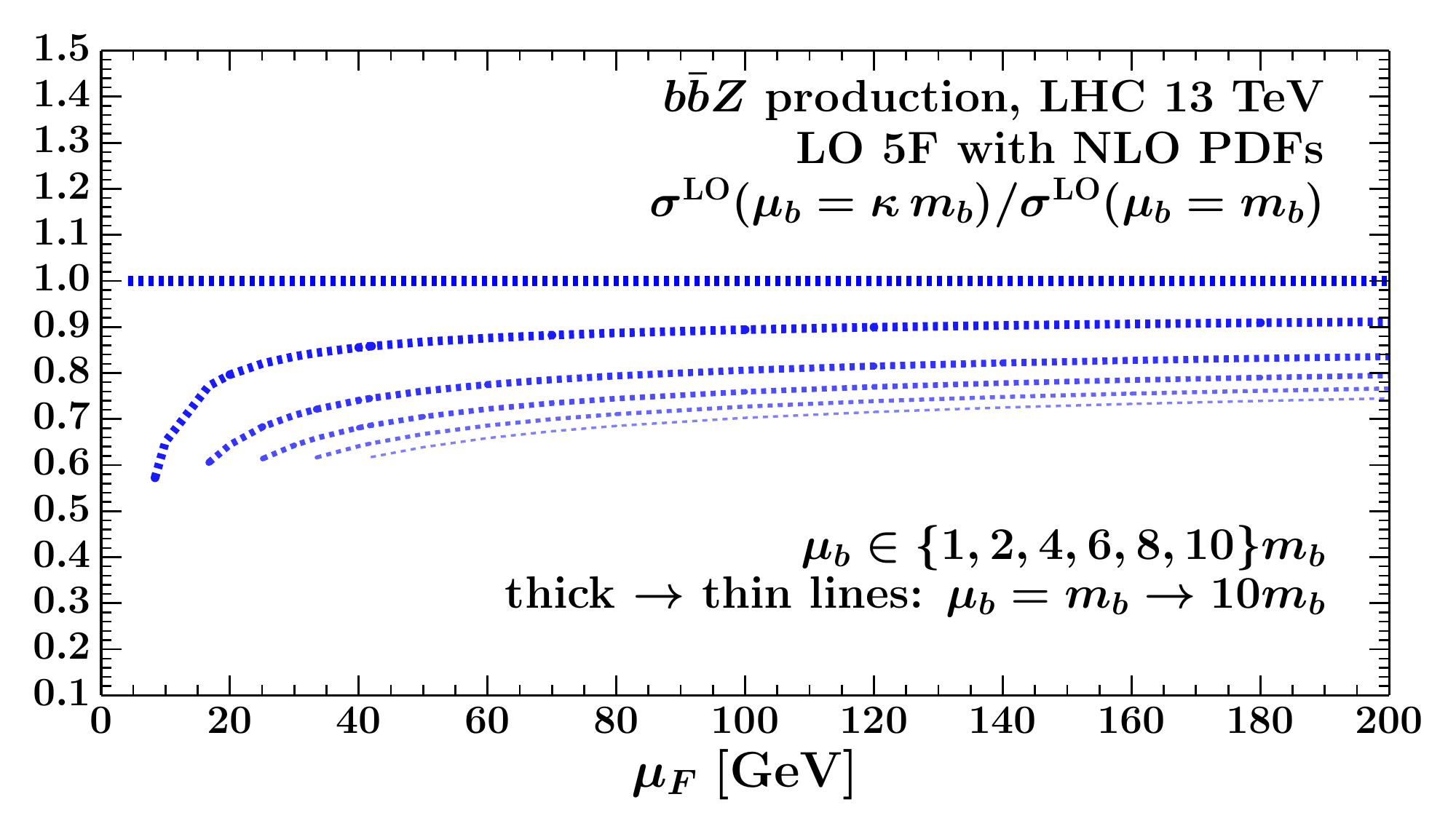}
\includegraphics[trim=0.3cm 0.0cm 0.5cm 0.5cm,clip,width=0.32\textwidth]{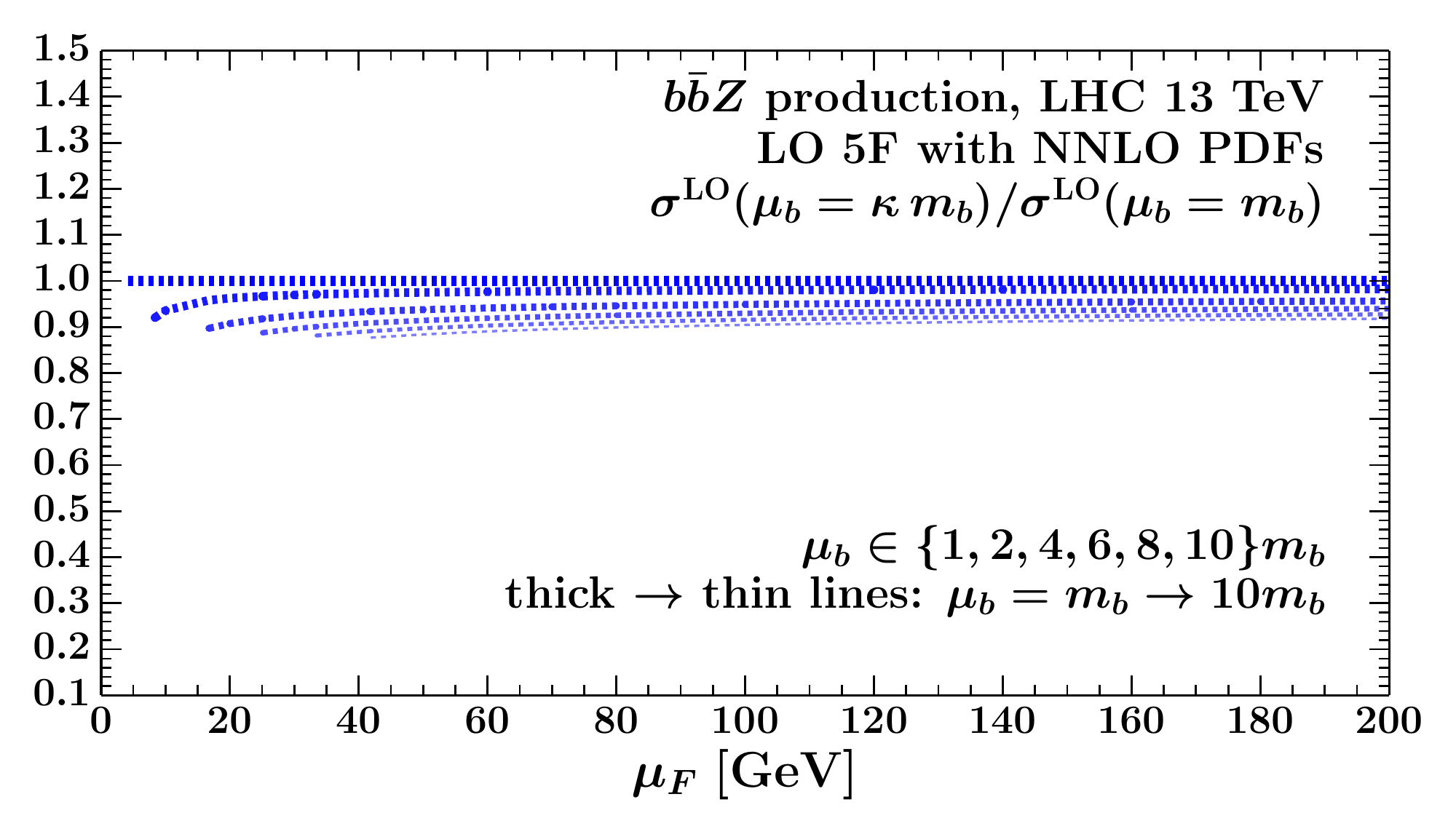}\\
\vskip -3.5mm
\hskip 51mm
\includegraphics[trim=0.3cm 0.0cm 0.5cm 0.5cm,clip,width=0.32\textwidth]{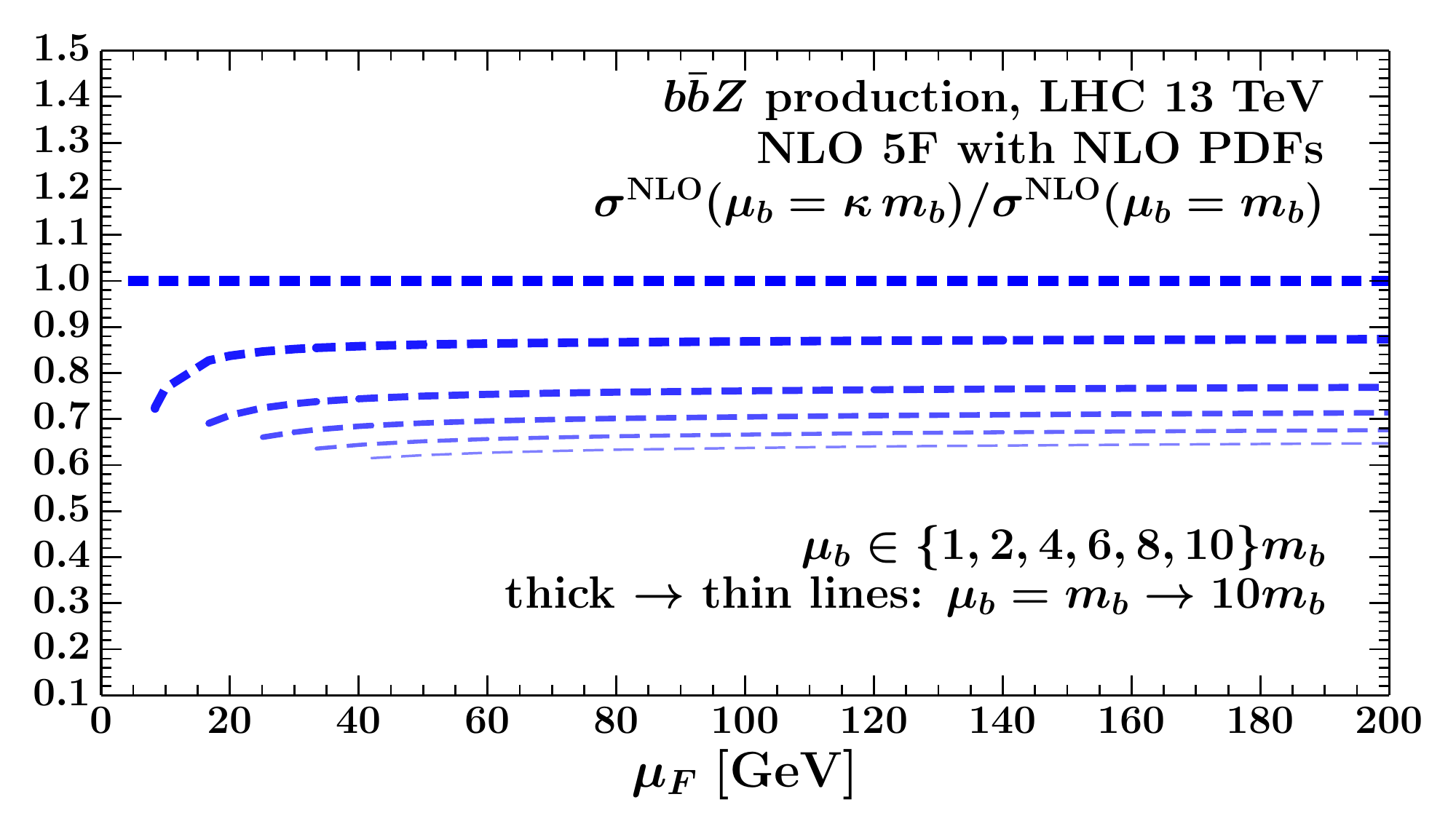}
\includegraphics[trim=0.3cm 0.0cm 0.5cm 0.5cm,clip,width=0.32\textwidth]{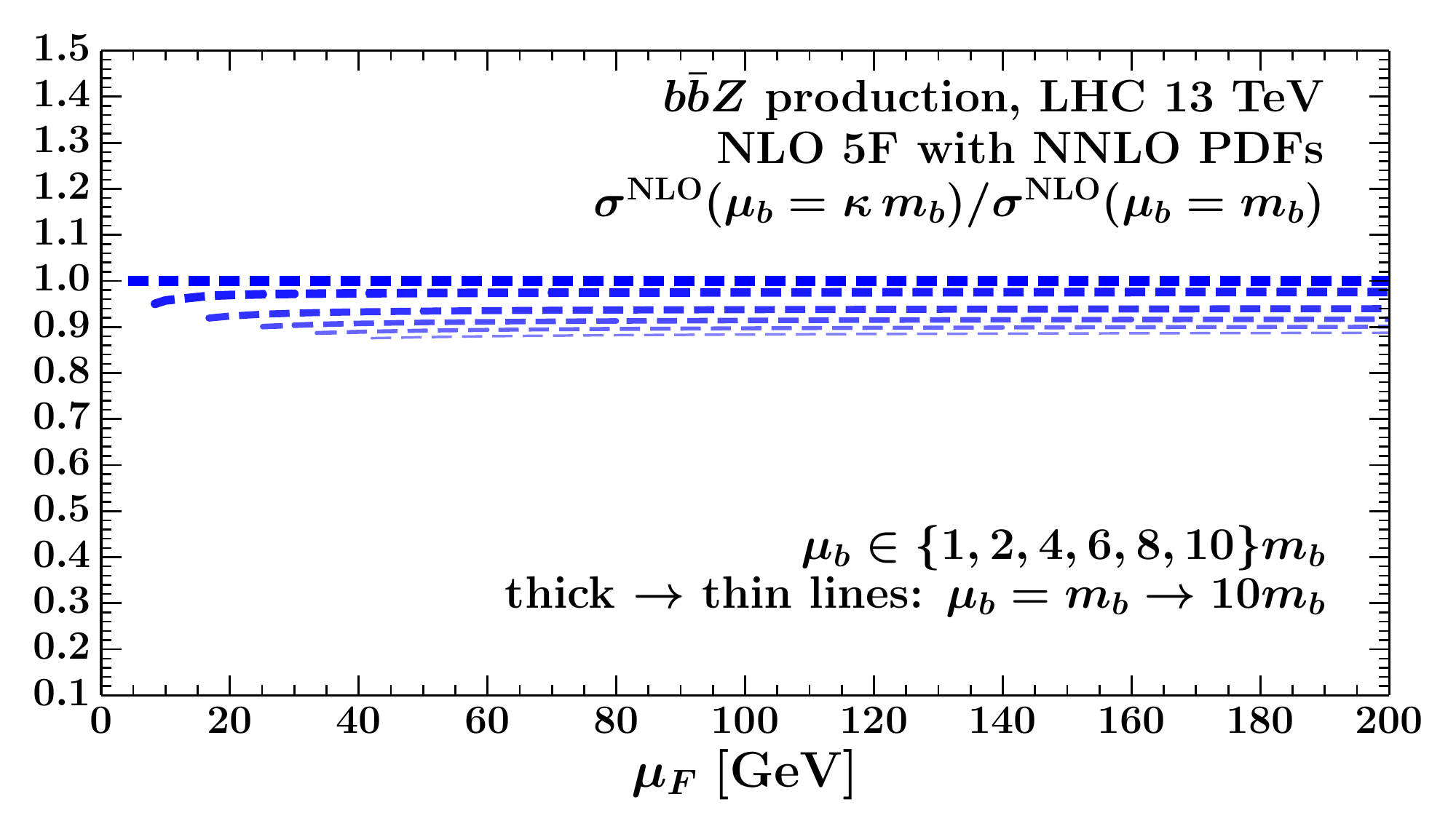} \\
\vskip -3.5mm
\hskip 102mm
\includegraphics[trim=0.3cm 0.0cm 0.5cm 0.5cm,clip,width=0.32\textwidth]{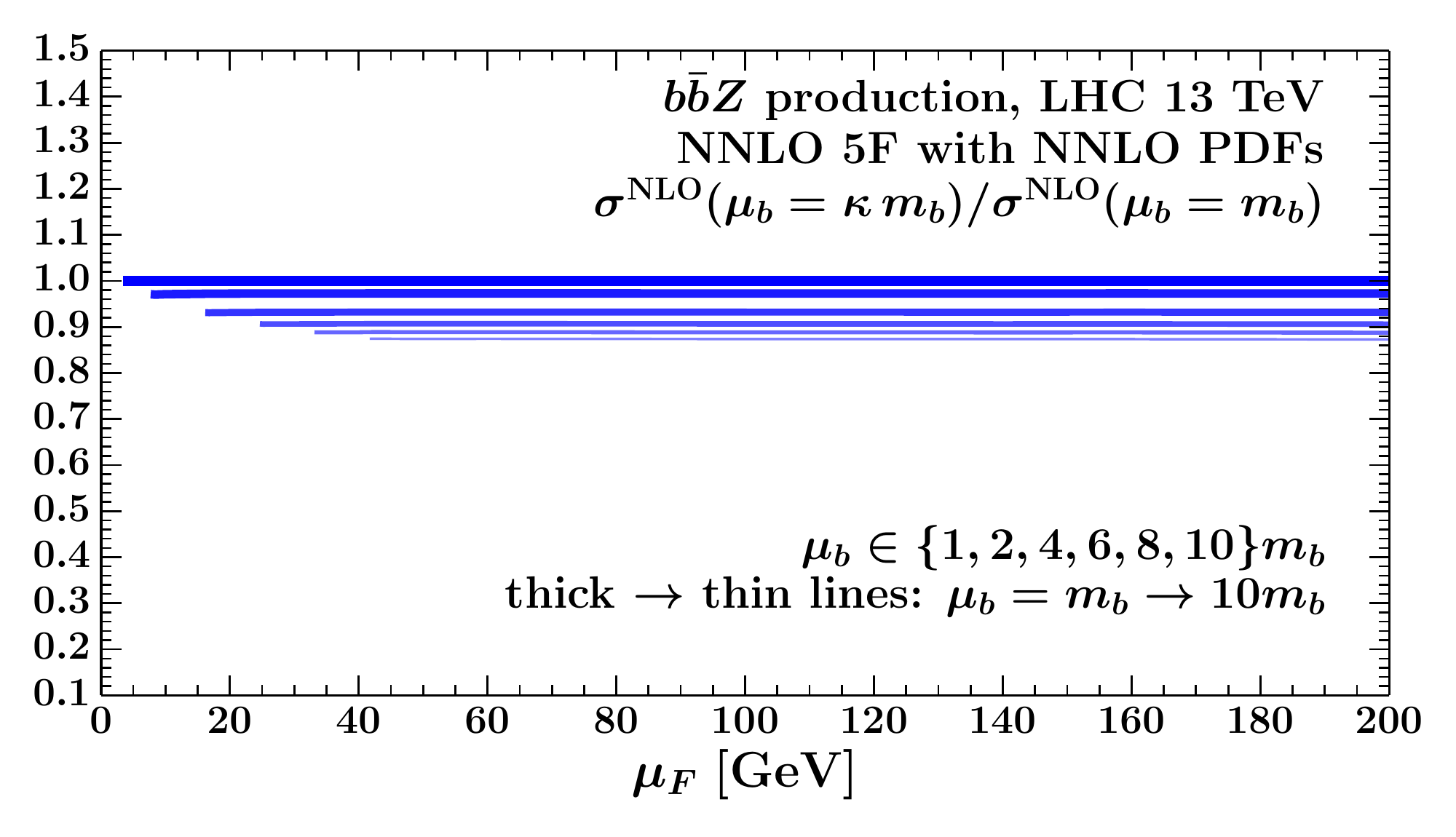} \\
\vskip -2mm
\caption{The ratio eq.~(\ref{eq:ratio}) as a function of $\mu_F$ for $b\bar{b}Z$ production at LHC 13 TeV.}
\label{fig:bbz-mub-dep}
\end{figure}

\begin{figure}[t]
\centering
\includegraphics[trim=0.45cm 0.0cm 0.5cm 0.1cm,clip,width=0.32\textwidth]{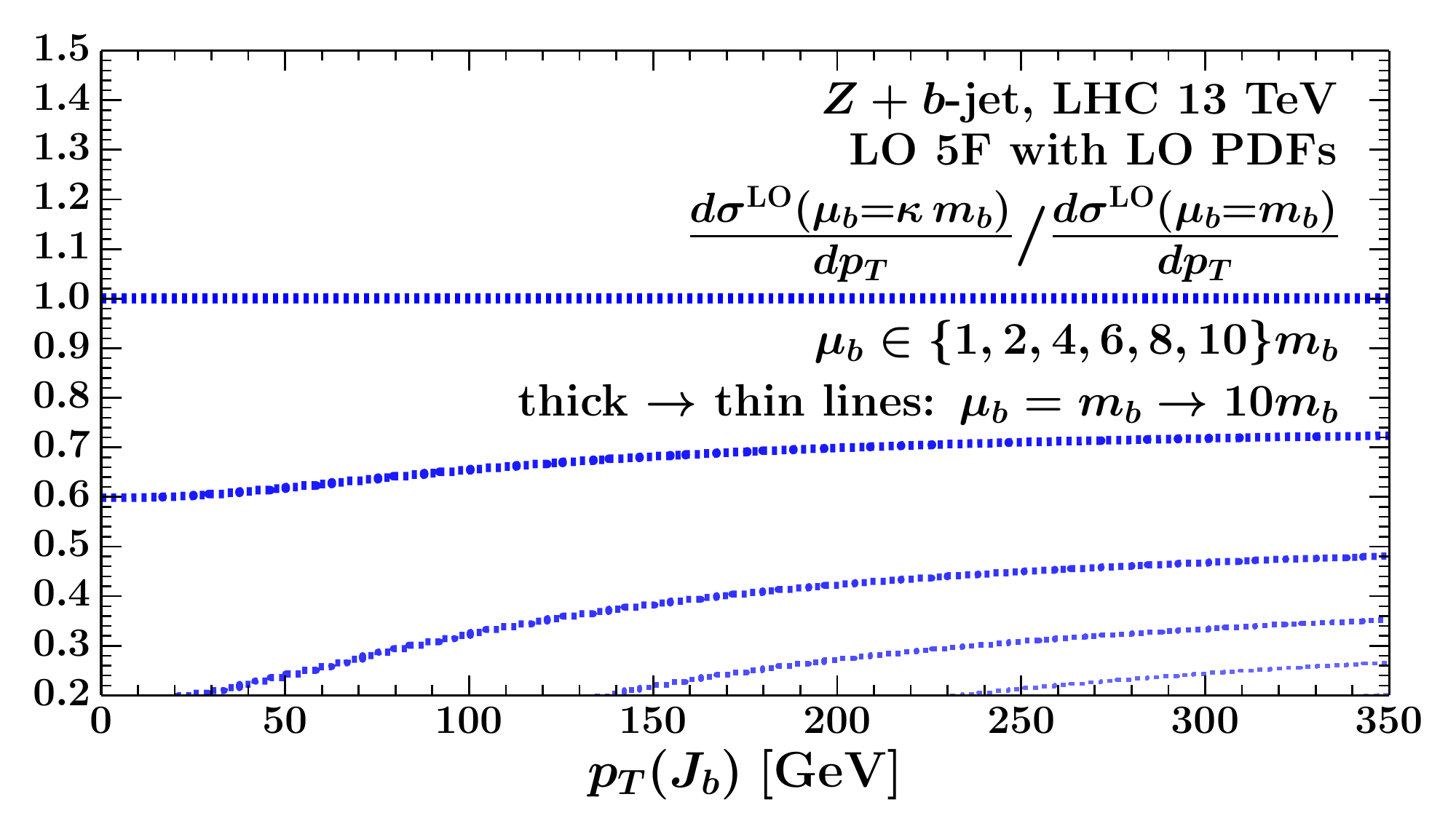}
\includegraphics[trim=0.45cm 0.0cm 0.5cm 0.1cm,clip,width=0.32\textwidth]{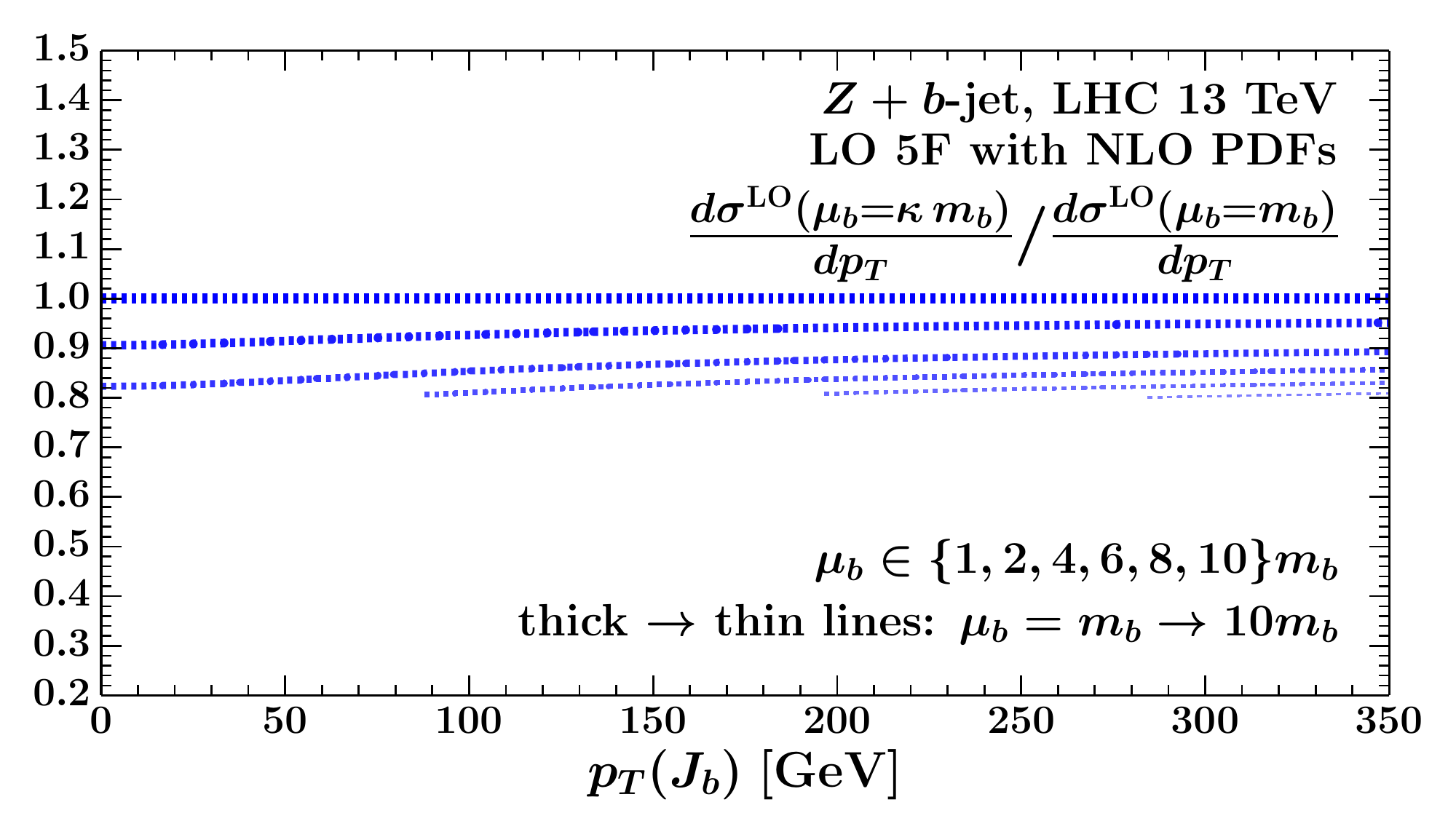}
\includegraphics[trim=0.45cm 0.0cm 0.5cm 0.1cm,clip,width=0.32\textwidth]{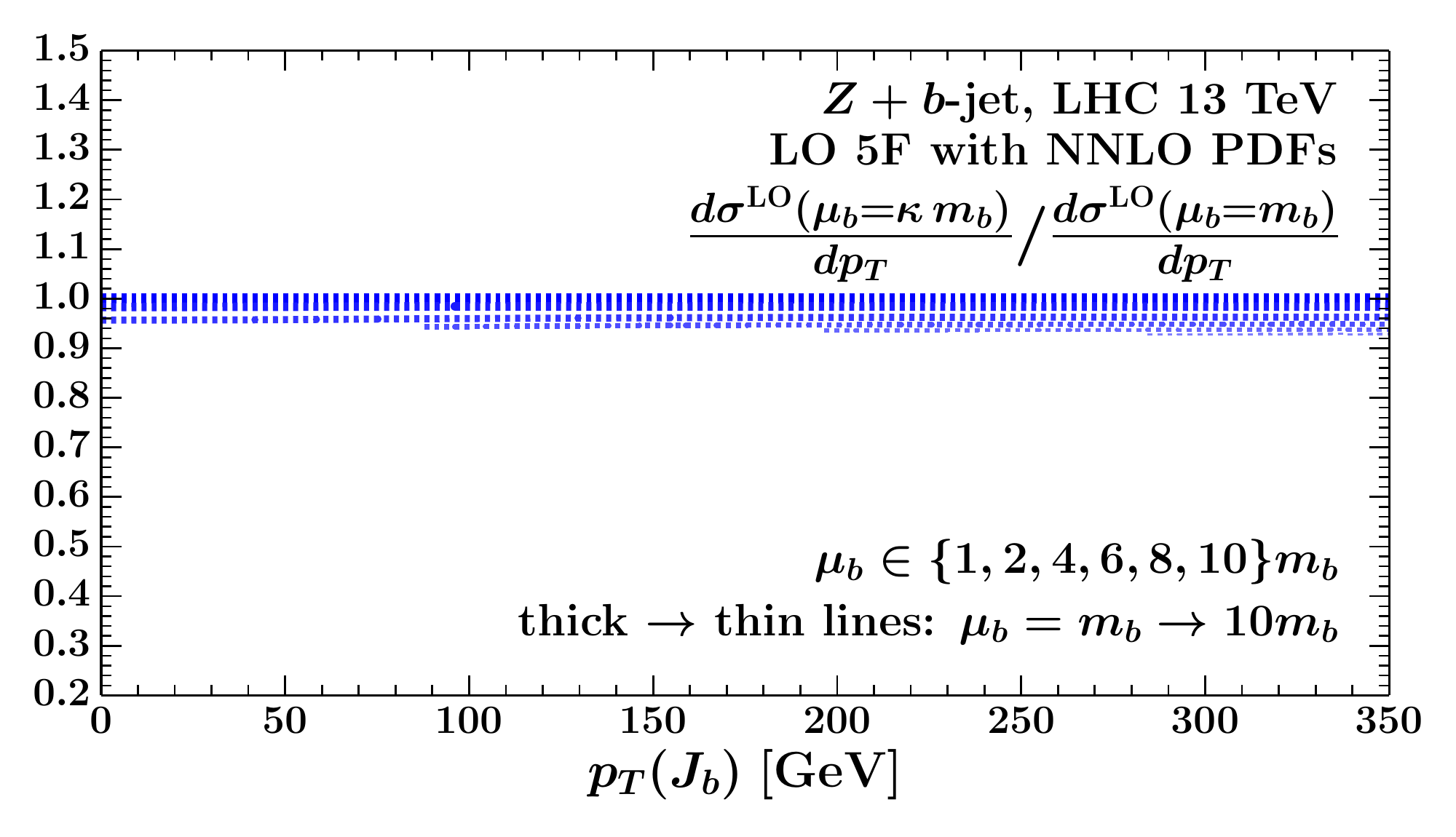} \\
\vskip -3.6mm
\hskip 51mm
\includegraphics[trim=0.45cm 0.0cm 0.5cm 0.1cm,clip,width=0.32\textwidth]{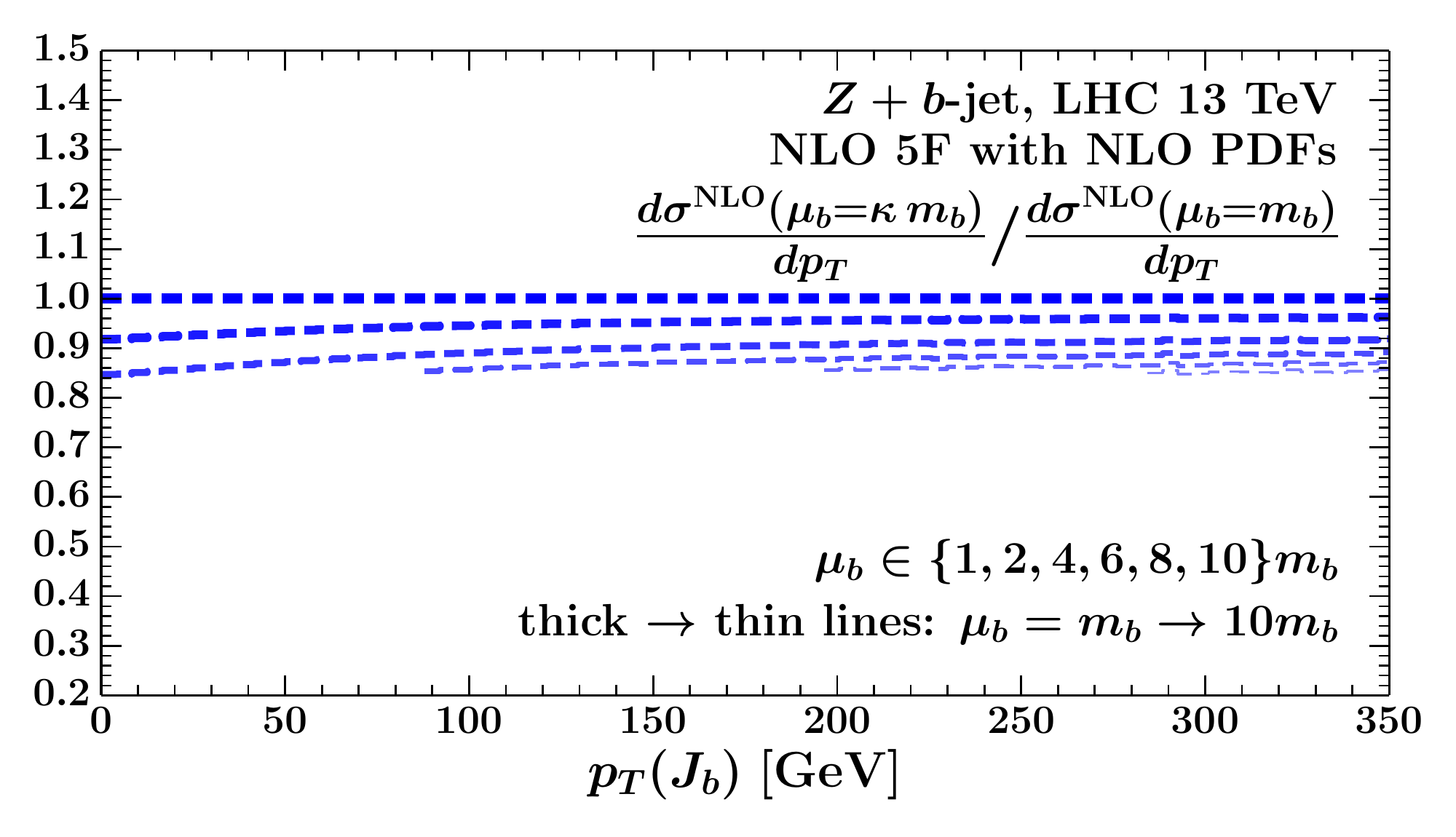}
\includegraphics[trim=0.45cm 0.0cm 0.5cm 0.1cm,clip,width=0.32\textwidth]{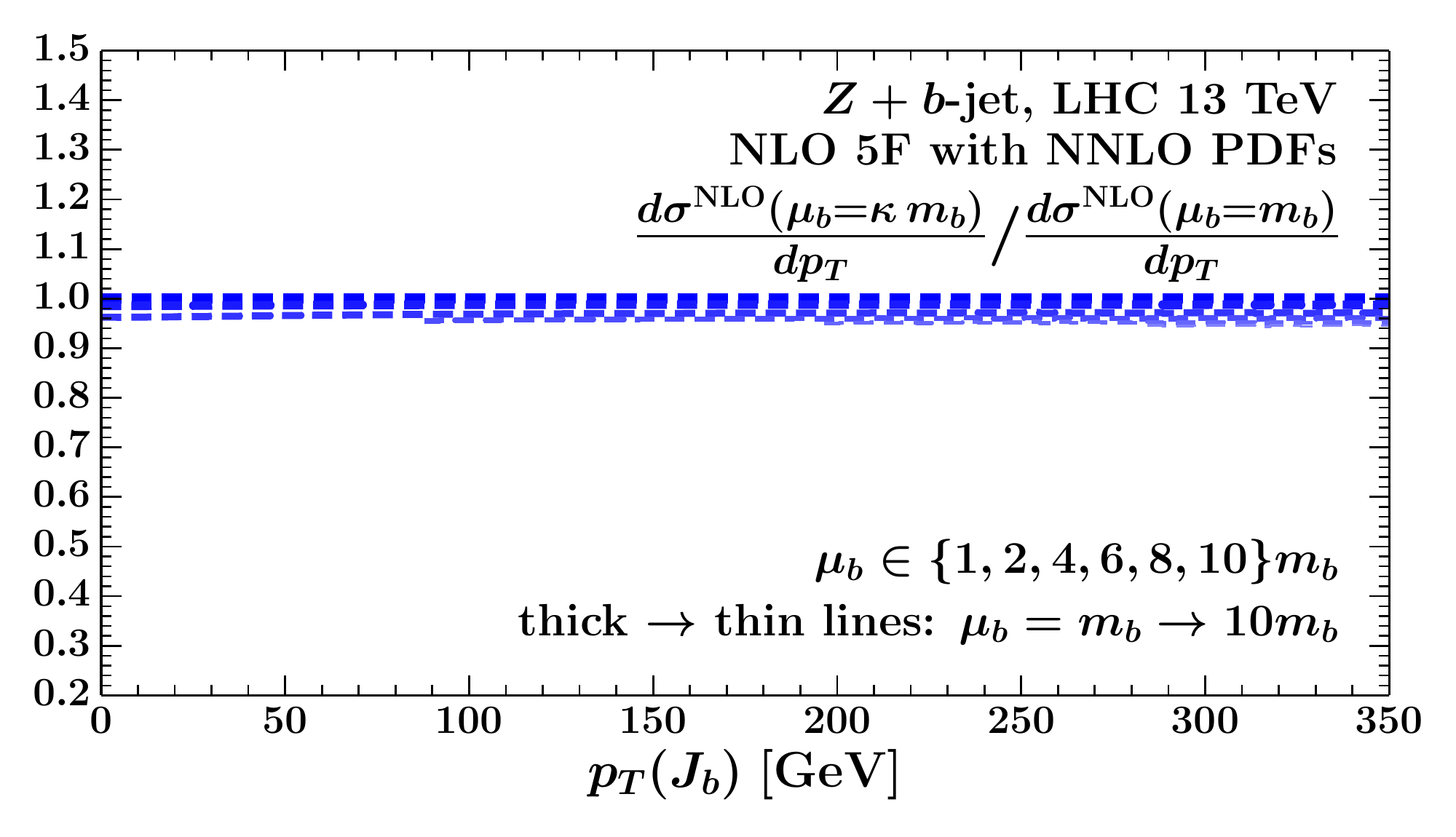} \\
\vskip -2mm
\caption{The ratio eq.~(\ref{eq:ratio}) as a function of $p_T(J_b)$ for $Z+b$-jet at LHC 13 TeV.}
\label{fig:zbj-mub-dep}
\end{figure}

In order to clarify the origin of the second feature mentioned above we perform further checks. In figs.~\ref{fig:tj-mub-dep}, \ref{fig:bbz-mub-dep}, \ref{fig:zbj-mub-dep} we plot the 5FS curves as the ratio:
\begin{equation}
{\sigma^{\rm 5F}(\mub=\k\cdot m) \over \sigma^{\rm 5F}(\mub=m)}\,,~~\k=(1,2,4,6,8,10)\,.
\label{eq:ratio}
\end{equation}
The differences in the accuracy of the matching conditions of the pdfs and $\as$ are particularly relevant for our study. Therefore, in order to disentangle effects due to higher order corrections to the coefficient functions and higher orders in the pdfs we plot all possible combinations of accuracies, i.e. (LO,LO), (LO,NLO), and (LO, NNLO) shown in first row, then (NLO,NLO) and (NLO,NNLO) in the second row and, when available, (NNLO,NNLO) in the third row. 

Strikingly, despite the very different nature of the three processes, there is an extreme similarity between all of them. Clearly, we observe a process-independent feature: by increasing the order of the perturbative cross-section the slope of the curves becomes smaller (i.e. they become less scale-dependent) while by increasing the order of the pdfs the curves become closer to each other. The improvement from the inclusion of higher order pdfs is very significant and, at NNLO, the dependence on the position of the HFMP is dramatically reduced irrespectively of the order of the perturbative cross-section.

\subsection{Discontinuities in cross-sections across HFMPs}

\begin{figure}[t]
\centering
\hskip-6pt
\includegraphics[trim=0.4cm 0.0cm 0.3cm 0.4cm,clip,width=0.33\textwidth]{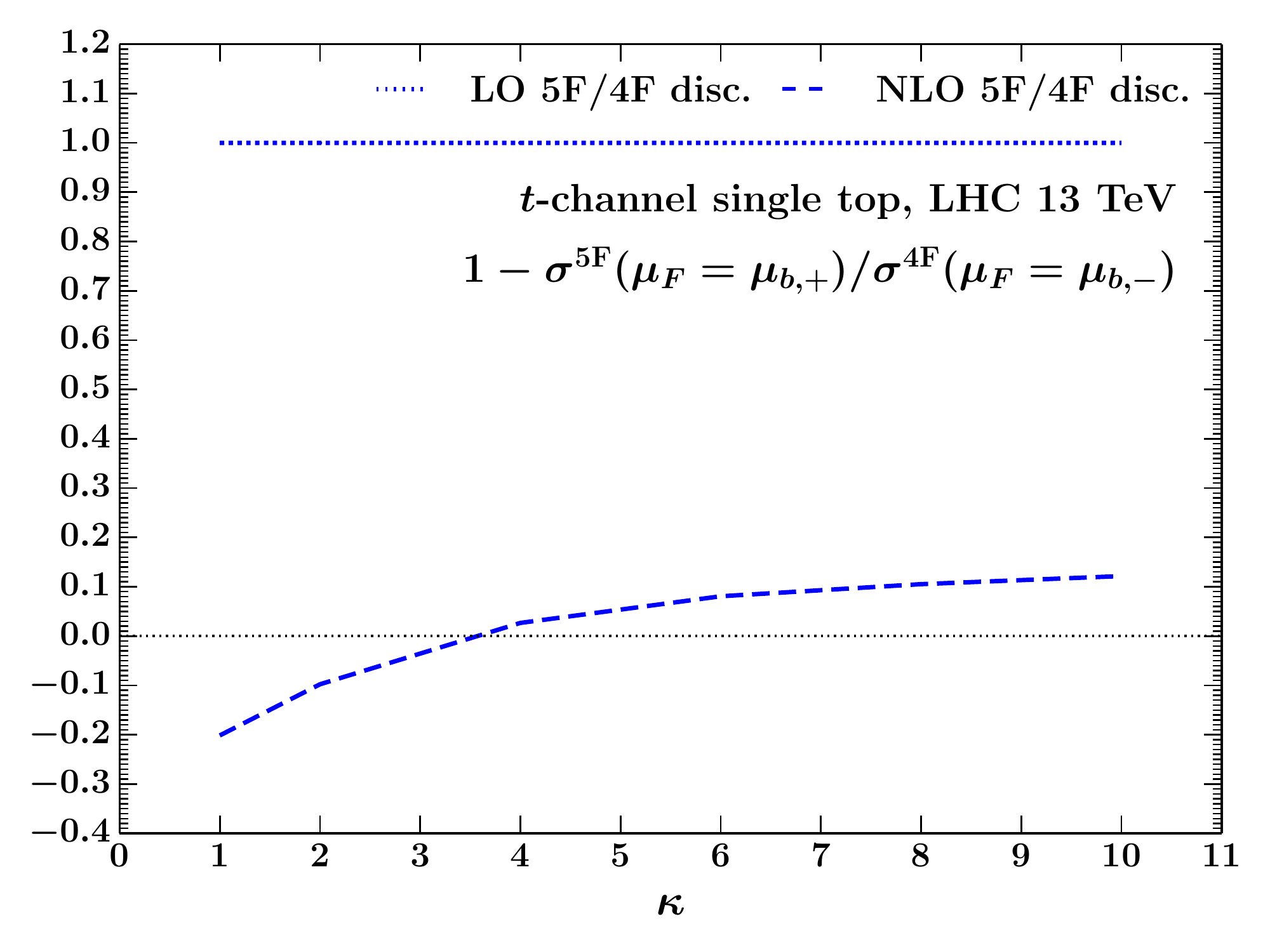}
\includegraphics[trim=0.4cm 0.0cm 0.3cm 0.4cm,clip,width=0.33\textwidth]{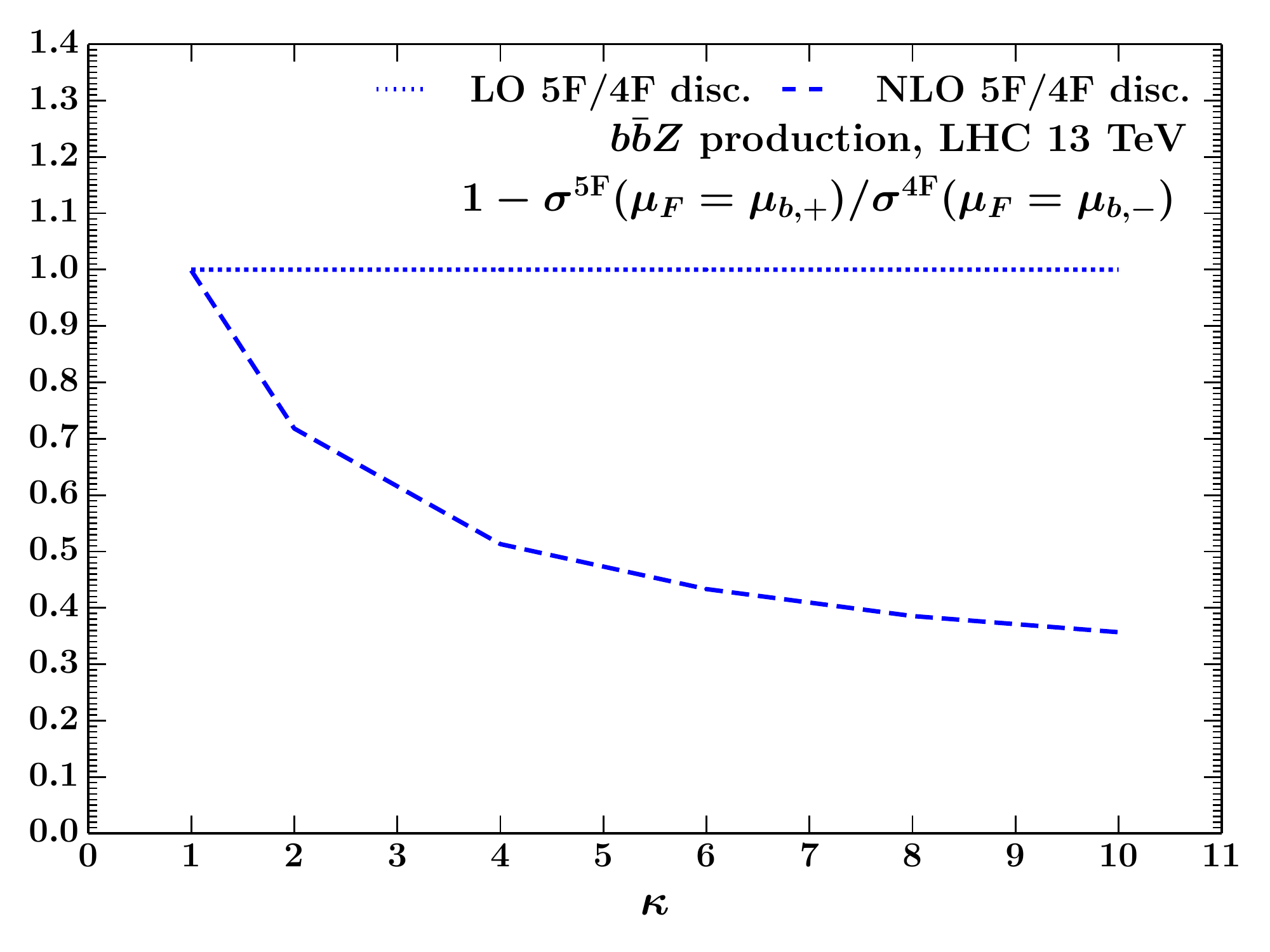}
\includegraphics[trim=0.4cm 0.0cm 0.3cm 0.4cm,clip,width=0.33\textwidth]{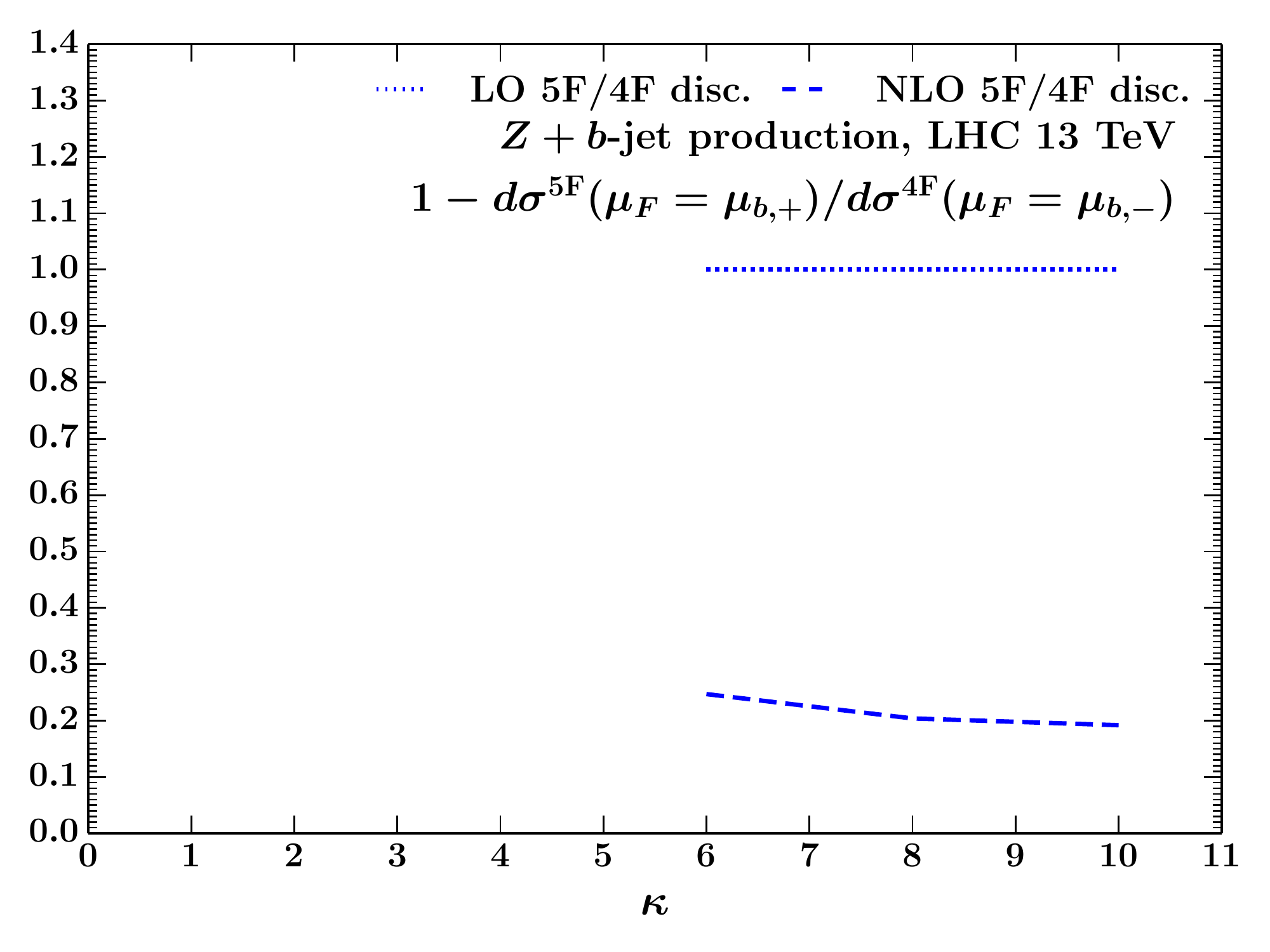}
\caption{Discontinuity between 5FS cross-section above threshold and 4FS cross-section below threshold for LO (dotted) and NLO (dashed) for $t$-channel single top cross-section (left), $b\bar{b}Z$ production cross-section (center) and the $p_T(J_b)$ differential distribution in $Z+b$ (right). All are for LHC at 13 TeV as a function of $\k$. The discontinuity is defined in eq.~(\ref{eq:disc}) and, in the ideal case, it should be zero.}
\label{fig:disc3procs}
\end{figure}

The size of the discontinuity of hadronic cross-sections during the transition between 4FS and 5FS at the scale $\mub$ is an important criterion for our work. We first consider the three processes already studied in sec.~\ref{sec:threshold-switch}. In fig.~\ref{fig:disc3procs} we show the discontinuities in the predicted cross-sections due to the transition from 4FS to 5FS, as a function of the matching point $\mub$. The discontinuity across threshold, as a function of $\mu_F$, is defined as 
\begin{equation}
{\rm Discontinuity} = 1 - \frac{\sigma^{\rm 5F}(M~{\rm fixed}, \mu_{F,R}=\mub+\epsilon)}{\sigma^{\rm 4F}(M~{\rm fixed}, \mu_{F,R}=\mub-\epsilon)} \,,
\label{eq:disc}
\end{equation}
where $M$ stands for the relevant mass parameter for each process ($m_Z$ or $m_t$) and $\epsilon$ is very small. In the discontinuity eq.~(\ref{eq:disc}) above (as well as in eq.~(\ref{eq:discthreshold}) below), the 5FS cross-section is evaluated just to the right of the matching point while the 4FS one is computed just to the left of it. Ideally, the discontinuity when going across the HFMP should be zero.

There are two important features in fig.~\ref{fig:disc3procs}. First, by going from LO to NLO the discontinuity decreases drastically. Second, we note that the 4FS--to--5FS discontinuity tends to decrease when the HFMP is increased. The details are process dependent but the trend is significant and generic. We, therefore, conclude that both the inclusion of higher order corrections in observables and the increase in the value of the HFMP lead to a decrease in the discontinuities in observables. We stress that in the case of the $p_T(J_b)$ distribution only three $b$-quark matching points are crossed as $p_T(J_b)$ sweeps the interval between zero and infinity. This is due to the specific functional form (\ref{eq:muFbjet}) adopted for the factorization scale. 

The processes just considered have two limitations as far as our study is concerned. First, the 4FS predictions for these processes are not known at NNLO. Second, as the matching point is varied in the range $\k=1$ to $\k=10$ the factorization scale $\muF$, which is kept equal to $\mub$, can significantly deviate from the natural scale 
\footnote{Although the scale choice is not the subject of our work, we would like to stress that for each hard-scattering process one can devise a possibly not unique ``natural" scale.}
in the respective process. This is not ideal since when the factorization and/or renormalization scales are taken to be very different from their natural value the convergence of the perturbation series gets affected. 

In order to demonstrate that the above two limitations do not alter our conclusions we extend our study as follows. We construct two families of (unphysical) processes for which both the 4FS and 5FS predictions can be derived at NNLO. These families of processes are constructed in such a way that for any value of the matching point $\mub$ the factorization scale is always close to the natural scale for that process. Specifically, we study the cross-sections for $Z$-like and $t\bar{t}$-like production. By $Z$-like we mean $Z$ production but with appropriately chosen $Z$ mass, such that $m_Z=\mu_b$ as we change $\mub$ from $m$ to $10m$. Same for $t\t$-like production but with $m_t=2\mub$. The factorization and renormalization scales are taken to be $\mu_F=\mu_R=\mub$ thus avoiding large ratios between the natural scales and $\mu_{F,R}$. 

The 4FS predictions for these processes are not known at NNLO but we devise an approximation which is sufficiently accurate for our purpose: we take the partonic-cross section to be the 5F NNLO one for massless $b$-quark and we evaluate it with a 4F $\alpha_s$ and convolute it with 4F pdfs. We expect this to be a good approximation because the extra diagrams present in the 5F NNLO partonic cross-sections are multiplied by vanishing pdfs and thus do not contribute. Terms proportional to the $b$-quark mass that originate from 4F diagrams with $b$-quark loops or $b$-quark pair emission are also missed in the massless-$b$ 5F diagrams but these terms should contribute little, especially at large scales. For these reasons we expect the error we make with this approximation (i.e. the difference to the full 4F NNLO results) to be at the sub-percent level. With the help of explicit calculations we have checked that at NLO the inclusive $t\bar{t}$ and $Z$ cross-sections are affected at the permille level.
\begin{figure}[t]
\centering
\includegraphics[trim=0.3cm 0.0cm 0.1cm 0.4cm,clip,width=0.48\textwidth]{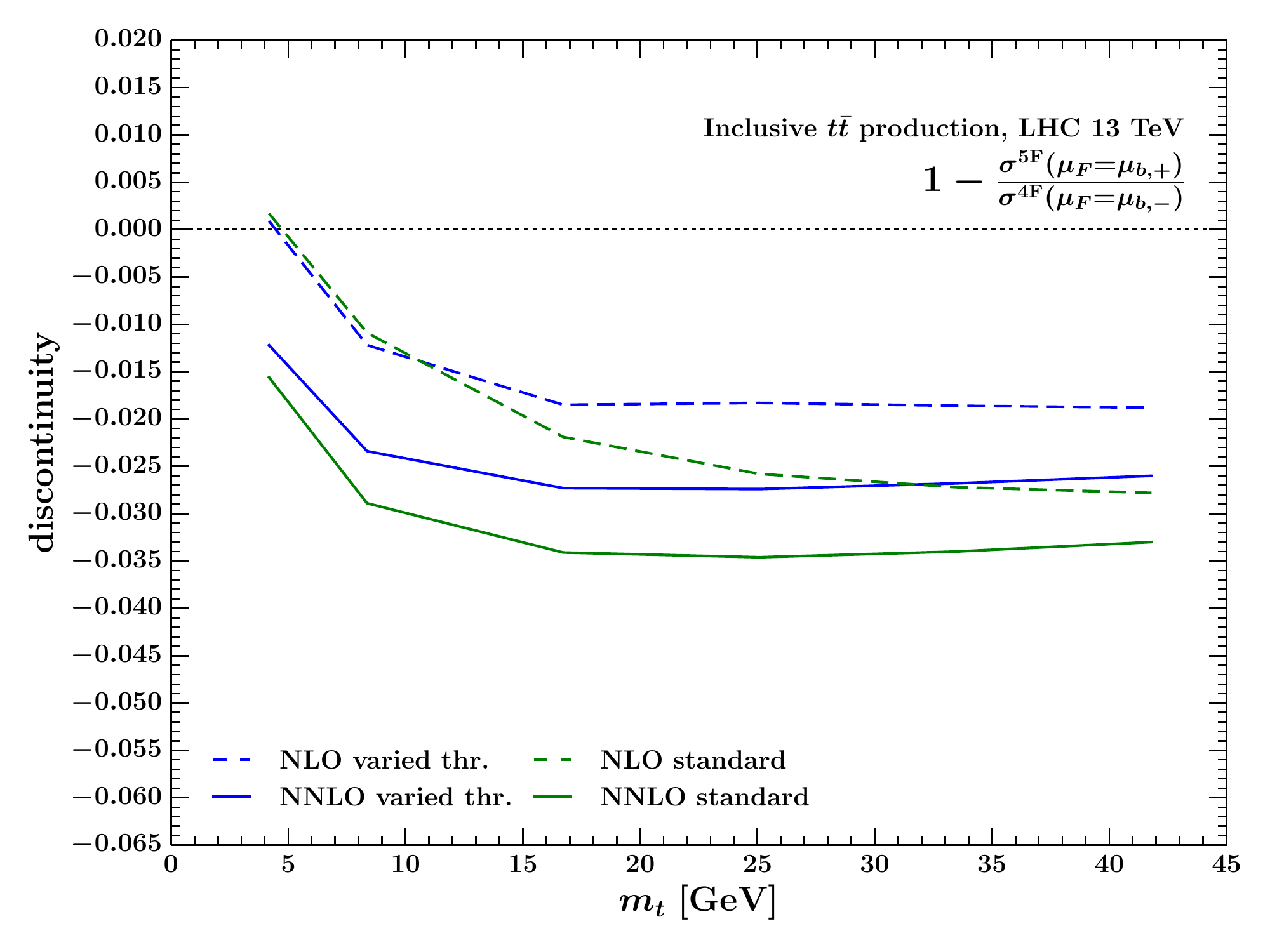}
\includegraphics[trim=0.3cm 0.0cm 0.1cm 0.4cm,clip,width=0.48\textwidth]{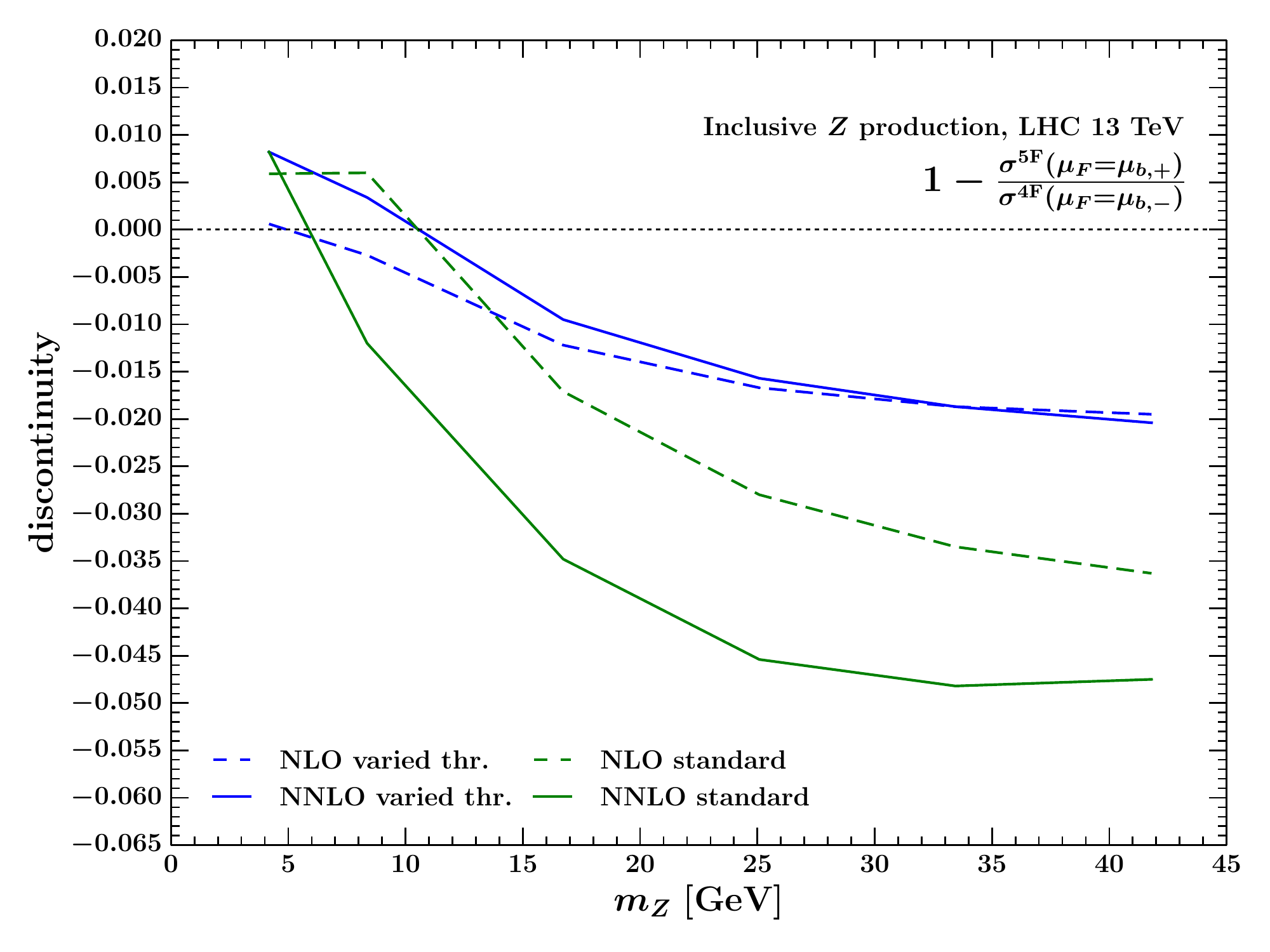}\\ 
\caption{Discontinuities eq.~(\ref{eq:discthreshold}) at NLO (dashed) and NNLO (solid) for $t\t$-like (left) and $Z$-like production (right). Both are for LHC at 13 TeV as a function of the mass $m_t=2\mub$ or $m_Z=\mub$. The discontinuity is computed in both the canonical approach (``Standard", in green) and the variable HFMP approach advocated here (``Variable", in blue).}
\label{fig:discprecprocs}
\end{figure}
\begin{table}[t]
\centering
\begin{tabular}{c| c c | c c | c c}
& \multicolumn{2}{c}{${\rm Discontinuity}_{\rm LO}$} & \multicolumn{2}{c}{${\rm Discontinuity}_{\rm NLO}$} & \multicolumn{2}{c}{${\rm Discontinuity}_{\rm NNLO}$} \\
$\k$ & Variable & Standard  & Variable & Standard & Variable & Standard \\
\hline
1  & -0.0020 & -0.0911 &  0.0006 &  0.0059 &  0.0082 &  0.0082 \\
2  & -0.0015 & -0.0699 & -0.0027 &  0.0060 &  0.0034 & -0.0120 \\
4  & -0.0011 & -0.0634 & -0.0122 & -0.0171 & -0.0095 & -0.0348 \\
6  & -0.0008 & -0.0615 & -0.0167 & -0.0280 & -0.0157 & -0.0454 \\
8  & -0.0007 & -0.0603 & -0.0187 & -0.0335 & -0.0187 & -0.0482 \\
10 & -0.0006 & -0.0593 & -0.0195 & -0.0363 & -0.0204 & -0.0475 \\
\hline
\end{tabular}
\caption{Discontinuity eq.~(\ref{eq:discthreshold}) for $Z$-like production at LHC 13 TeV at LO, NLO and NNLO in the variable HFMP approach (``Variable") advocated here and in the canonical (``Standard") approach. Here $m_Z=\mub$ and $\mu_R=\mu_F=\mub \pm \epsilon$, as appropriate.}
\label{tab:zdisc}
\end{table}
\begin{table}[t]
\centering
\begin{tabular}{c| c c | c c | c c}
& \multicolumn{2}{c}{${\rm Discontinuity}_{\rm LO}$} & \multicolumn{2}{c}{${\rm Discontinuity}_{\rm NLO}$} & \multicolumn{2}{c}{${\rm Discontinuity}_{\rm NNLO}$} \\
$\k$ & Variable & Standard  & Variable & Standard & Variable & Standard \\
\hline
1  & -0.0011 & -0.0011 &  0.0009 &  0.0017 & -0.0122 & -0.0156 \\
2  &  0.0003 & -0.0029 & -0.0122 & -0.0109 & -0.0234 & -0.0289 \\
4  &  0.0010 & -0.0043 & -0.0185 & -0.0219 & -0.0273 & -0.0341 \\
6  &  0.0011 & -0.0035 & -0.0183 & -0.0258 & -0.0274 & -0.0346 \\
8  &  0.0013 & -0.0021 & -0.0186 & -0.0272 & -0.0268 & -0.0340 \\
10 &  0.0014 & -0.0007 & -0.0188 & -0.0278 & -0.0260 & -0.0330 \\
\hline
\end{tabular}
\caption{As in table~\ref{tab:zdisc} but for $t\t$-like production with $m_t=2\mub$.}
\label{tab:ttdisc}
\end{table}

The definition of the discontinuity is adapted from eq.~(\ref{eq:disc}):
\begin{equation}
\label{eq:discthreshold}
{\rm Discontinuity} =
\begin{cases}
 1- {\sigma^{\rm 5F}(M \,=\, \mu_{F,R}-\epsilon \,=\, \kappa \cdot m \,;\,\mub= m) \over \sigma^{\rm 4F}(M\,=\,\mu_{F,R}+\epsilon\,=\,\kappa \cdot m)}, & \text{for `standard'} \\[10pt]
 1- {\sigma^{\rm 5F}(M \,=\, \mu_{F,R}-\epsilon \,=\, \kappa \cdot m \,;\,\mub=\kappa \cdot m) \over \sigma^{\rm 4F}(M\,=\,\mu_{F,R}+\epsilon\,=\,\kappa \cdot m)}, & \text{for `variable'}\,, \\
\end{cases}
\end{equation}
with $M=m_Z$ or $m_t$, as appropriate for the process.

The results for $Z$-like and $t\t$-like production are given in table~\ref{tab:zdisc} and table~\ref{tab:ttdisc}, respectively. Both are plotted in fig.~\ref{fig:discprecprocs}. The discontinuities are given at LO, NLO and NNLO in the variable HFMP approach advocated here (``Variable") as well as in the canonical approach (``Standard"). The numbers in the canonical approach are derived in the following way: we take standard 4FS and 5FS pdf sets (with $\mub=m$, as usual) and evaluate the ratio (\ref{eq:discthreshold}) at the points $\k\cdot m$. In other words we extend the 4FS pdf set from scales $\sim m$ all the way up to $\k\cdot m$ which reflects the current practice within the canonical approach. 

The way to read fig.~\ref{fig:discprecprocs} is to recall that in a given process the value of $M$ is fixed (with $M$ being $m_Z$ or $m_t$). Thus, one should choose between computing the cross-section for that process in either the standard or variable approaches and at LO, NLO or NNLO. 

It is evident from fig.~\ref{fig:discprecprocs} that the variable $\mub$ approach advocated here has smaller discontinuities than the standard approach for the whole range of masses $M$ considered here and for any perturbative order (except the $Z$-like cross-section for small values of $M$ at NLO where numerical effects play a role). The discontinuity at NLO is smaller than the NNLO one for $t\t$-like production with small masses. However, as the mass $m_t$ increases the NNLO discontinuity becomes competitive with the NLO one. For $Z$-like production, the NNLO discontinuity in the variable approach is smallest for the whole range of $m_Z$ values.

\begin{table}[t]
\centering
\begin{tabular}{c|c|c|c}
$\k$ & $\sigma_{\rm LO}$ [pb$\times 10^4$] & $\sigma_{\rm NLO}$ [pb$\times 10^4$] & $\sigma_{\rm NNLO}$ [pb$\times 10^4$]\\
\hline
1  & 4.42964 \phantom{(+0.7\%)} & 5.42333 \phantom{(+0.7\%)} & 5.64074 \phantom{(+0.7\%)} \\
2  & 4.46018 (+0.7\%)           & 5.43158 (+0.1\%)           & 5.62619 (-0.3\%)           \\
4  & 4.51340 (+1.9\%)           & 5.40903 (-0.3\%)           & 5.60047 (-0.7\%)           \\
6  & 4.55424 (+2.8\%)           & 5.38918 (-0.6\%)           & 5.58349 (-1.0\%)           \\
8  & 4.58731 (+3.6\%)           & 5.37355 (-0.9\%)           & 5.57117 (-1.2\%)           \\
10 & 4.61520 (+4.2\%)           & 5.36088 (-1.2\%)           & 5.56158 (-1.4\%)           \\
\hline
\end{tabular}
\caption{Dependence of the total $Z$ cross section at LHC 13 TeV on the threshold scale $\mub=\k\cdot m$ (recall that $\k=1$ represents the standard choice in all publicly available pdf sets). Shown is the 5FS cross-section predicted at LO, NLO and NNLO for $m_Z=\mu_R=\mu_F=$ 91.1876 GeV.}
\label{tab:Zincl}
\end{table}
\begin{table}[t]
\centering
\vskip 1mm
\begin{tabular}{c|c|c|c}
$\k$ & $\sigma_{\rm LO}$ [pb] & $\sigma_{\rm NLO}$ [pb] & $\sigma_{\rm NNLO}$ [pb]\\
\hline
1  & 560.86 \phantom{(+0.7\%)} & 735.21 \phantom{(+0.7\%)} & 806.15 \phantom{(+0.7\%)} \\
2  & 566.28 (+1.0\%)           & 736.49 (+0.2\%)           & 807.50 (+0.2\%)           \\
4  & 570.59 (+1.7\%)           & 739.52 (+0.6\%)           & 809.22 (+0.4\%)           \\
6  & 572.63 (+2.1\%)           & 741.78 (+0.9\%)           & 810.33 (+0.5\%)           \\
8  & 573.86 (+2.3\%)           & 743.53 (+1.1\%)           & 811.14 (+0.6\%)           \\
10 & 574.70 (+2.5\%)           & 744.95 (+1.3\%)           & 811.78 (+0.7\%)           \\
\hline
\end{tabular}
\caption{As in table \ref{tab:Zincl} but for the $t\bar{t}$ total cross-section with $m_t=\mu_R=\mu_F=$ 173.3 GeV.}
\label{tab:ttincl}
\end{table}
\begin{table}[t]
\centering
\vskip 1mm
\begin{tabular}{c|c|c|c}
$\k$ & $\sigma_{\rm LO}$ [pb] & $\sigma_{\rm NLO}$ [pb] & $\sigma_{\rm NNLO}$ [pb]\\
\hline
1  & 18.375 \phantom{(+0.7\%)} & 35.055 \phantom{(+0.7\%)} & 44.423 \phantom{(+0.7\%)} \\
2  & 18.836 (+2.5\%)           & 35.327 (+0.8\%)           & 44.466 (+0.1\%)           \\
4  & 19.332 (+5.2\%)           & 35.442 (+1.1\%)           & 44.480 (+0.1\%)           \\
6  & 19.635 (+6.7\%)           & 35.466 (+1.2\%)           & 44.481 (+0.1\%)           \\
8  & 19.855 (+8.1\%)           & 35.469 (+1.2\%)           & 44.478 (+0.1\%)           \\
10 & 20.028 (+9.0\%)           & 35.465 (+1.2\%)           & 44.475 (+0.1\%)           \\
\hline
 \end{tabular}
\caption{As in table \ref{tab:Zincl} but for the $ggH$ total cross-section with $m_H=2\mu_R=2\mu_F=$ 125.0 GeV.}
\label{tab:Hincl}
\end{table}
\begin{table}[t]
\centering
\vskip 1mm
\begin{tabular}{c| c c c |c c c}
$\k$ & $\sigma_{\rm LO}$ [pb] & $\sigma_{\rm NLO}$ [pb] & $\sigma_{\rm NNLO}$ [pb] & $R^{\rm LO}_{t/\bar{t}}$ &  $R^{\rm NLO}_{t/\bar{t}}$ & $R^{\rm NNLO}_{t/\bar{t}}$\\
\hline
1  & 119.19 \phantom{(+0.7\%)} & 138.28 \phantom{(+0.7\%)} & 139.90 \phantom{(+0.7\%)}  & 1.654 & 1.660 & 1.638 \\
2  &  90.26 (-24.2\%)          & 130.78 (-5.4\%)           & 138.48 (-1.0\%)    & 1.668 & 1.658 & 1.641       \\
4  &  62.22 (-47.8\%)          & 124.10 (-10.2\%)          & 136.30 (-2.6\%)   & 1.680 & 1.662 & 1.644       \\
6  &  46.30 (-61.2\%)          & 120.34 (-13.0\%)          & 134.90 (-3.6\%)   & 1.687 & 1.666 & 1.645        \\
8  &  35.23 (-70.4\%)          & 117.69 (-14.9\%)          & 133.86 (-4.3\%)   & 1.691 & 1.670 & 1.647       \\
10 &  26.78 (-77.5\%)          & 115.63 (-16.4\%)          & 133.03 (-4.9\%)  & 1.694 & 1.673 & 1.649         \\
\hline
\end{tabular}
\caption{As in table \ref{tab:Zincl} but for the $t$-channel single-top total cross-section and the ratio $R_{t/\bar{t}}$ of single top versus single antitop cross-sections with $m_t=$ 173.3 GeV and $\mu_{F,R}=m_t/2$.}
\label{tab:1tincl}
\end{table}

\subsection{Effect of changing the value of $\mub$ on standard LHC candles}

Another important test for our approach is how increasing the value of the HFMP affects standard precision LHC candles. We note that these processes have typical scales that are much larger than any of the $b$-quark matching points considered here. In table~\ref{tab:Zincl} we show the $Z$ cross section at LO, NLO and NNLO, by using the full range of NNPDF3.0-based sets with varying $\mub$ constructed by us. Similarly, in table~\ref{tab:ttincl} we show the results for the $t\t$ cross-section, in table~\ref{tab:Hincl} for the Higgs inclusive cross-section in gluon fusion while in table~\ref{tab:1tincl} we show the $t$-channel single top cross-section and the ratio of single top versus single antitop cross-sections, $R_{t/\bar{t}}$. All results are for LHC at 13 TeV.

\begin{figure}[t]
\centering
\includegraphics[width=0.55\textwidth]{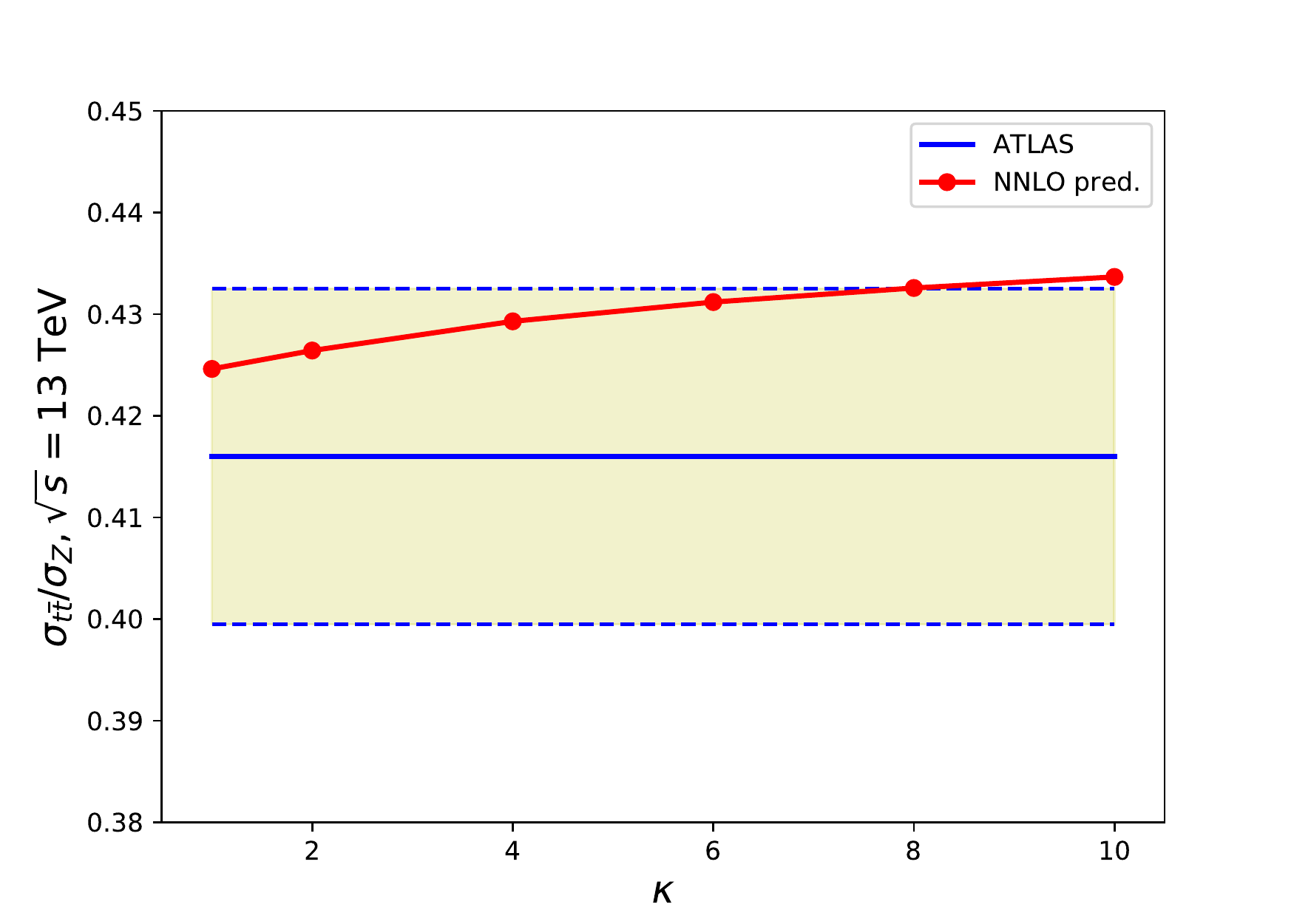}
\caption{Ratio of total inclusive $t\t$ and $Z$ cross-sections for $\sqrt{s}=13$ TeV for a series of $\mub = \kappa m_b$ values versus the ATLAS measurement~\cite{Aaboud:2016zpd} (shaded band).}
\label{fig:tt_z}
\end{figure}

We observe that for all processes the changes between the canonical approach $\k=1$ and any of the variable ones with $\k$ as large as 10 at NNLO are small compared to the theoretical and experimental errors. We note that the increased order of the prediction significantly decreases this difference. The most sizable effect is observed in single top, where for $\mub=10m$ we have a NNLO prediction that we estimate to be almost 5\% lower than the current prediction. This change is significantly larger than the NNLO scale error (which is around 1\%) but the two are compatible within the experimental error (which currently is below 10\%). The long-term prospective for measuring the single top cross-section at the LHC points towards 5\% precision for this observable. This means that it may not be easy to use single top LHC data to discriminate between these two approaches. An alternative may be the measurement of the ratio $\sigma(t\t)/\sigma(Z)$ which, as can be seen in fig.~\ref{fig:tt_z},  is currently known with accuracy of about 2.8\%~\cite{Aaboud:2016zpd}. Future improvements in this ratio may become the leading candidate for distinguishing the predictions based on these two approaches.

\section{Conclusions}

In this work we advocate a new approach to constructing heavy-flavor pdfs, namely, a standard ZM-VFNS but with heavy flavor matching point $\mub$ which is taken to be significantly higher than the conventional value $\mub=m$. We extensively test our proposal on a range of NLO and NNLO precision observables at the LHC. We find that our approach is competitive with the current state-of-the-art GM-VFNS approaches. Its main advantage over existing GM-VFN schemes is its transparency and simplicity; it is straightforward to formulate and implement in practice for any process at hadron colliders and it avoids the need for adding by hand rescaling or damping factors. Our proposed approach typically leads to smaller discontinuities in observables across the heavy flavor matching point compared to conventional approaches. It is straightforward to implement in practice  thanks to the existence of public tools like {\tt xFitter}~\cite{Alekhin:2014irh},  {\tt APFEL}~\cite{Bertone:2013vaa} and LHAPDF~\cite{Buckley:2014ana}.

We demonstrate that our proposal satisfies all requirements for constructing a {\it good} pdf set: first, it maintains collinear resummation provided the HFMP $\mub$ is not chosen to be too large; we consider the range $\mub=5m-10m$ to be optimal. Second, power corrections ${\cal O}(m)$ of the heavy quark mass, which constitute the main problem in constructing heavy-flavor pdfs, are under control. Despite the fact that in the $\nf=\nl+1$ scheme we only use purely massless coefficient functions (which greatly simplifies scheme's implementation) the neglected power corrections can be made negligible, by construction. We have verified this explicitly by a direct comparison with FONLL predictions for the case of DIS production of bottom (at NLO and NNLO) and charm (at NNLO); see appendix \ref{app:BGMPUfits} for details.

We pay close attention to the issue of continuity of observables when one crosses the HFMP. As the HFMP is moved away from the point $\mub=m$ and both $\as$ and pdfs become more discontinuous across that point, one may naively expect that observables may also become more discontinuous (which would be bad). Our analysis shows precisely the opposite: typically, using pdfs of higher orders or with higher HFMPs, leads to smaller discontinuities in observables. We suspect that pdfs with full N$^3$LO accuracy will further improve continuity, mainly as a result of higher-order corrections to the heavy-flavor matching conditions. The results of ref.~\cite{Forte:2013mda} suggest that large shifts in predictions compared to N$^2$LO should not be expected. The significant cancellation between various contributions we observe is in line with the expectation that observables should be continuous to all orders. This demonstration of the self-consistency of the theory is, in our opinion, not as apparent in other schemes.

For the sake of simplicity in our consideration we have mostly focused on the case of a single heavy flavor which we have taken to be the bottom. However, this need not be the case; all heavy flavors (charm, bottom and top) can be treated this way and their corresponding matching points may be set independently. Our ideas are particularly well suited for applications in top production. No calculations for top production exist in the 6F scheme,
\footnote{The original FONLL approach \cite{Cacciari:1998it} would work in 6FS but to the best of our knowledge it has never been applied to top production.}
despite the fact that top quarks with $p_T$ in the TeV range are already routinely measured at the LHC. This fixed-flavor-like treatment clearly contradicts the spirit of all existing GM-VFNS. On the contrary, our proposal indicates that, especially when working with the factorization scale proposed in ref.~\cite{Czakon:2016dgf}, top production should be described within the 5FS for $p_T$'s as large as 3 TeV if $\kappa_t=10$ is chosen. 

Similarly, the recently introduced pdfs with full SM content at very-high energies \cite{Bauer:2017isx} represent another natural application for the ideas introduced in the present work.

Our proposal may be useful also in the context of fragmentation functions, especially when treating heavy quark contributions to the fragmentation of light hadrons (this problem was recently considered in ref.~\cite{Epele:2016gup} within a FONLL-inspired GM-VFNS framework). While typically the uncertainties in fragmentation function analyses are larger than in pdf ones, recent advances in this subject have introduced NNLO QCD precision into global fragmentation function fits \cite{Anderle:2015lqa,Bertone:2017tyb} which suggest that a more refined treatment of HFMPs in fragmentation functions may also be beneficial.

Finally, we would like to emphasize that in order to fully explore the phenomenological implications of our proposal, a more complete analysis along the lines of the global fits performed by the pdf fitting collaborations will be required. This will allow one to precisely determine the impact of our proposal on precision observables like $Z$ and single top production where we see differences with respect to standard approaches that are significant yet not sufficient at present to differentiate between the two approaches. We hope that future work will ultimately help clarify the proper interpretation of such differences, i.e. if they should be considered as due to difference in approach or as a genuine uncertainty within existing approaches not fully exhibited until now.

\begin{acknowledgments}
V.B. is particularly grateful to Fred Olness for illuminating discussions on the role of the HFMPs in pdf fits. M.U. is grateful to Davide Napoletano for their discussions on the FONLL scheme. A.M. thanks the Department of Physics at Princeton University for hospitality during the completion of this work. This work was initiated at the KITP workshop ``LHC Run II and the Precision Frontier" which is supported by NSF PHY11-25915. V.B. is supported by the European Research Council Starting Grant ``PDF4BSM". The work of A.M. and A.P. is supported by the UK STFC grants ST/L002760/1 and ST/K004883/1 and by the European Research Council Consolidator Grant ``NNLOforLHC2". M.U. is supported by a Royal Society Dorothy Hodgkin Research Fellowship and this work has been partially supported by the STFC consolidated grant ST/P000681/1.
\end{acknowledgments}

\appendix
\section{Example: charm and bottom fits from DIS data}\label{app:BGMPUfits}

In this appendix we explicitly demonstrate in the context of inclusive DIS that the power corrections ${\cal O}(m)$ indeed become negligible as the HFMP is increased. We perform fits to inclusive DIS data for bottom and charm production at NLO and NNLO. The size of the power corrections can be established by comparing pdf fits based on our approach and on the FONLL approach which is designed to contain the bulk of those power corrections.

Specifically, we perform a series of pdf fits by implementing the method proposed in this paper and using the open-source fitting code {\tt xFitter} (former {\tt HERAFitter})~\cite{Alekhin:2014irh} interfaced to the {\tt APFEL} code~\cite{Bertone:2013vaa}. The dataset included in these fits comprises the combined HERA1+2 H1 and ZEUS inclusive DIS cross-section data~\cite{Abramowicz:2015mha}, the combined H1 and ZEUS charm production cross-section measurements~\cite{Abramowicz:1900rp} and the separate bottom production cross sections from H1~\cite{Aaron:2009af} and ZEUS~\cite{Abramowicz:2014zub}. This dataset provides sufficient information for a reliable determination of pdfs. Furthermore, thanks to the inclusion of heavy-quark tagged data, it is also sensitive to
heavy-quark mass effects which allows us to assess the impact of different prescriptions for describing heavy flavors.

\begin{figure}[t]
\centering
\includegraphics[width=0.5\textwidth,angle=-90]{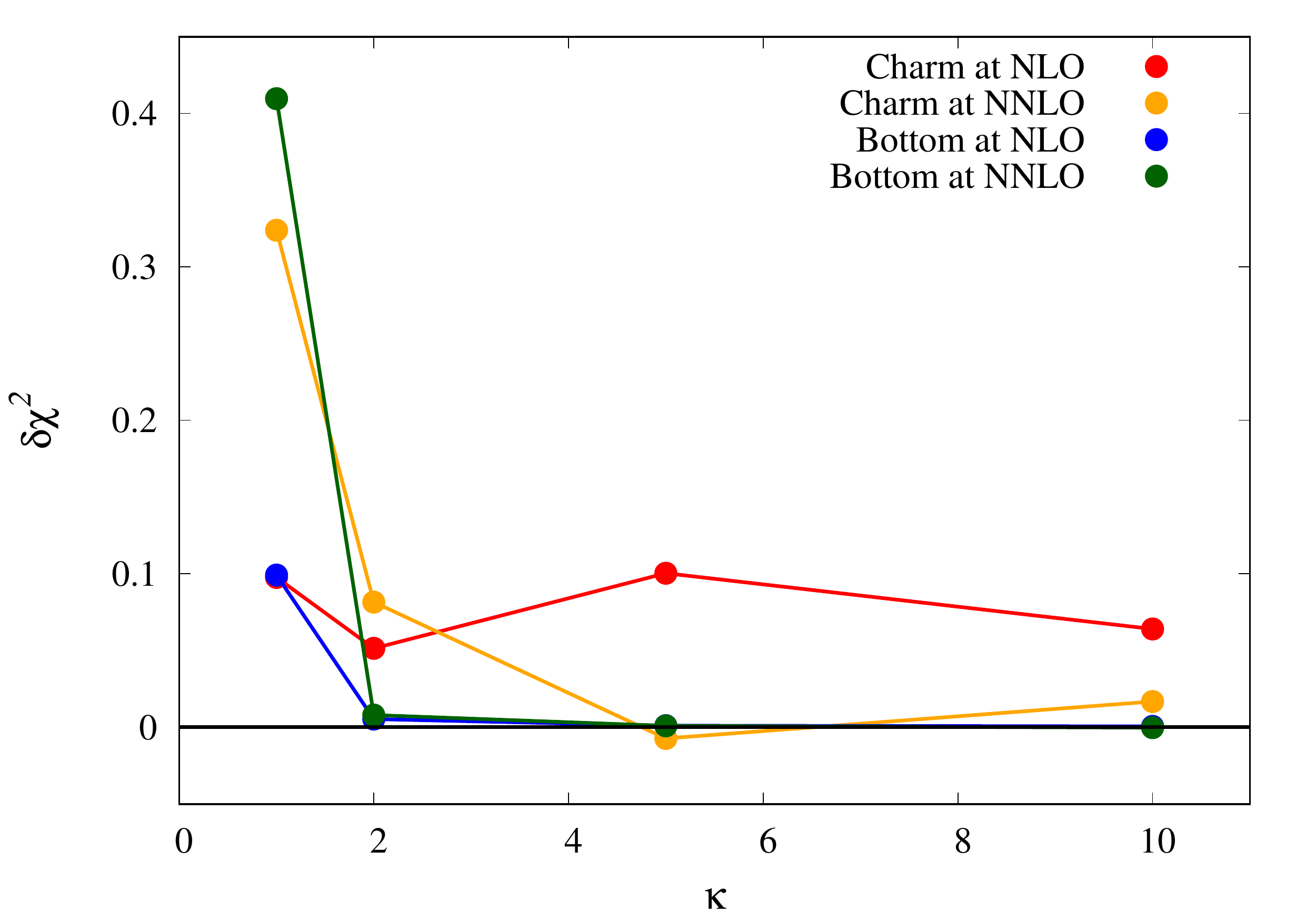}
\caption{Behavior of $\delta\chi^2$ as defined in
  Eq.~(\ref{eq:Deltachi2}) as a function of the HFMP rescaling
  parameter $\kappa$ for pdf fits at NLO and NNLO in which both
  the charm and the bottom HFMPs are displaced by a factor $\kappa$ relative to
  their mass.}
\label{fig:chi2}
\end{figure}

In order to validate our method, we perform fits both at NLO and NNLO moving separately the charm and bottom HFMPs $\mu_{c,b}$ by a factor $\kappa=1,2,5,10$ with respect to the masses $m_{c,b}$. We then compare to the corresponding fits performed in the FONLL scheme. In the case of charm, given that $m_b/m_c\sim 3$, in all fits we set $\mu_b=5m_b$ in such a way that the bottom HFMP is always sufficiently above the charm one even for $\kappa_c = 10$. The aim of these fits is to show that, as $\kappa$ increases, the fits using our method and the FONLL scheme become closer. In order to assess the difference between our method and the FONLL scheme, in fig.~\ref{fig:chi2} we display the quantity:
\begin{equation}
\label{eq:Deltachi2}
\delta\chi^2(\kappa) = \frac{\chi^2(\mbox{this work}) - \chi^2(\mbox{FONLL})}{\chi^2(\mbox{FONLL})}\,,
\end{equation}
as a function of the threshold rescaling parameter $\kappa$ when the charm threshold is moved at NLO (red curve) and NNLO (orange curve) and when the bottom threshold is moved at NLO (blue curve) and NNLO (green curve). 
\begin{table}[t]
  \begin{center}
    \begin{tabular}{lllcccc}
      \toprule
      \multirow{2}{*}{Heavy quark} & \multirow{2}{*}{Pert. order} & \multirow{2}{*}{Scheme} & \multicolumn{4}{c}{$\chi^2$ / d.o.f.} \\ 
      & & & $\kappa=1$ & $\kappa=2$ & $\kappa=5$ & $\kappa=10$ \\ 
      \midrule
      \multirow{4}{*}{Charm} & \multirow{2}{*}{NLO} & FONLL &1.144&1.179&1.166&1.227 \\
                 &         & This work &1.255&1.239&1.283&1.305 \\
                 & \multirow{2}{*}{NNLO} & FONLL &1.207&1.194&1.214&1.226 \\
                 &           & This work &1.598&1.292&1.205&1.246 \\
      \midrule
      \multirow{4}{*}{Bottom} & \multirow{2}{*}{NLO} & FONLL &1.148&1.143&1.144&1.146 \\
                 &         & This work &1.262&1.149&1.145&1.146 \\
                 & \multirow{2}{*}{NNLO} & FONLL &1.204&1.208&1.207&1.212 \\
                 &           & This work &1.697&1.218&1.208&1.212 \\
      \bottomrule
    \end{tabular}
  \end{center}
  \caption{Values of the total $\chi^2$ normalized to the number of
    degrees of freedom for the fits with different values of $\kappa$
    for the charm and bottom HFMPs, at NLO and NNLO, derived using
    the FONLL scheme and the proposal in this work.}
\label{tab:chi2}
\end{table}

First we observe that, as expected, the FONLL scheme at $\kappa=1$ provides a much better description than the conventional ZM-VFNS. As the value of $\kappa$ increases, however, the difference between our prescription and the FONLL scheme starts to decrease. Notably, in the case of bottom, a low value of $\kappa=2$ is sufficient to bring $\delta\chi^2$ close to zero both at NLO and NNLO. This is partly due to the fact that the value $\as(\mub)$ is sufficiently small. 

The picture for charm is more complicated. At NLO the FONLL fit is systematically better than our prescription even for large values of $\kappa$. The large value of $\as(\mu_c)$ plays a role here since likely it enhances the discontinuity around the HFMP. As can be seen from table~\ref{tab:chi2}, however, one should keep in mind that the quality of the FONLL fit is significantly degraded for NLO charm compared to the other cases we consider. This most likely indicates that the perturbative description of charm is simply problematic at low orders. The situation is greatly improved at NNLO where the convergence between the fits derived in our method and in the FONLL scheme is reached around $\kappa\simeq 5$, fully in line with our expectations.

In table~\ref{tab:chi2} we give the values of the total $\chi^2$ normalized to the number of degrees of freedom for each of the fits discussed above. 
In order to assess the significance of deviations, it is useful to keep in mind that all fits discussed here have been performed with the same dataset and using the same pdf parametrization. Therefore, the number of degrees of freedom is common to all fits and equals 1207. This means that variations of the order of a few permill at the level of the normalized $\chi^2$'s are statistically significant.
\footnote{We thank Pavel Nadolsky for a discussion on this point.}
For the case of charm at NLO it is clear that going from $\kappa=1$ to $\kappa=10$ the $\chi^2$ deteriorates significantly, by about 7\%, for FONLL and about 4\% for our fits. As a results of this we believe that the $\chi^2$ values in the first two lines
of table~\ref{tab:chi2} do not provide a basis for assessing the goodness of our prescription; see also the related discussion in ref.~\cite{Bertone:2017ehk}.

The variation of the $\chi^2$ between $\kappa = 1$ and $\kappa = 10$ for charm at NNLO is much smaller, about 1.5\% for FONLL. This reduced sensitivity to the charm threshold position allows one to reliably estimate the quality of the method proposed in this work. Indeed, as mentioned above, the fits based on our prescription approach the FONLL ones for $\kappa=5$ which is fully in line with our expectations of $\kappa$ being in the interval $\kappa=5-10$.

As far as the bottom is concerned, the picture at NLO and NNLO is essentially the same as for the charm at NNLO. In particular, the fits in the FONLL scheme are very stable upon displacement of the HFMP. Correspondingly, the quality of the fit within our method quickly approaches that of the FONLL scheme as $\kappa$ increases and already at $\kappa = 2$ the two are essentially equivalently good.

%
\begin{figure}[t]
\centering
\includegraphics[width=0.33\textwidth,angle=-90]{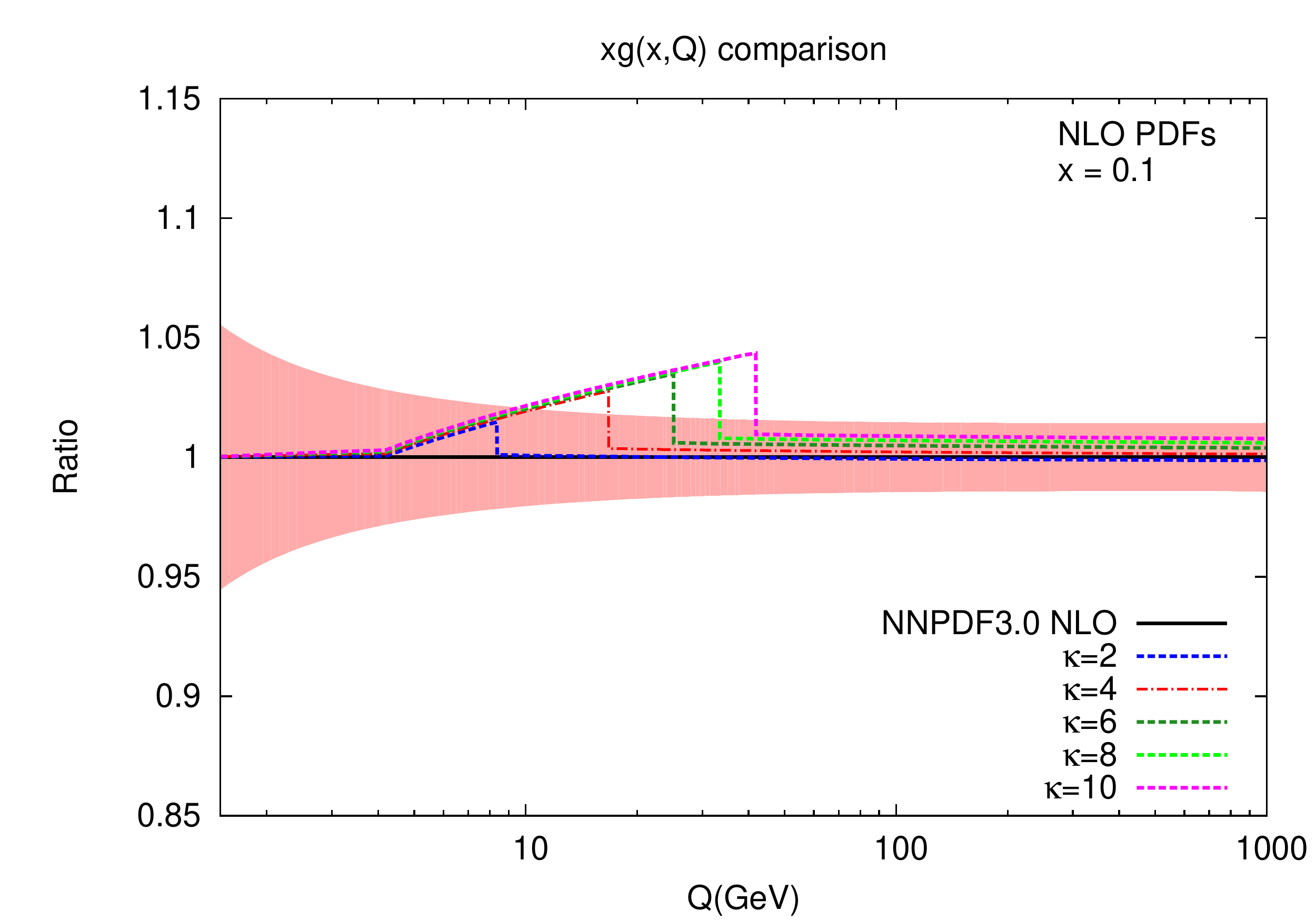}
\includegraphics[width=0.33\textwidth,angle=-90]{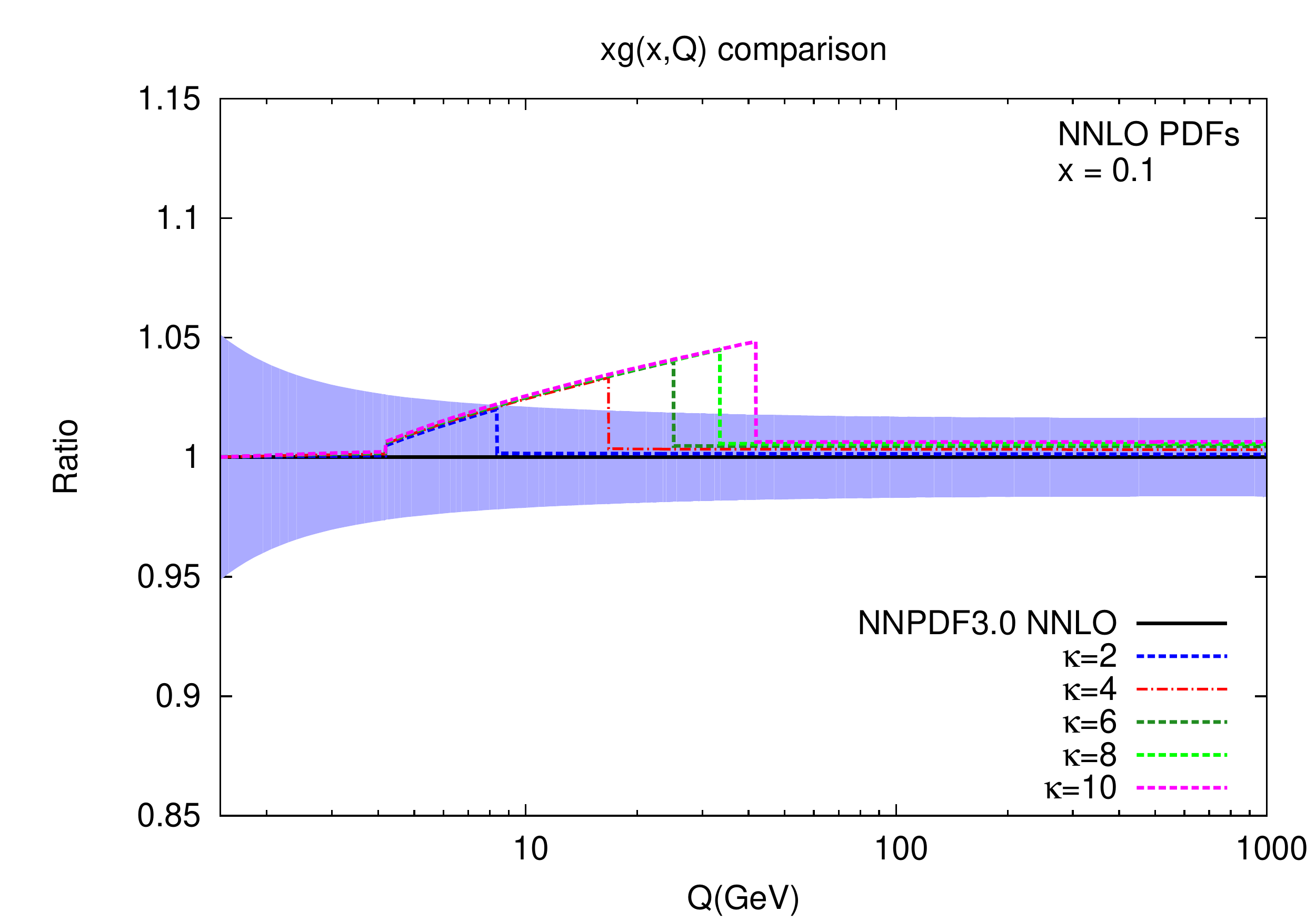}
\includegraphics[width=0.33\textwidth,angle=-90]{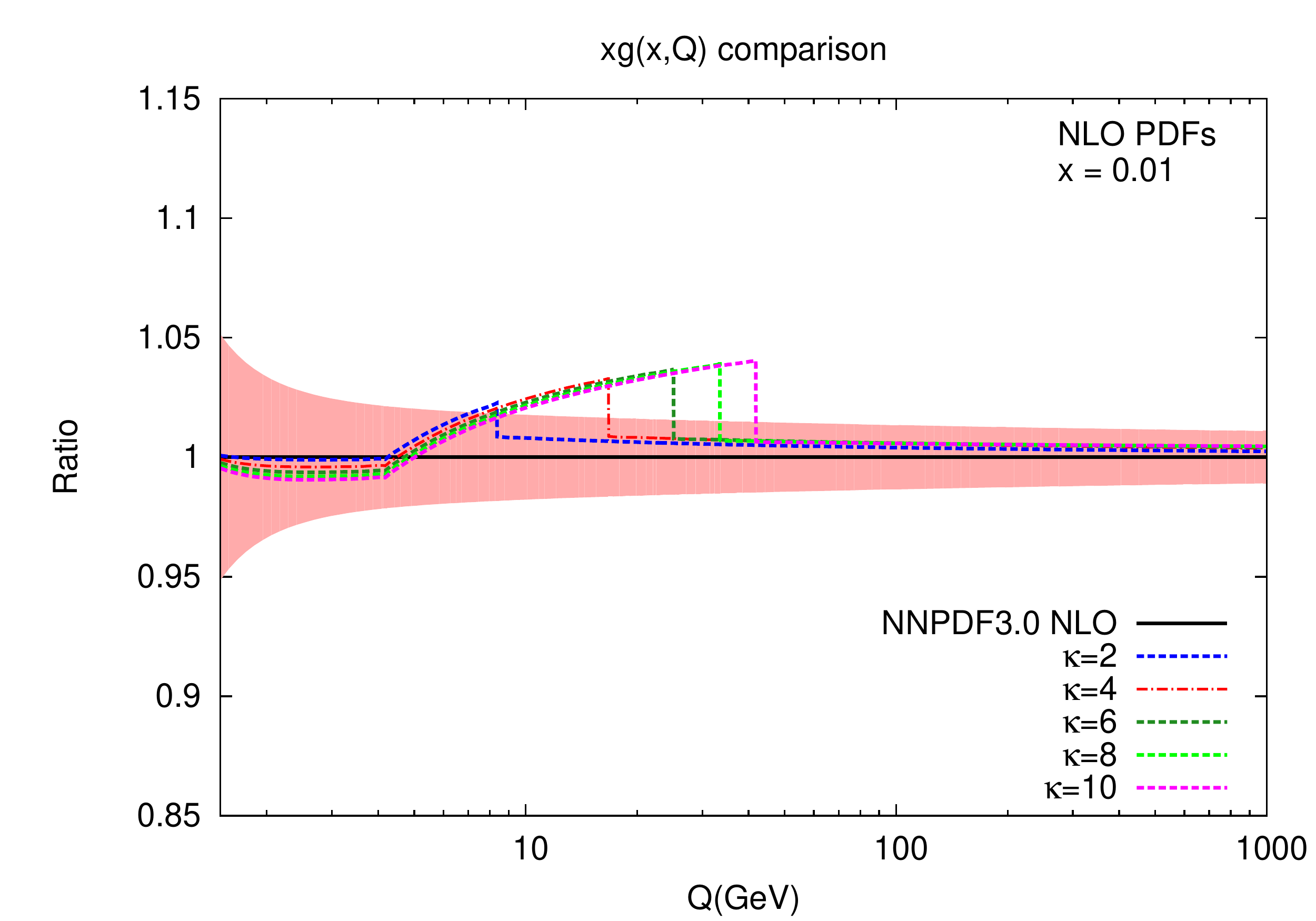}
\includegraphics[width=0.33\textwidth,angle=-90]{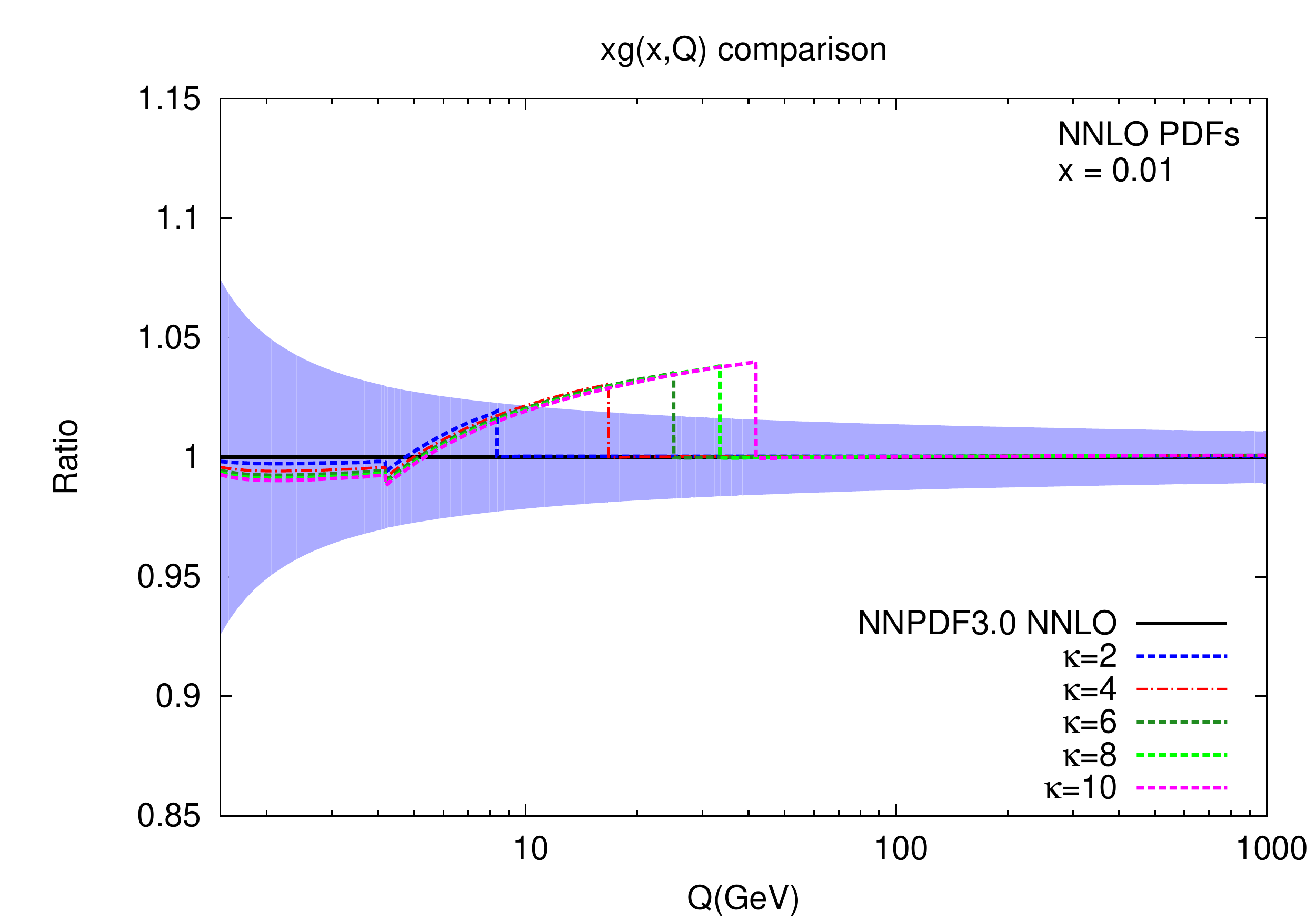}
\includegraphics[width=0.33\textwidth,angle=-90]{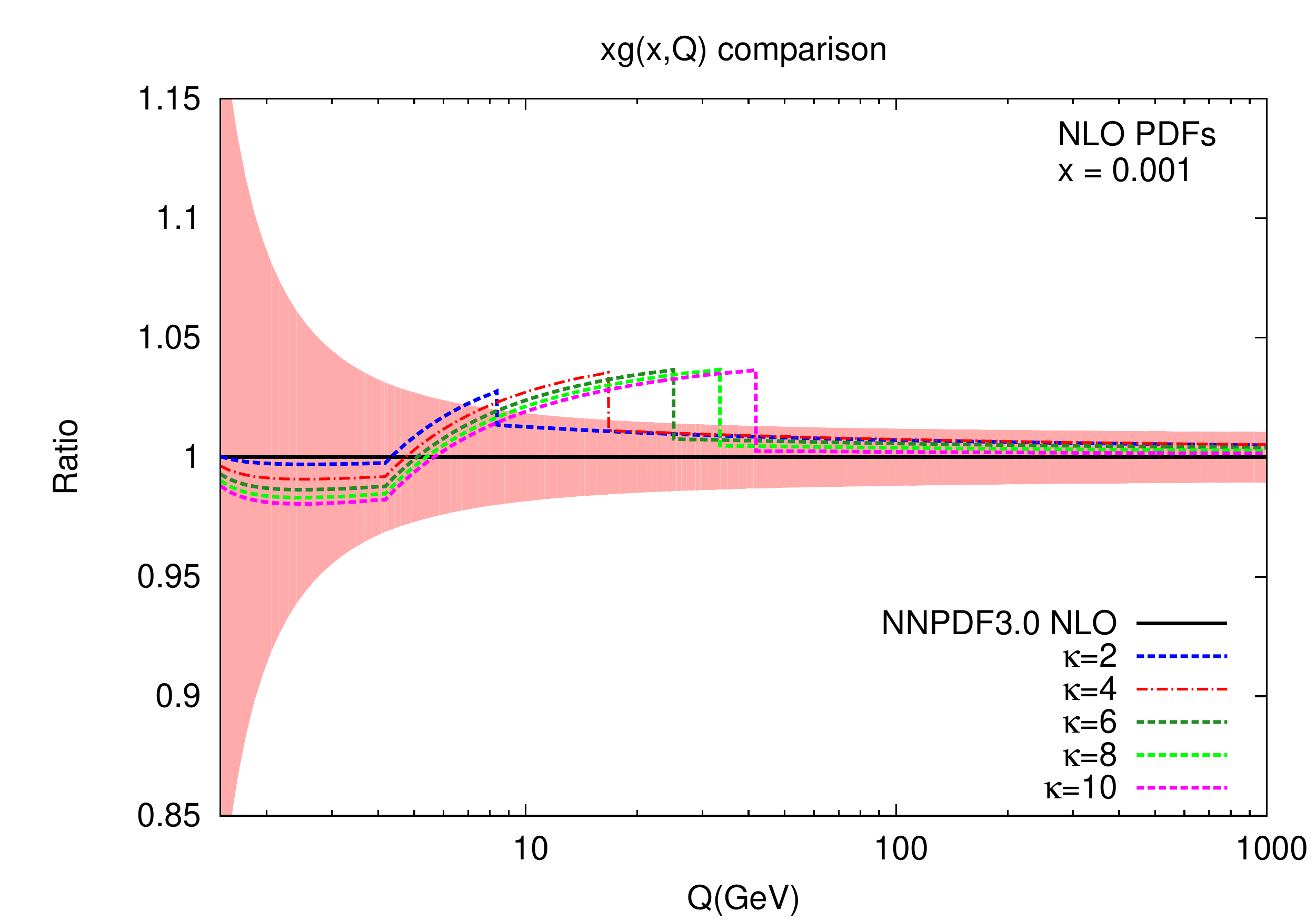}
\includegraphics[width=0.33\textwidth,angle=-90]{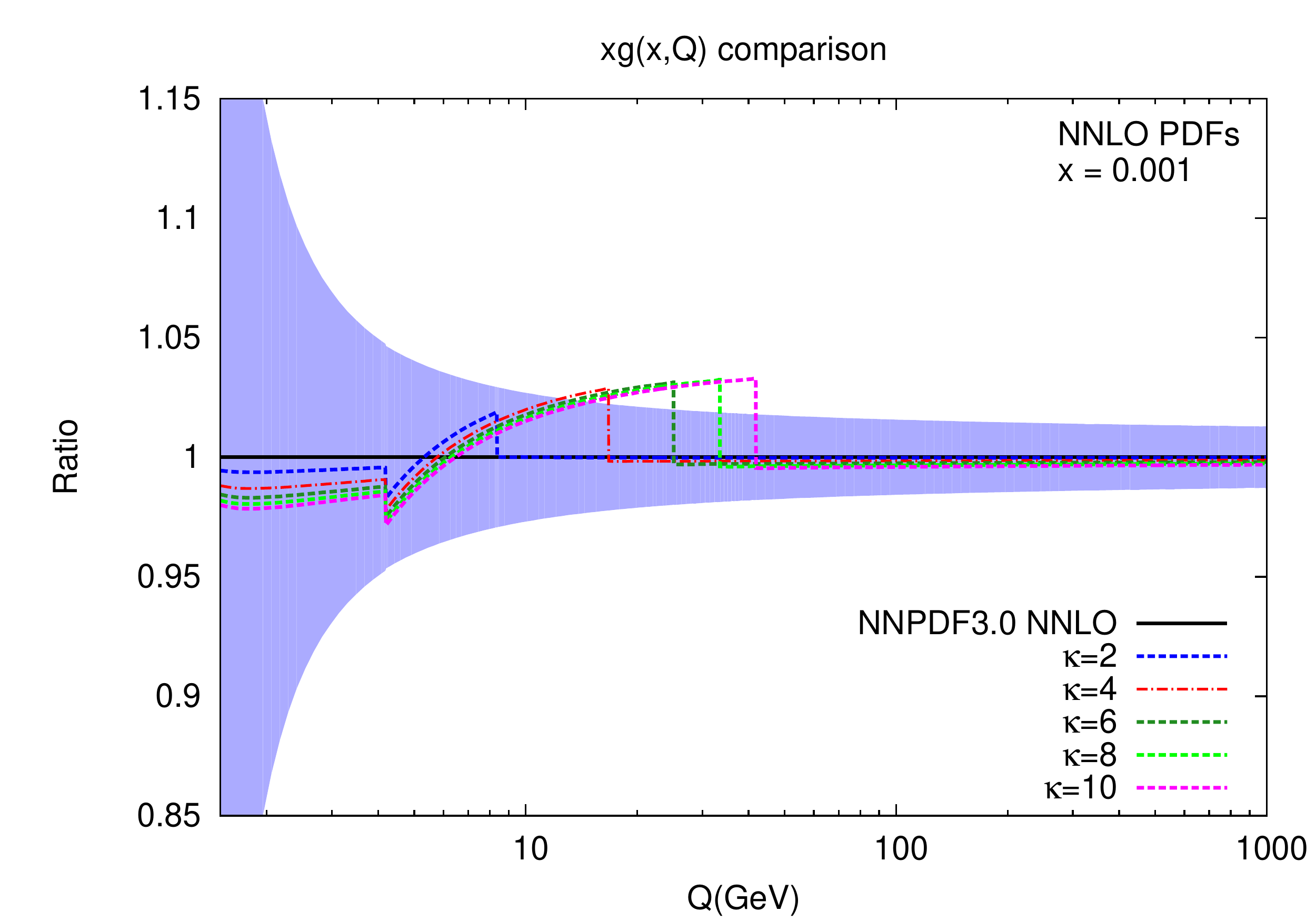}
\includegraphics[width=0.33\textwidth,angle=-90]{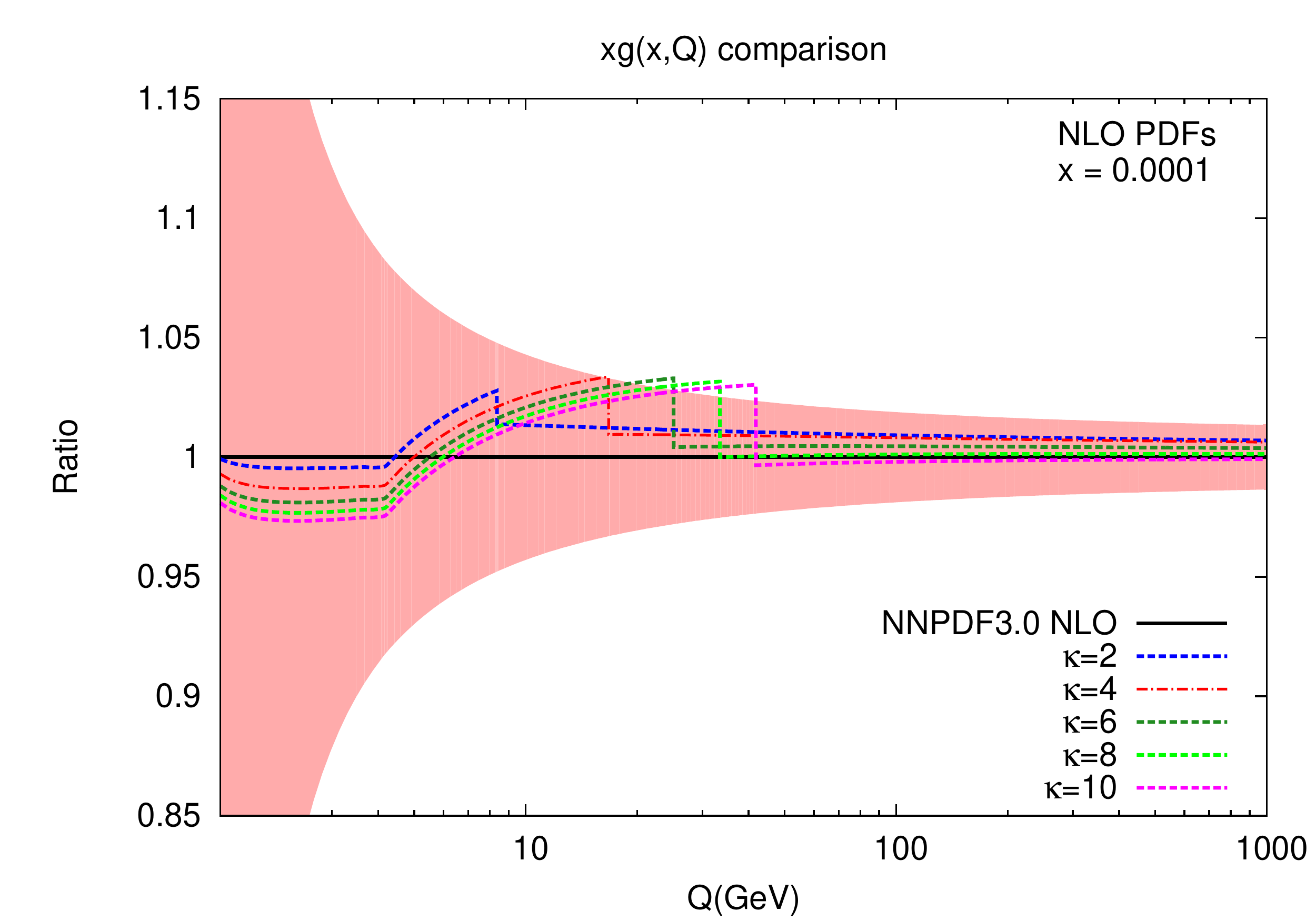}
\includegraphics[width=0.33\textwidth,angle=-90]{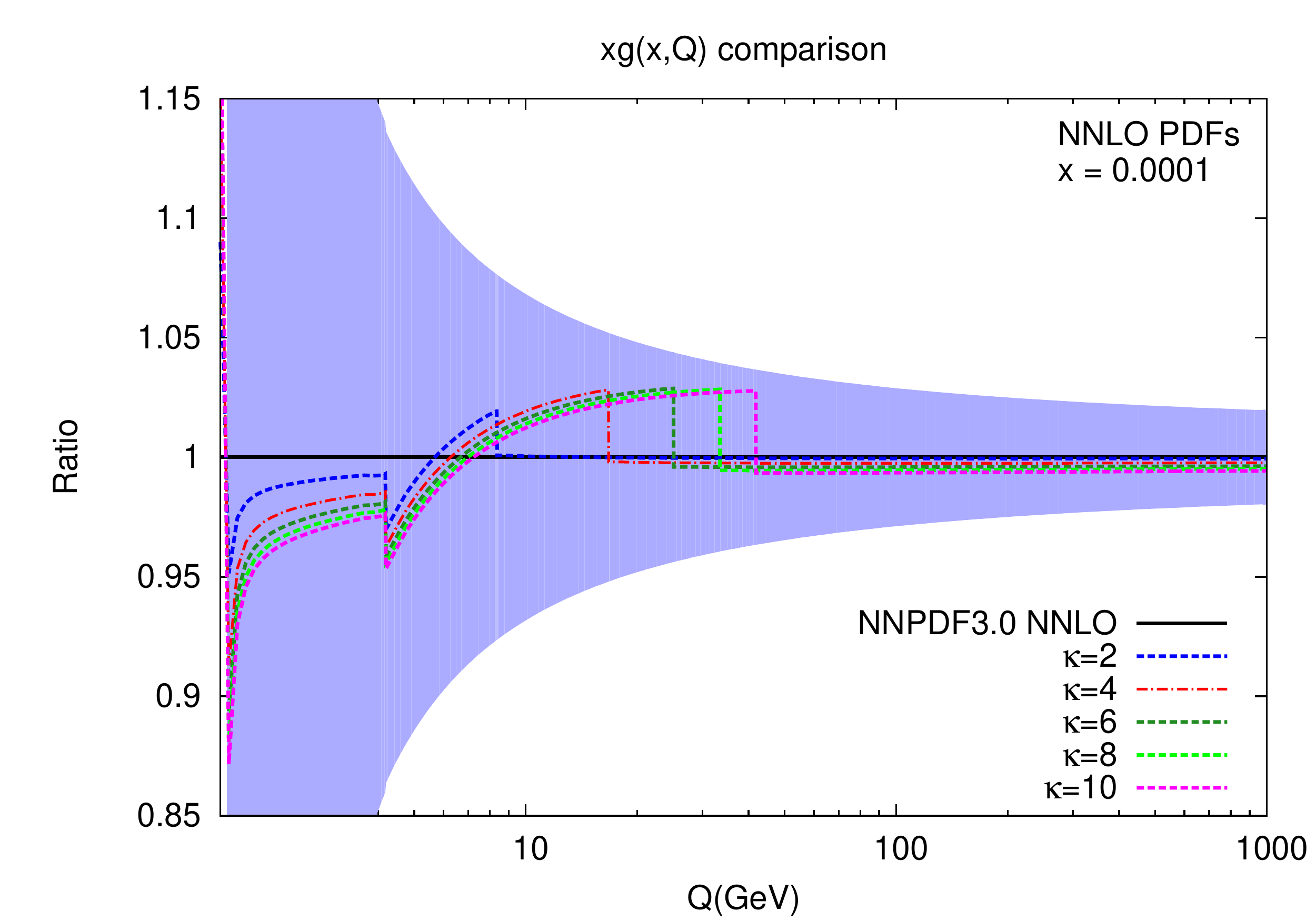}
\caption{Ratio between the gluon pdf for $\k=$2,4,6,8,10 and the standard $\k=$1 at NLO (left) and NNLO (right) as a function of the factorization scale $Q$ and for four values of the partonic fraction $x$.}
\label{fig:gluonpdf}
\end{figure}
\begin{figure}[t]
\centering
\includegraphics[width=0.33\textwidth,angle=-90]{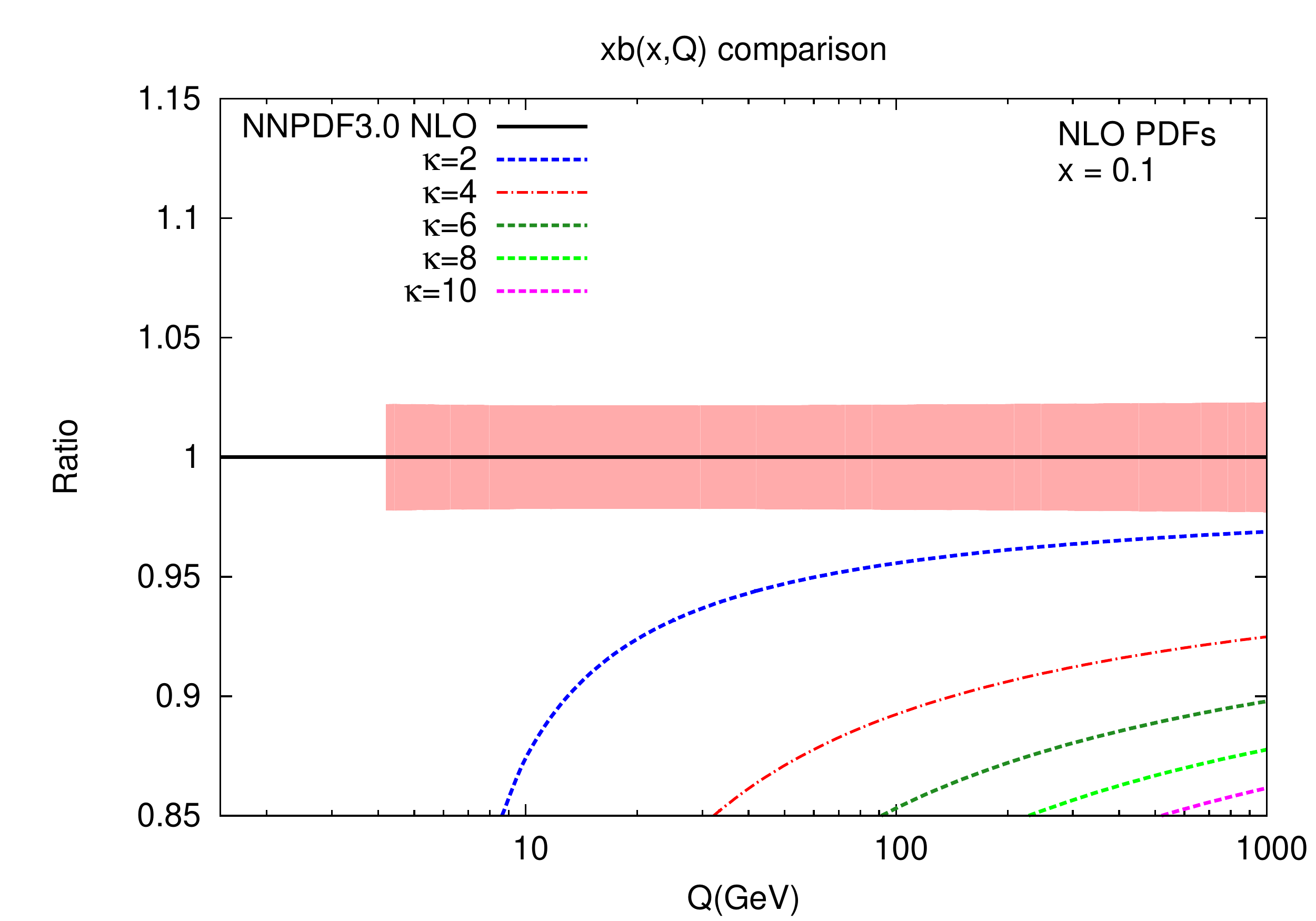}
\includegraphics[width=0.33\textwidth,angle=-90]{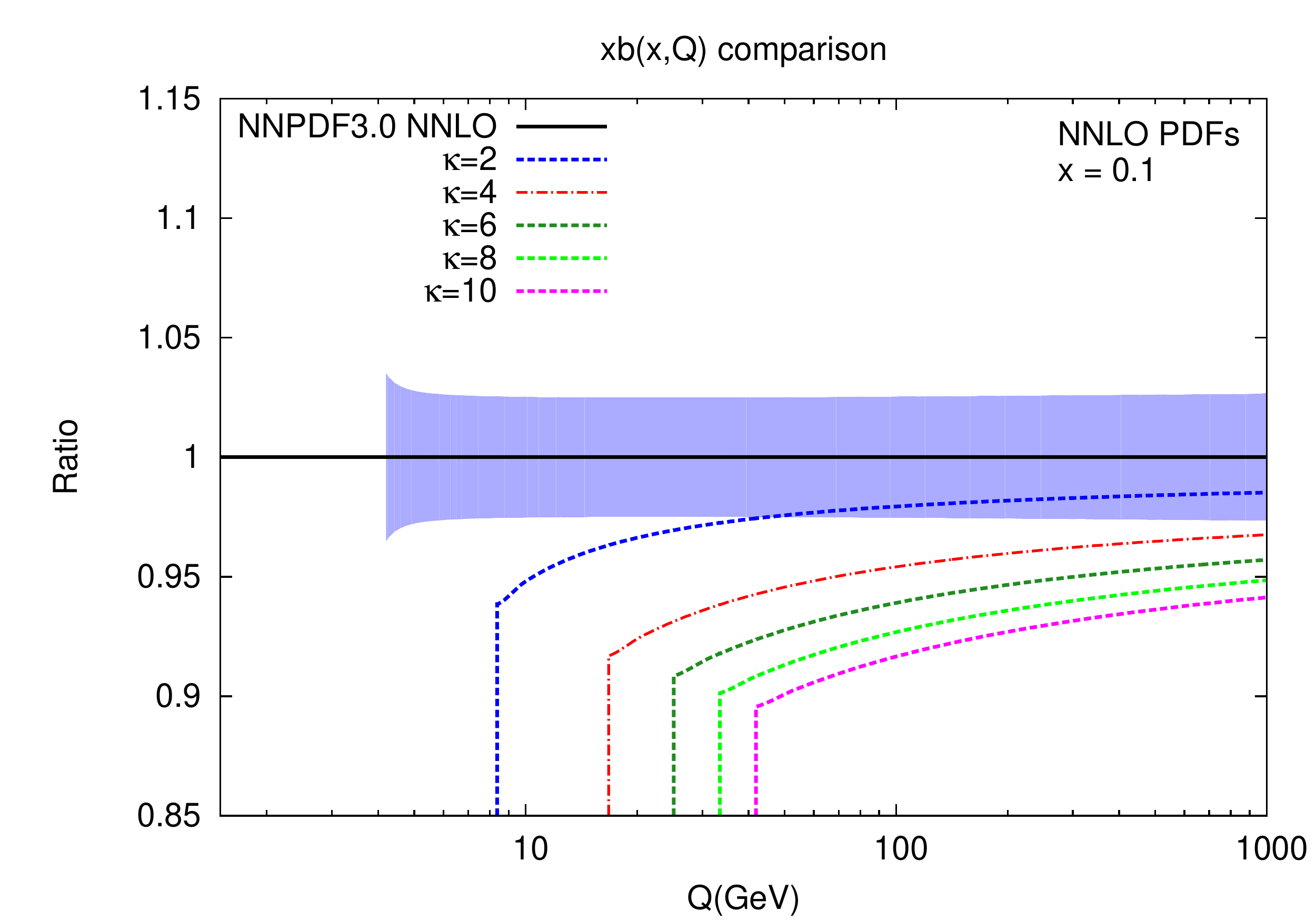}
\includegraphics[width=0.33\textwidth,angle=-90]{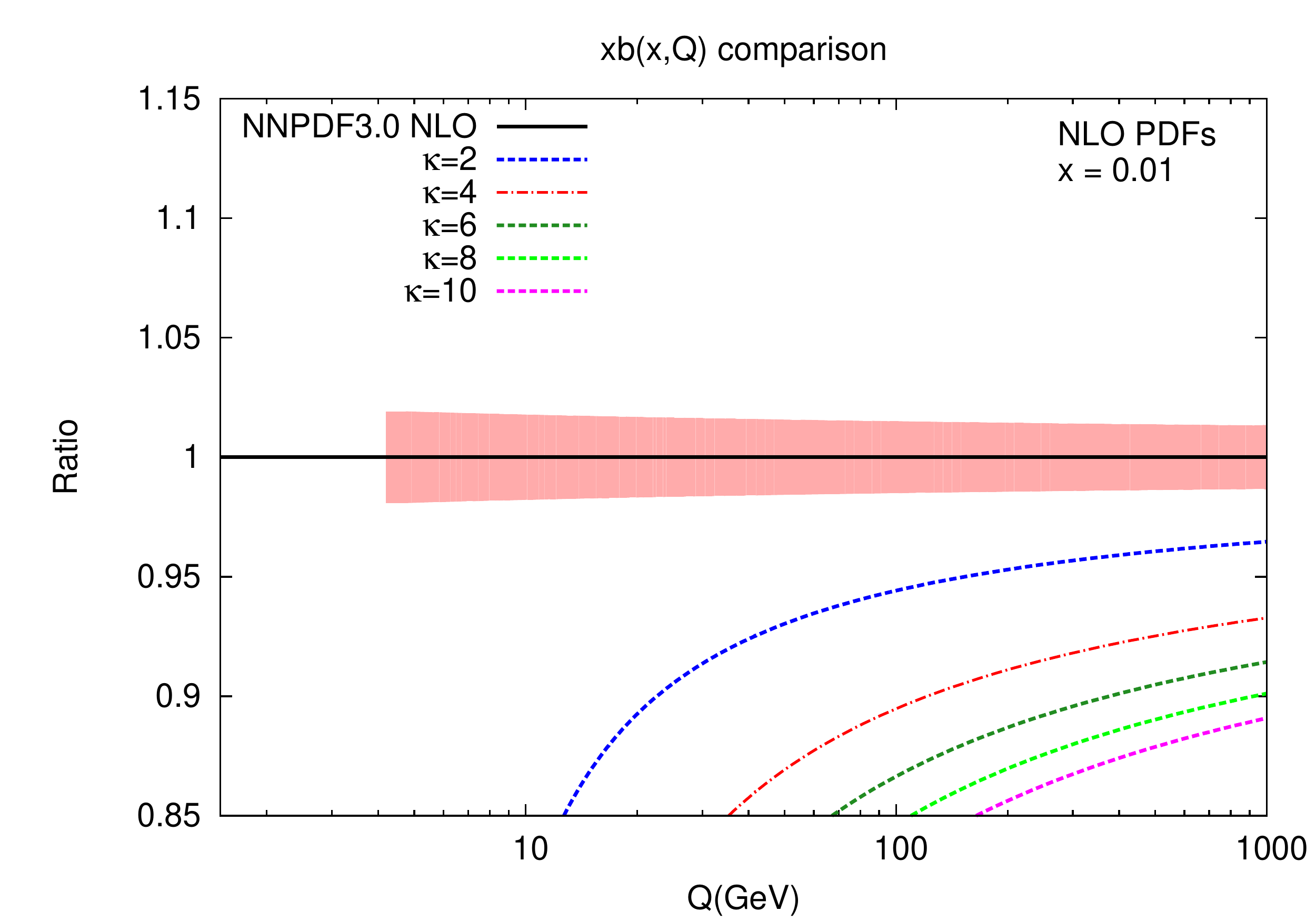}
\includegraphics[width=0.33\textwidth,angle=-90]{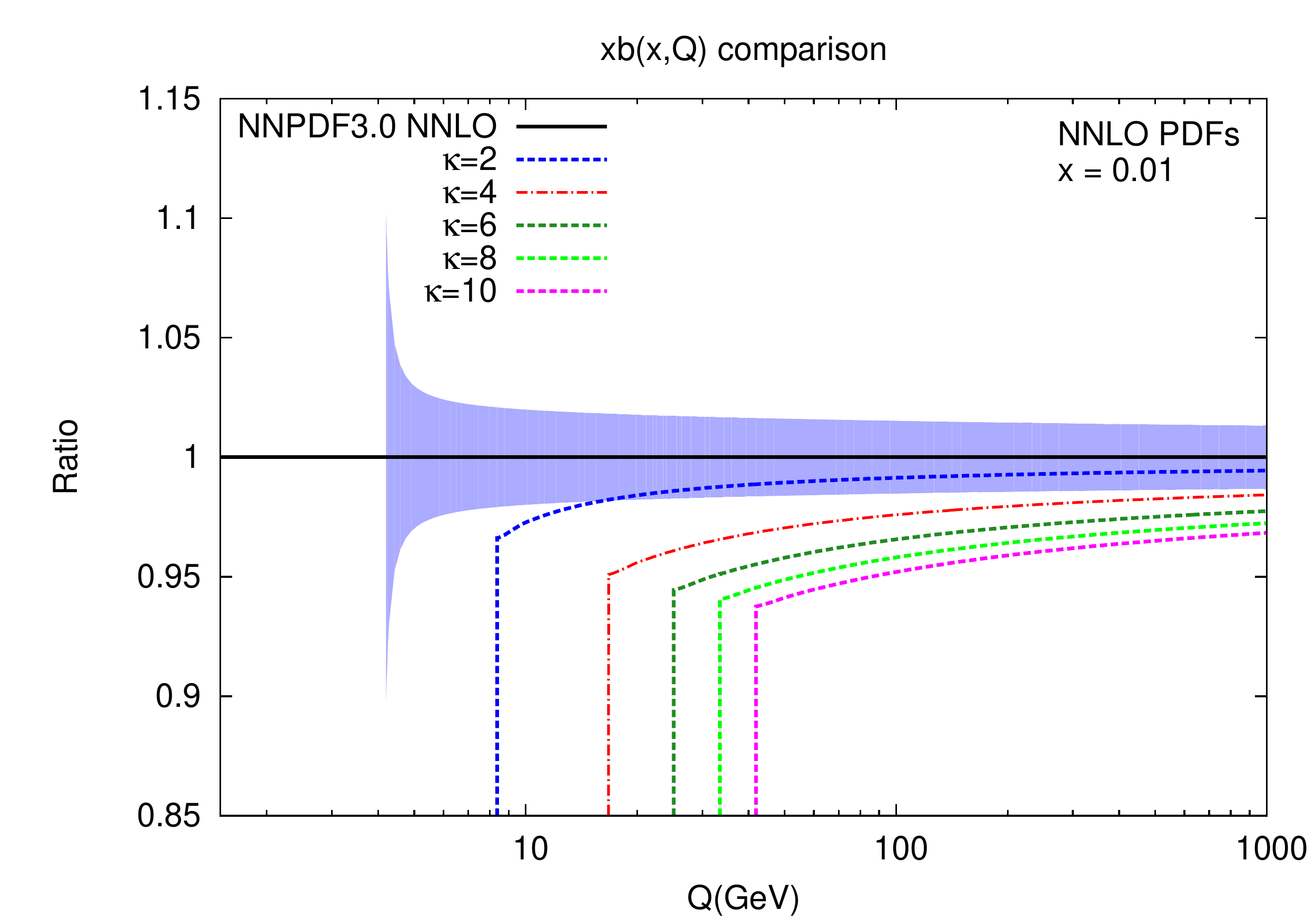}
\includegraphics[width=0.33\textwidth,angle=-90]{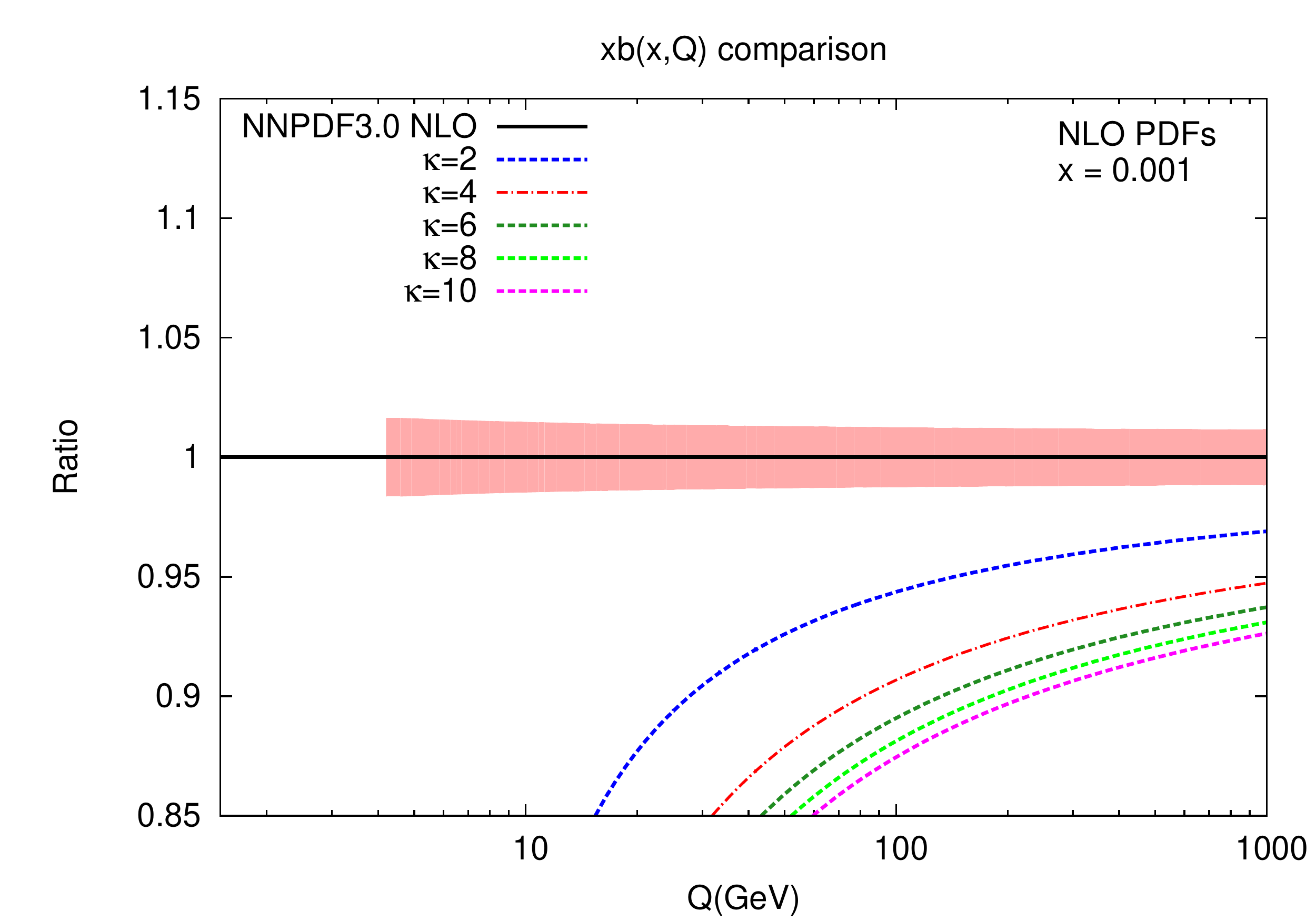}
\includegraphics[width=0.33\textwidth,angle=-90]{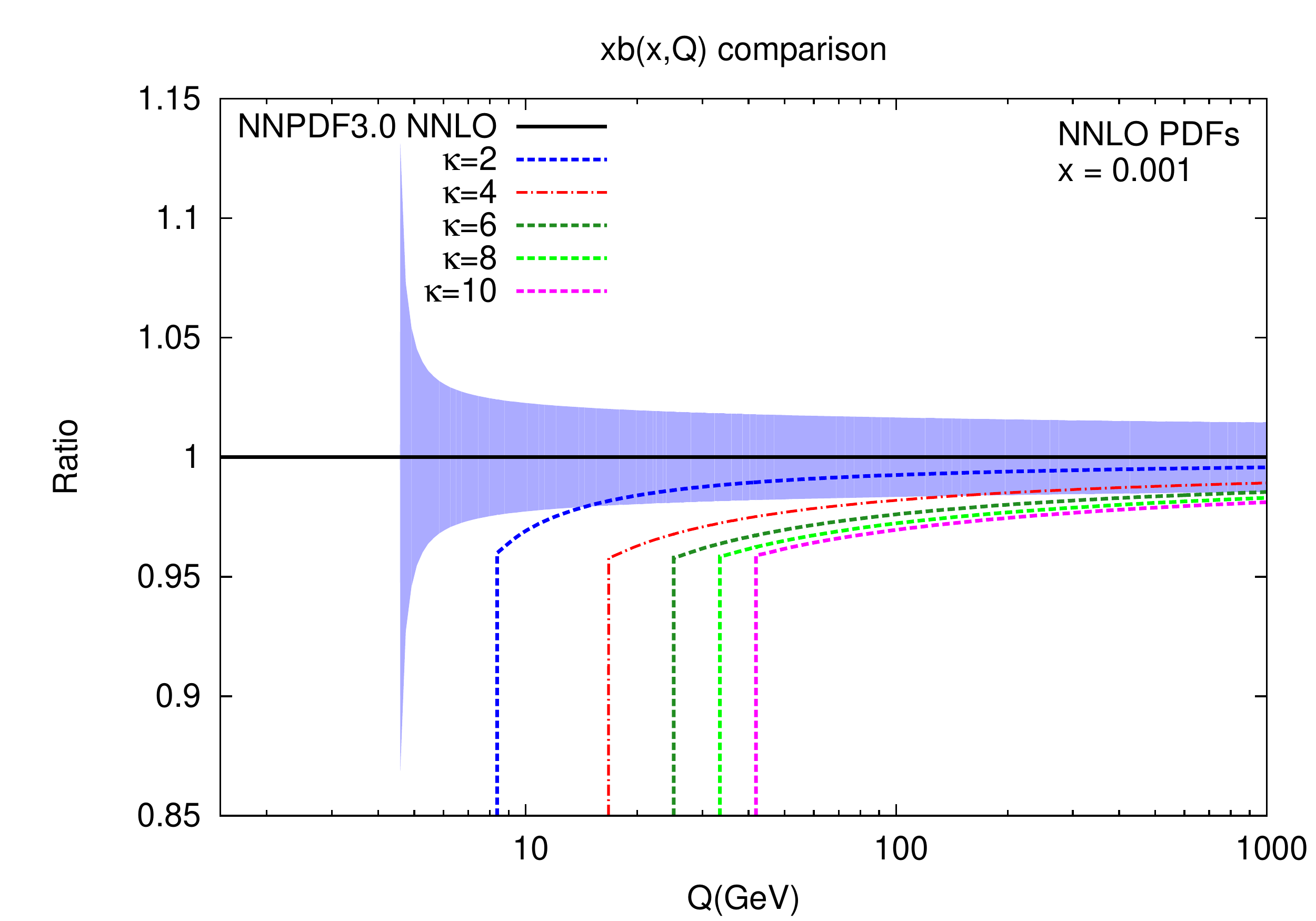}
\includegraphics[width=0.33\textwidth,angle=-90]{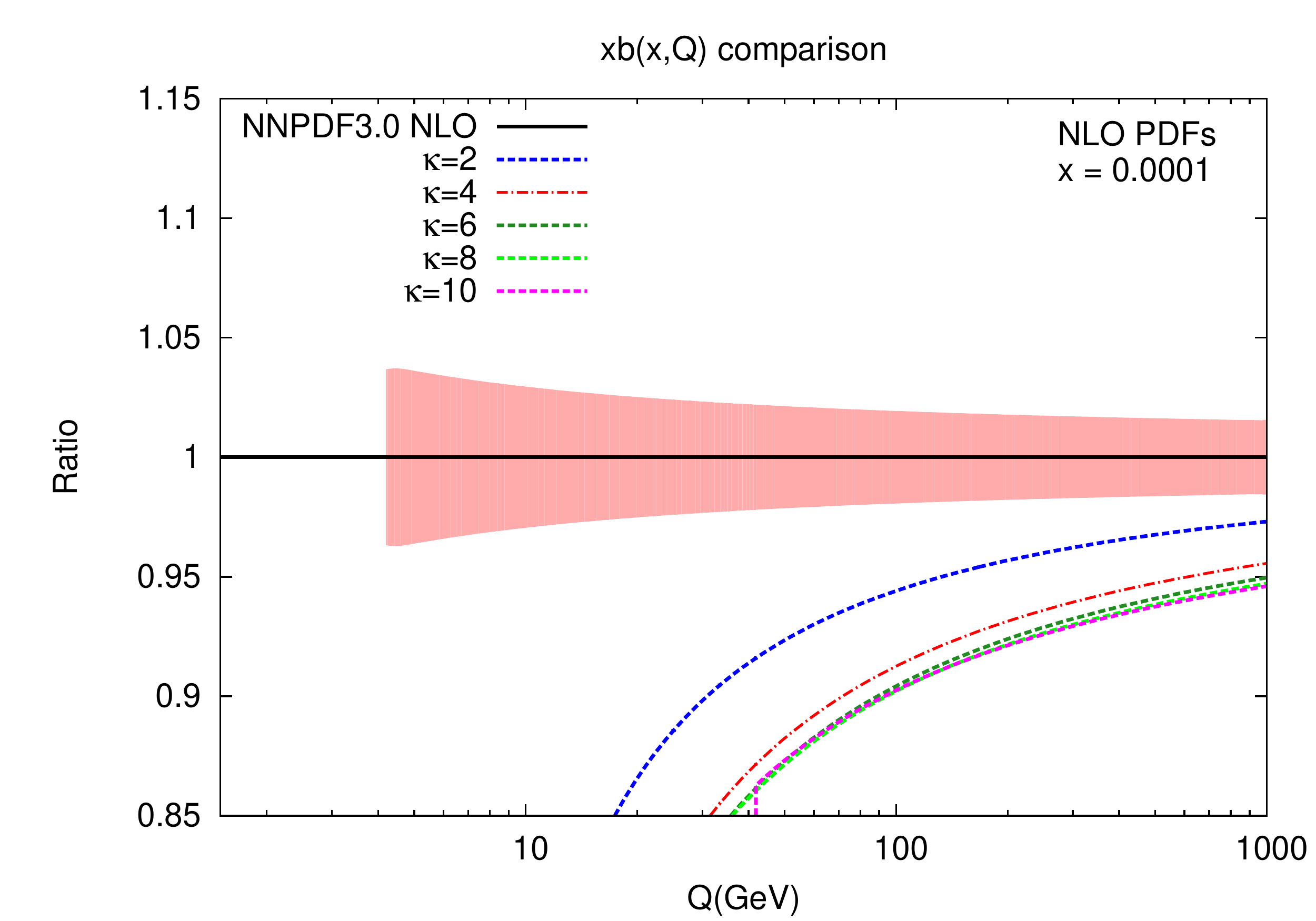}
\includegraphics[width=0.33\textwidth,angle=-90]{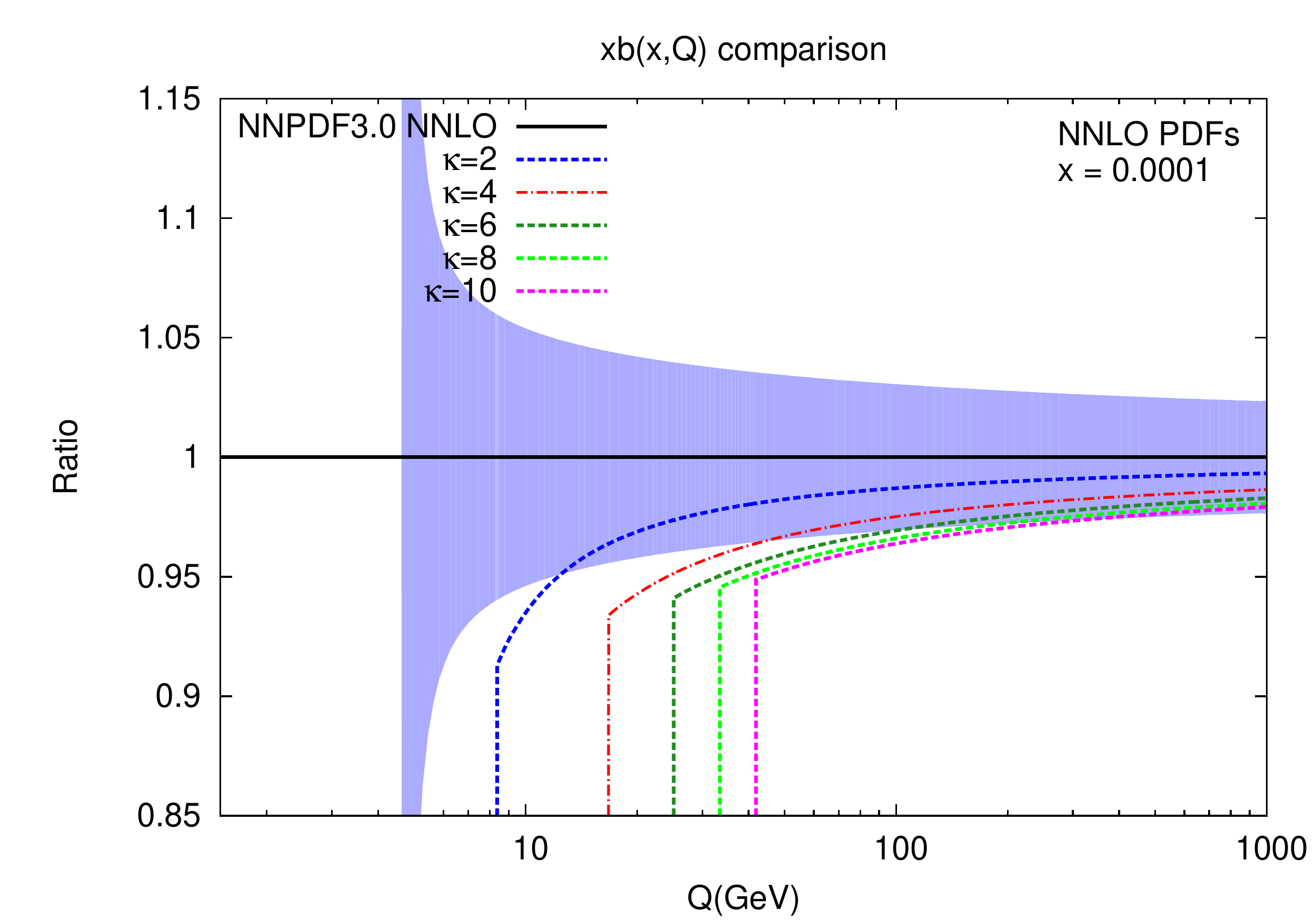}
\caption{As in fig.~\ref{fig:gluonpdf} but for the bottom pdf.}
\label{fig:bottompdf}
\end{figure}
\begin{figure}[h]
\centering
\includegraphics[width=0.33\textwidth,angle=-90]{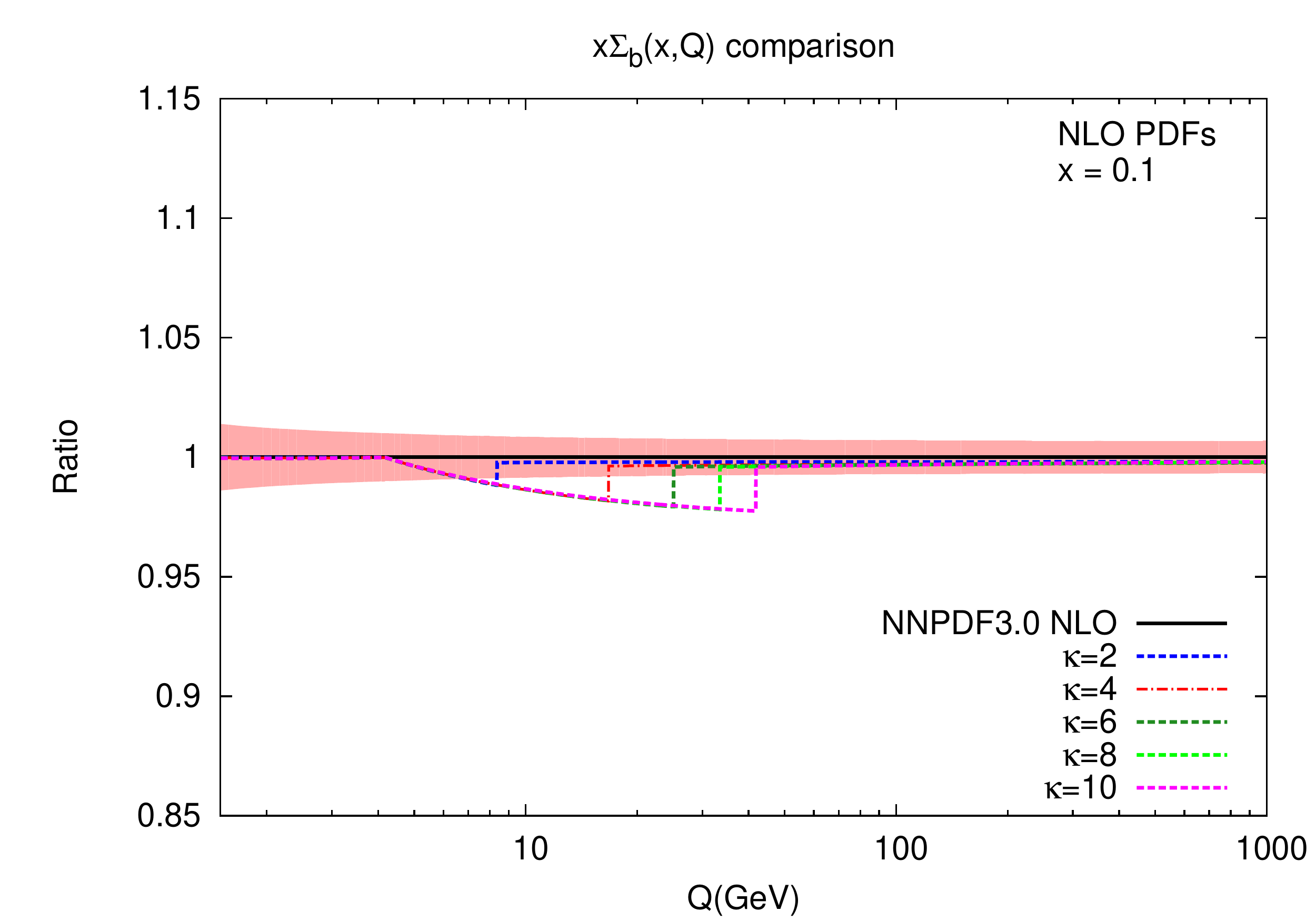}
\includegraphics[width=0.33\textwidth,angle=-90]{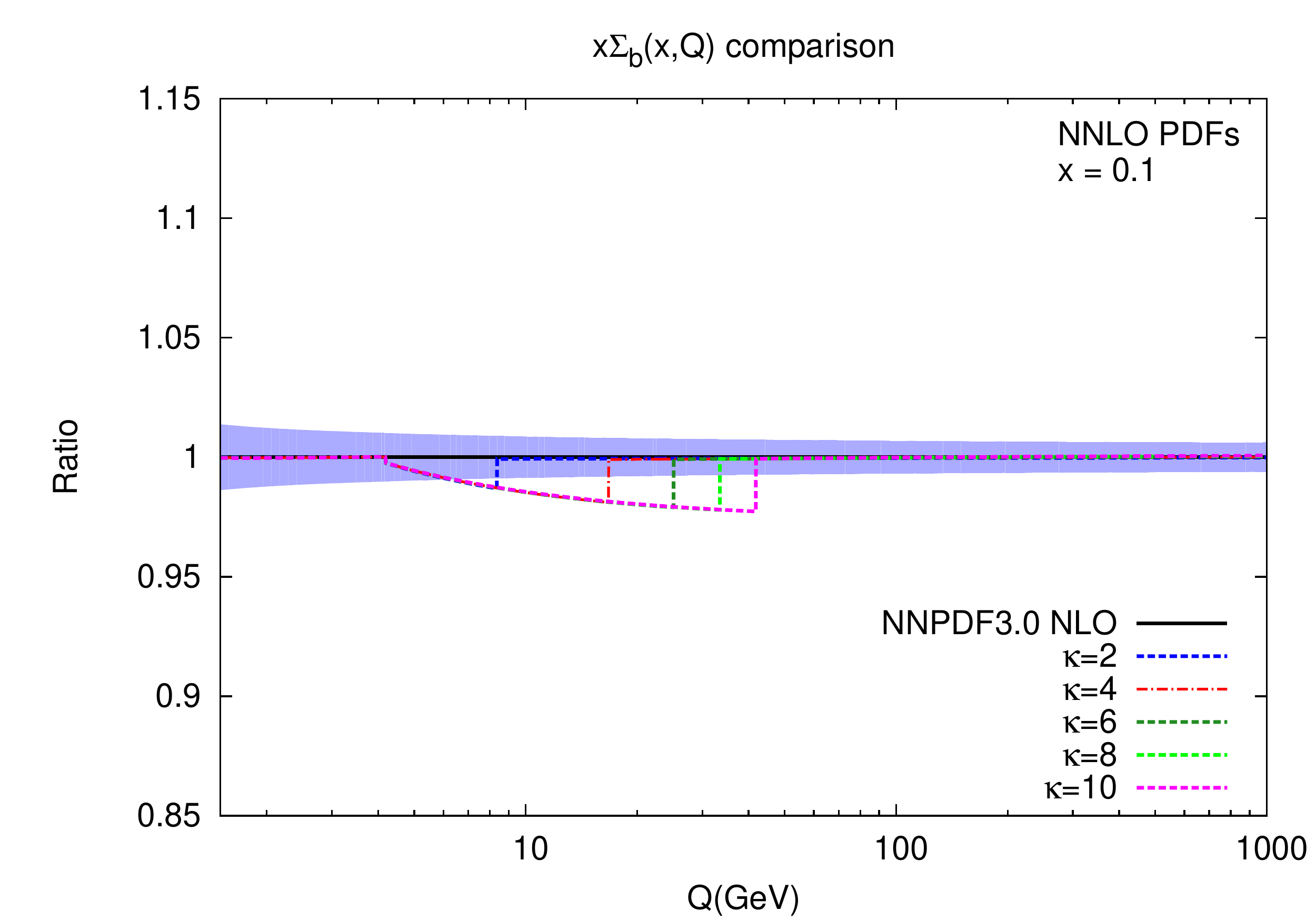}
\includegraphics[width=0.33\textwidth,angle=-90]{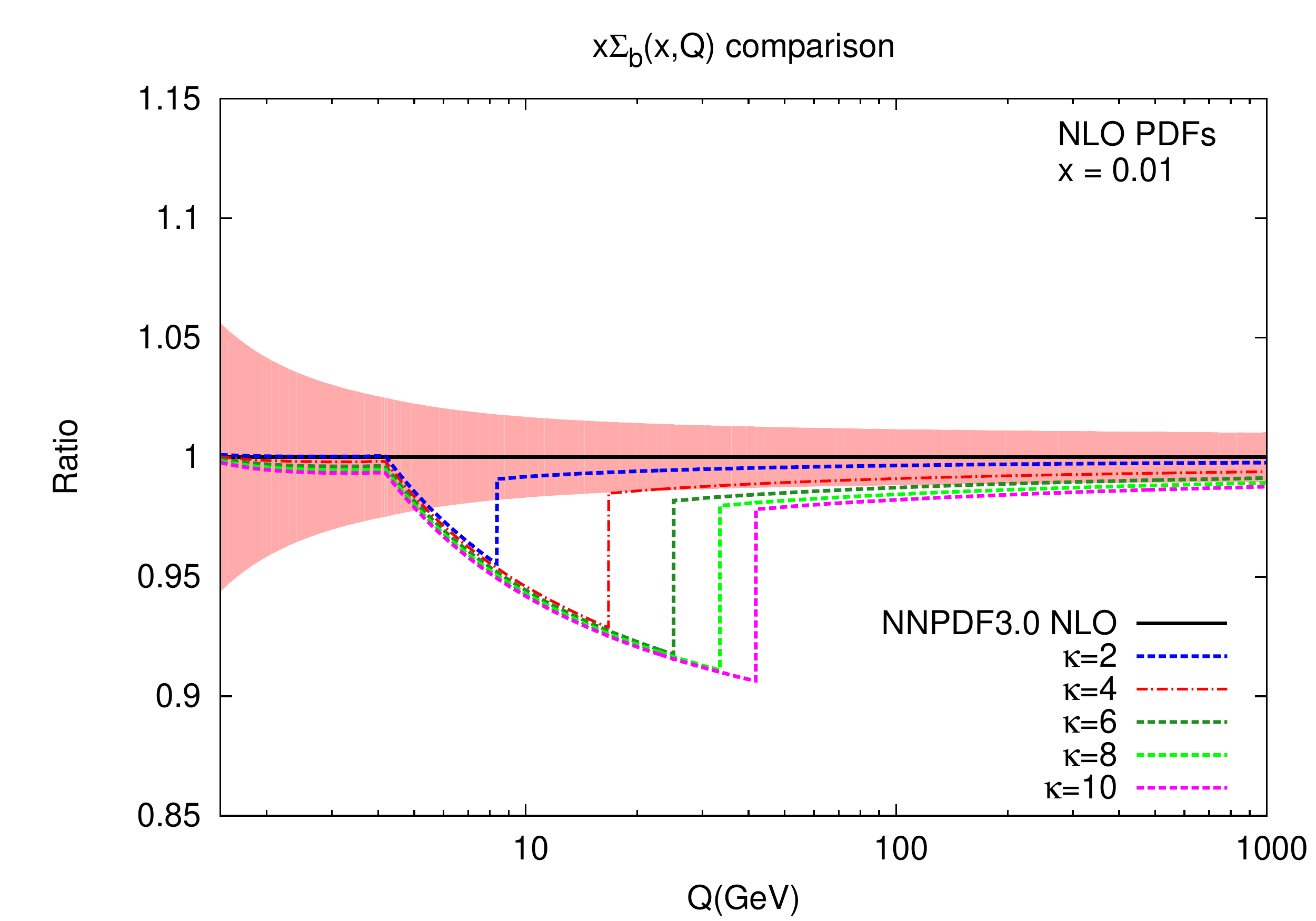}
\includegraphics[width=0.33\textwidth,angle=-90]{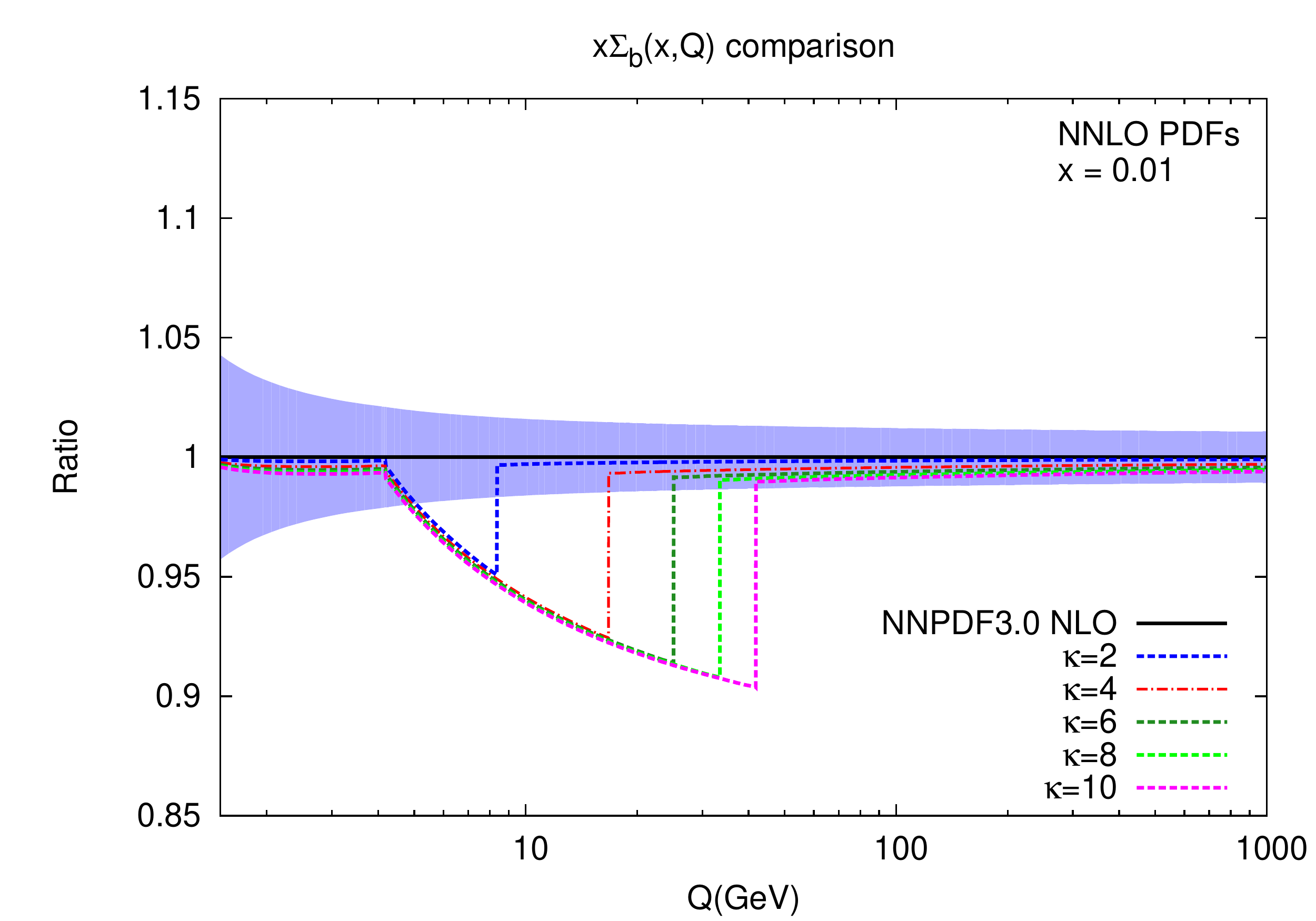}
\includegraphics[width=0.33\textwidth,angle=-90]{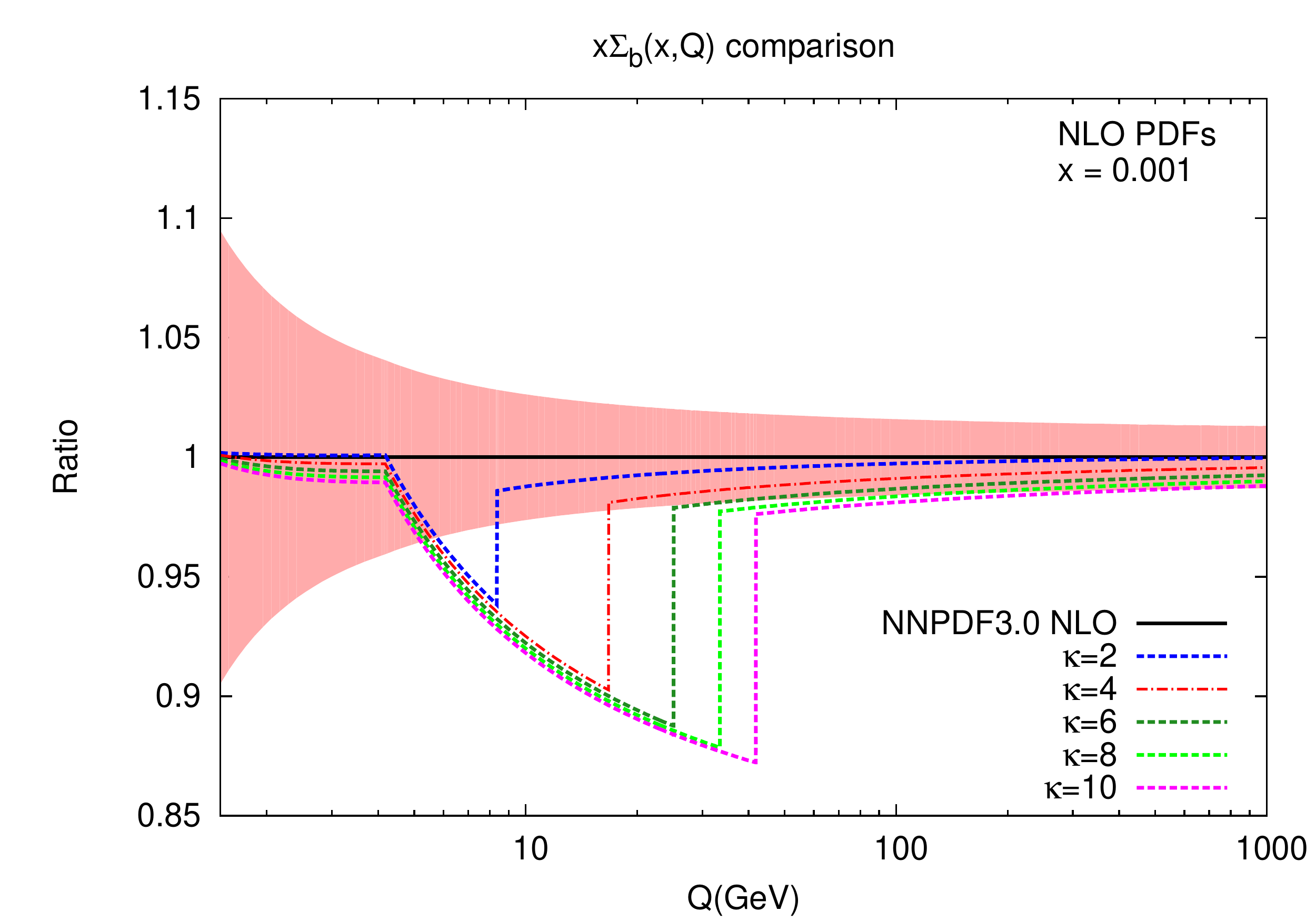}
\includegraphics[width=0.33\textwidth,angle=-90]{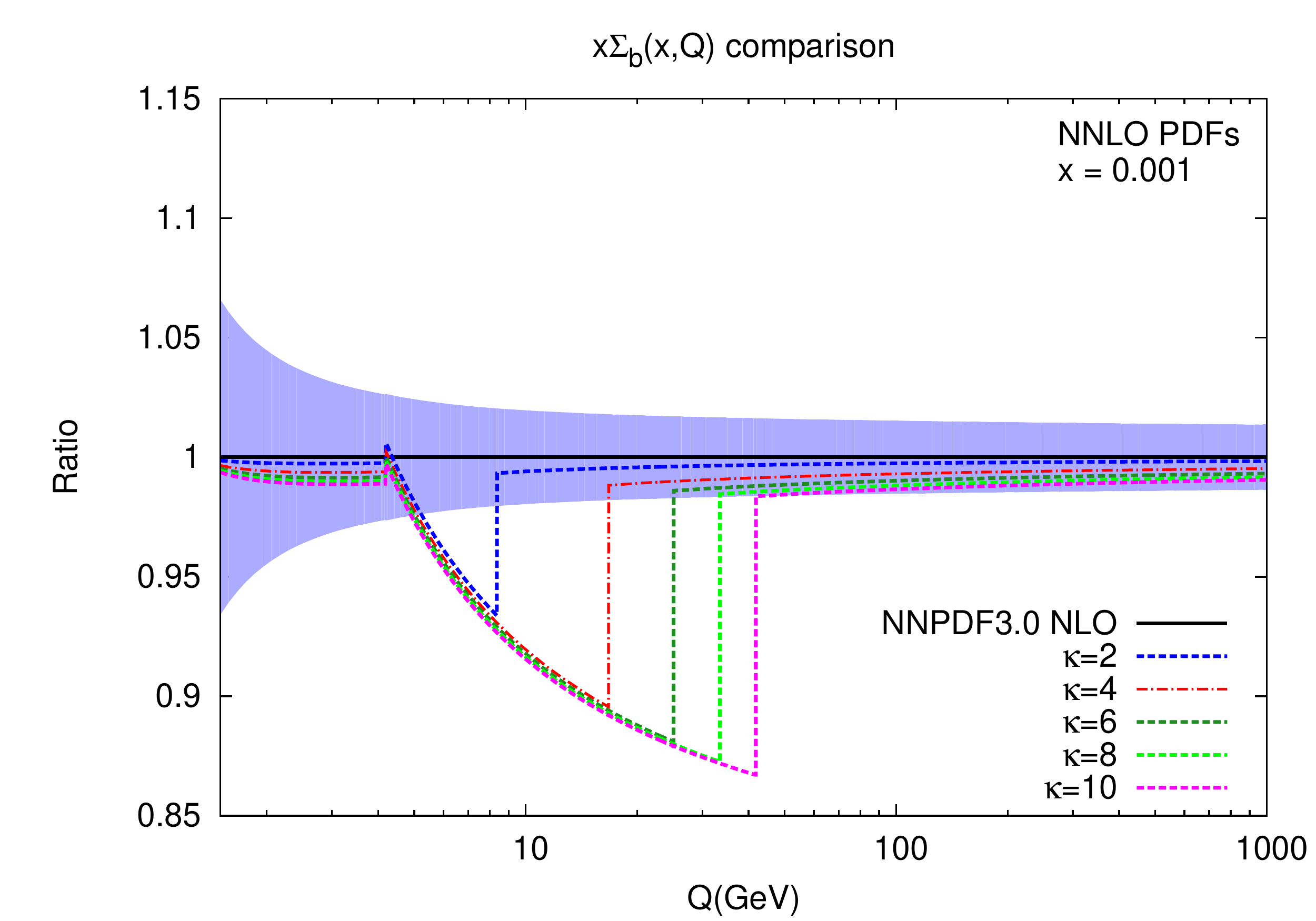}
\includegraphics[width=0.33\textwidth,angle=-90]{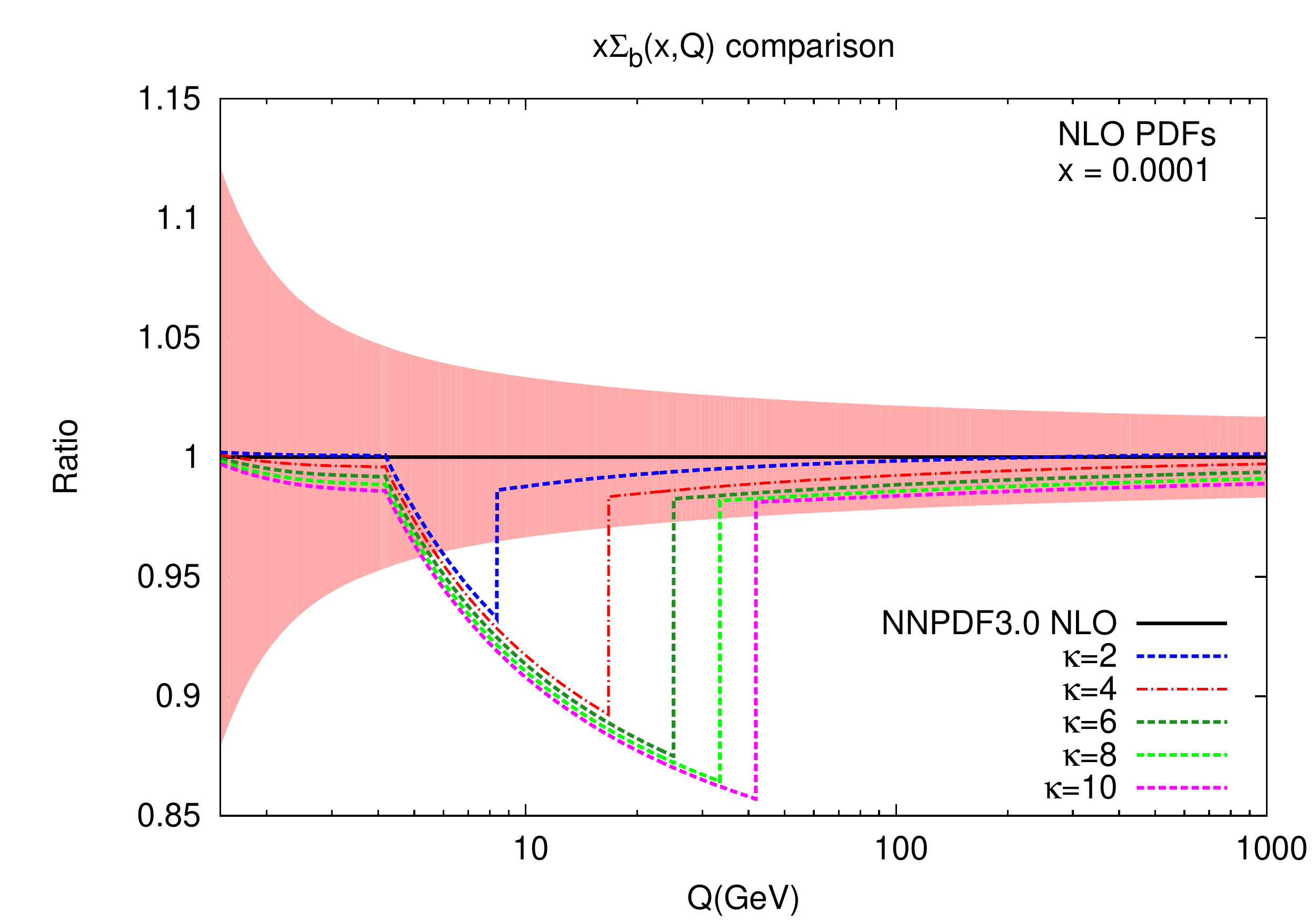}
\includegraphics[width=0.33\textwidth,angle=-90]{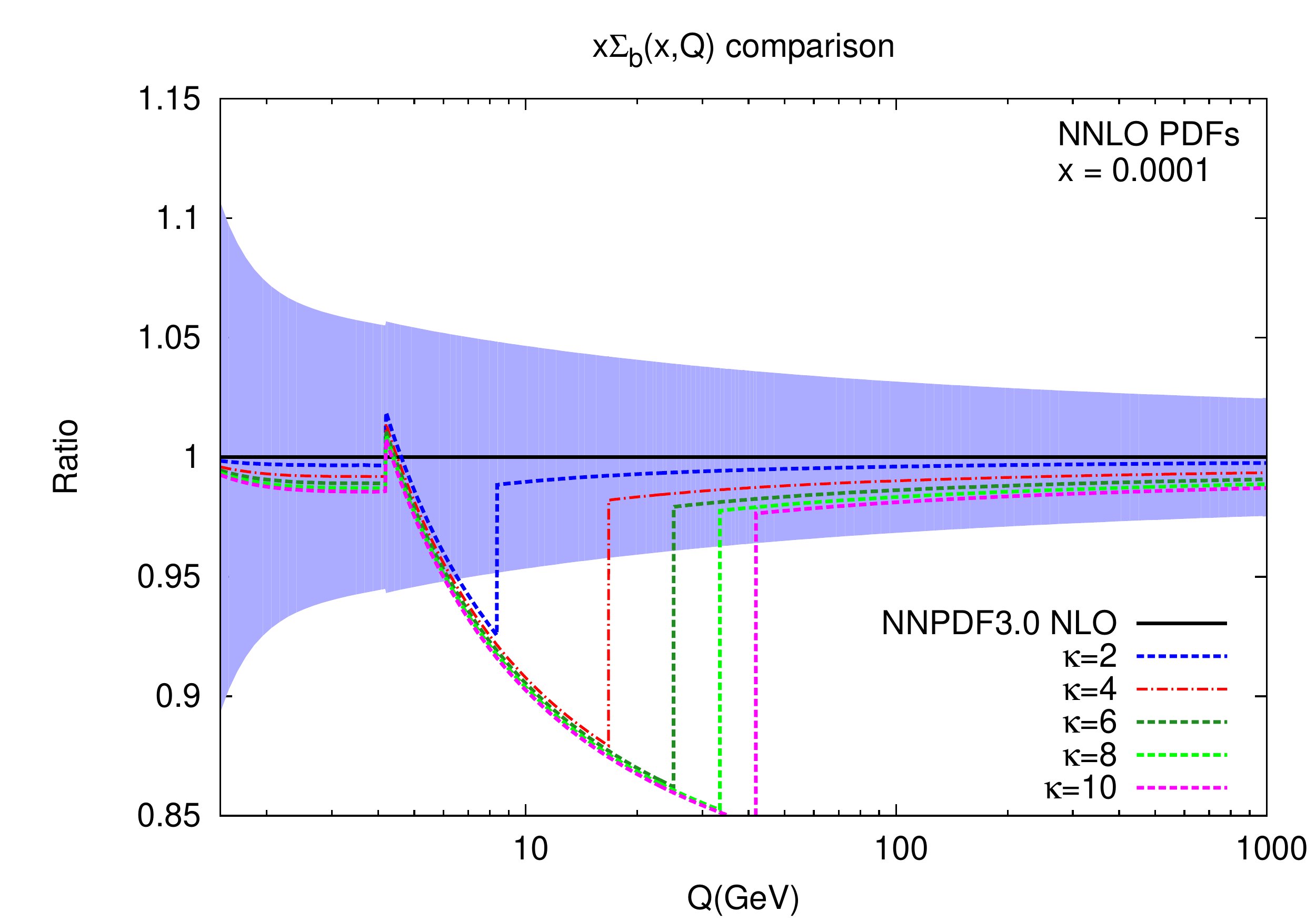}
\caption{As in fig.~\ref{fig:gluonpdf} but for $\Sigma=u+\bar{u}+d+\bar{d}+s+\bar{s}+c+\bar{c}+b+\bar{b}$.}
\label{fig:singletb}
\end{figure}
\begin{figure}[h]
\centering
\includegraphics[width=0.33\textwidth,angle=-90]{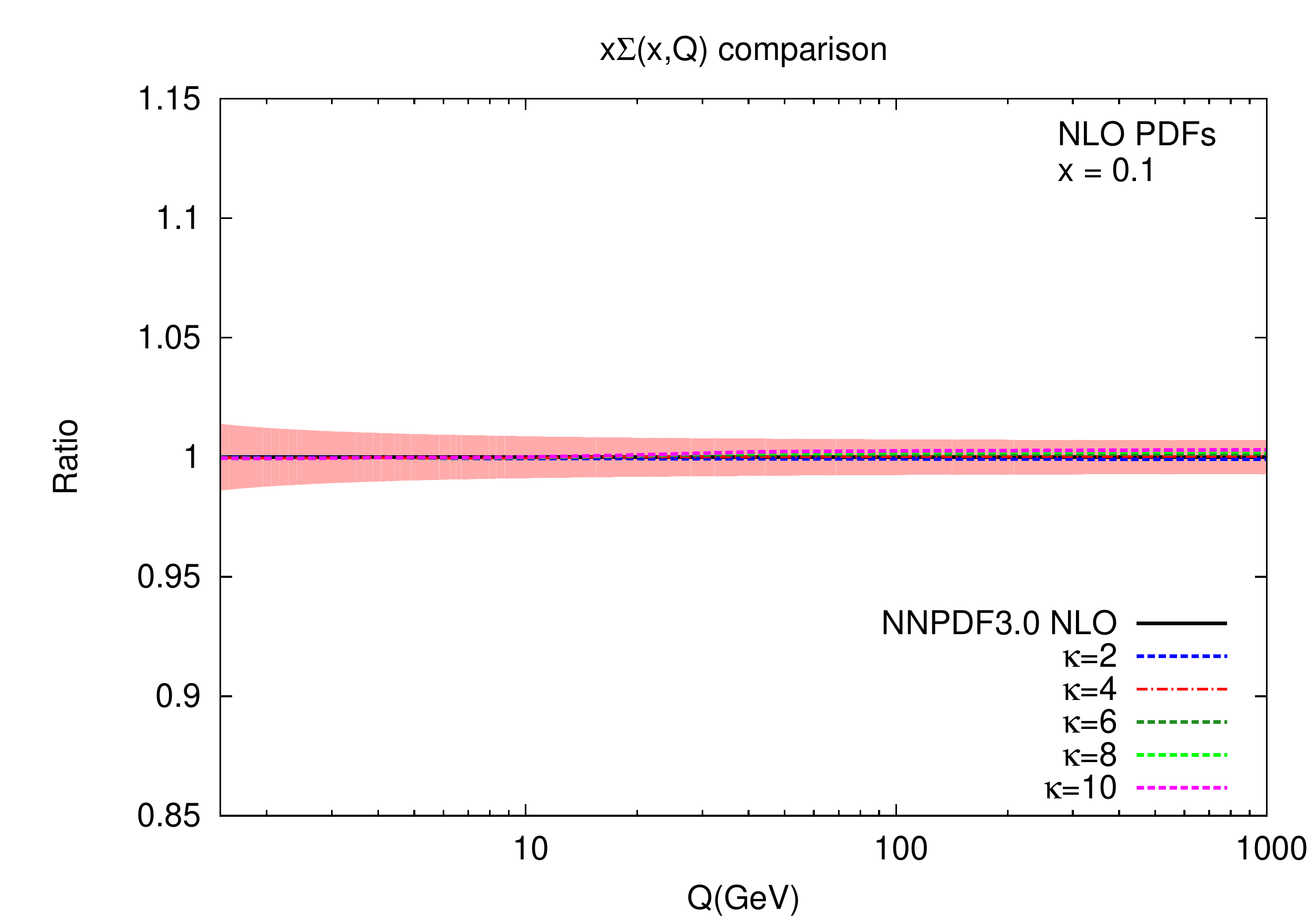}
\includegraphics[width=0.33\textwidth,angle=-90]{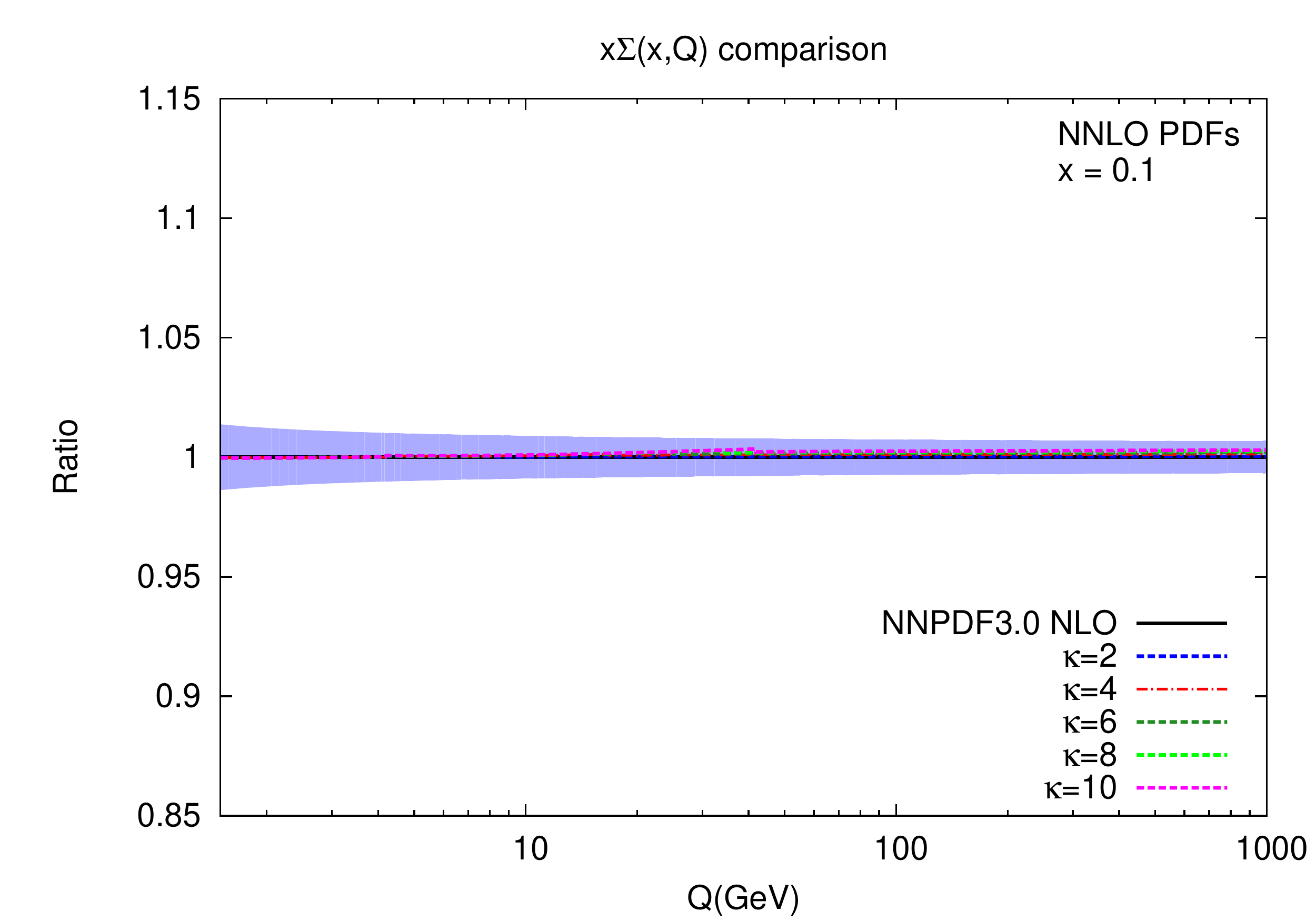}
\includegraphics[width=0.33\textwidth,angle=-90]{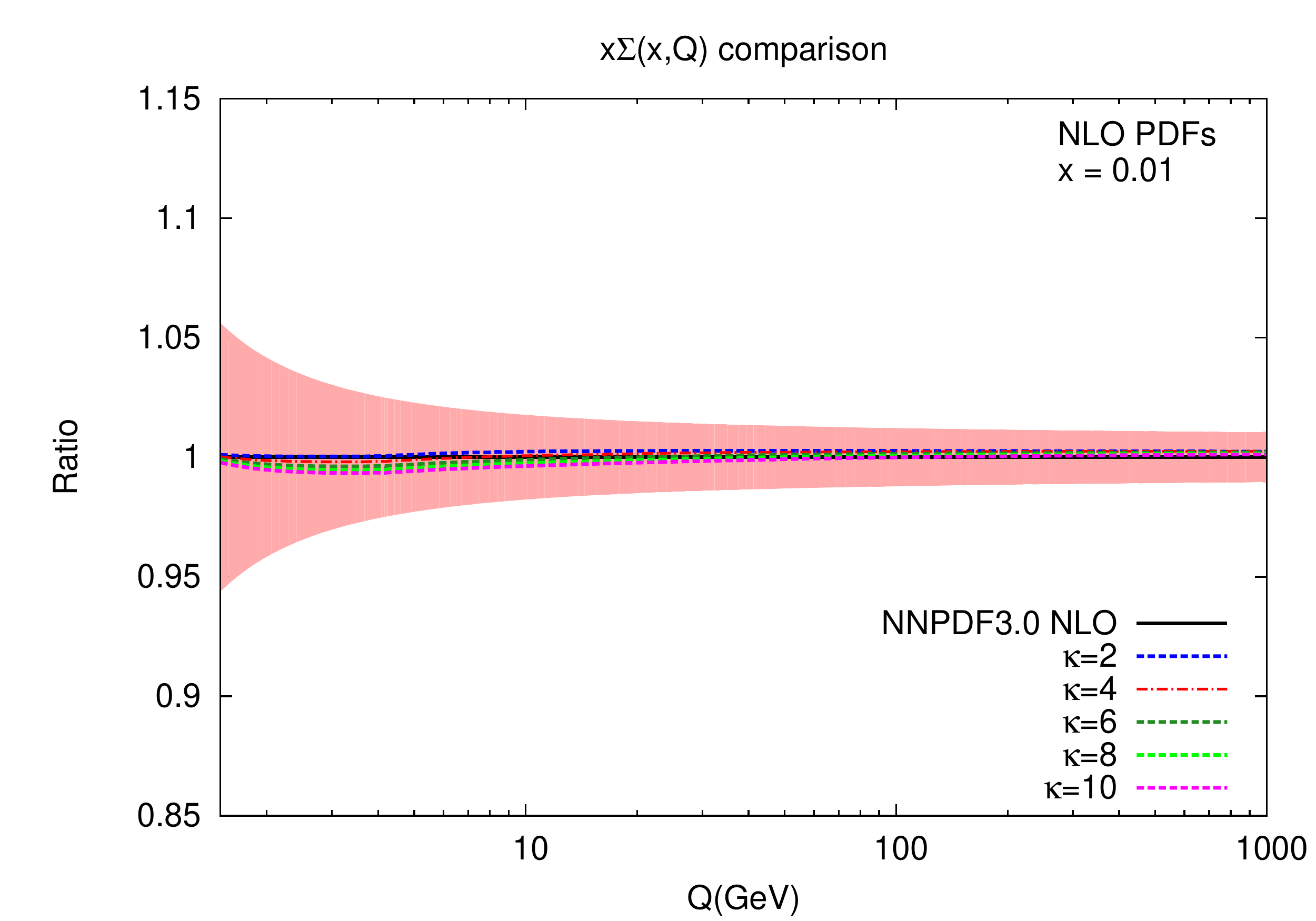}
\includegraphics[width=0.33\textwidth,angle=-90]{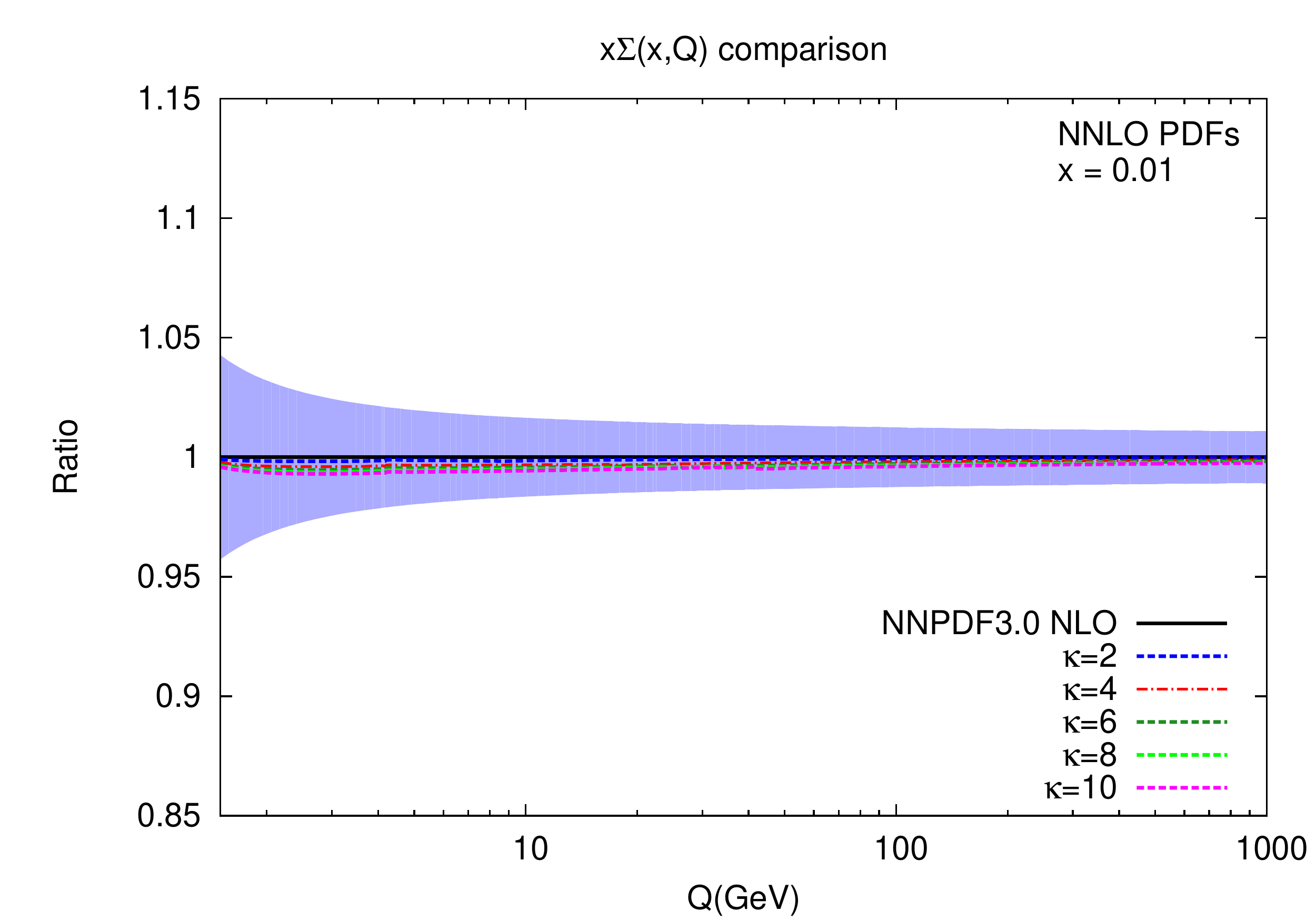}
\includegraphics[width=0.33\textwidth,angle=-90]{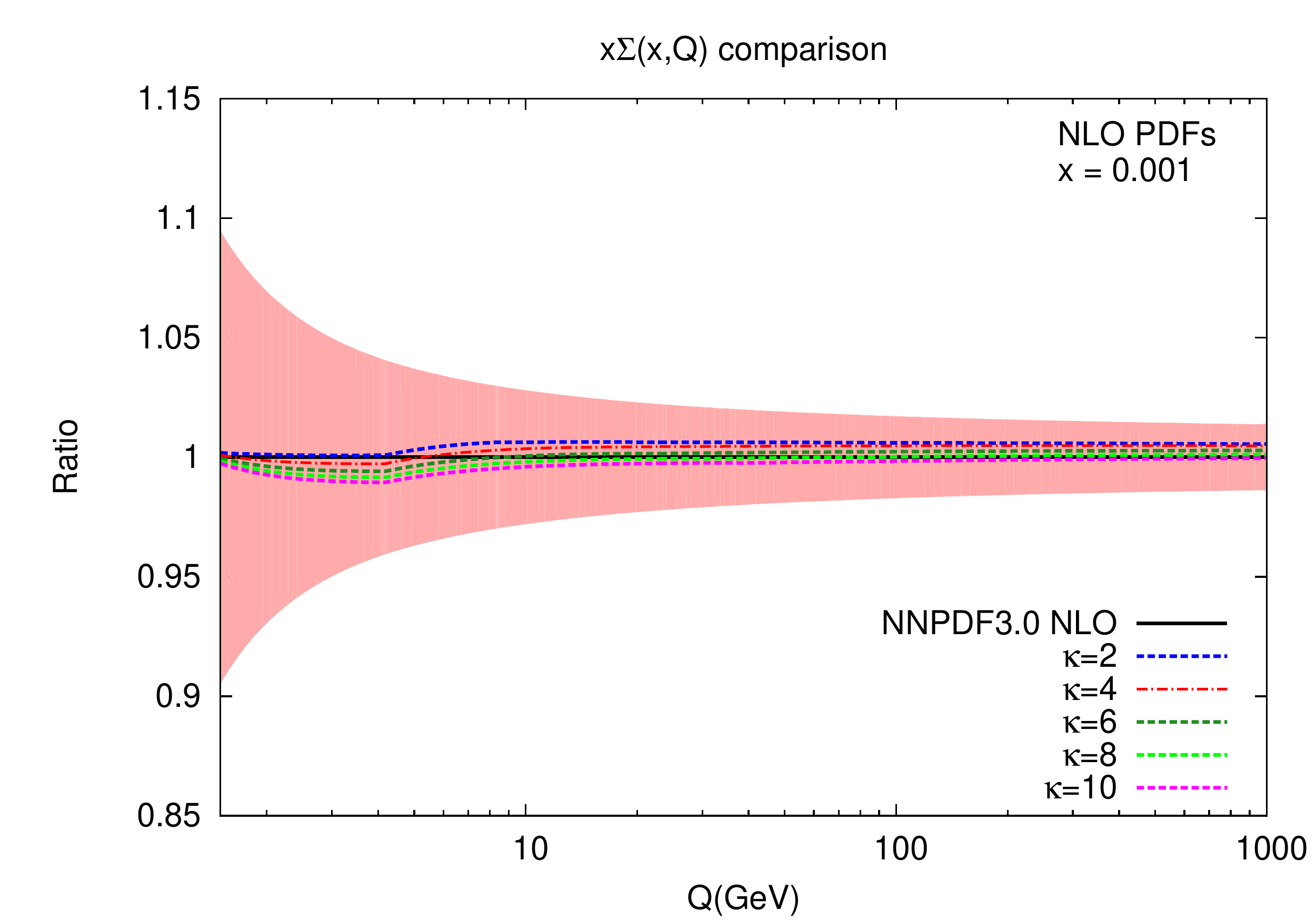}
\includegraphics[width=0.33\textwidth,angle=-90]{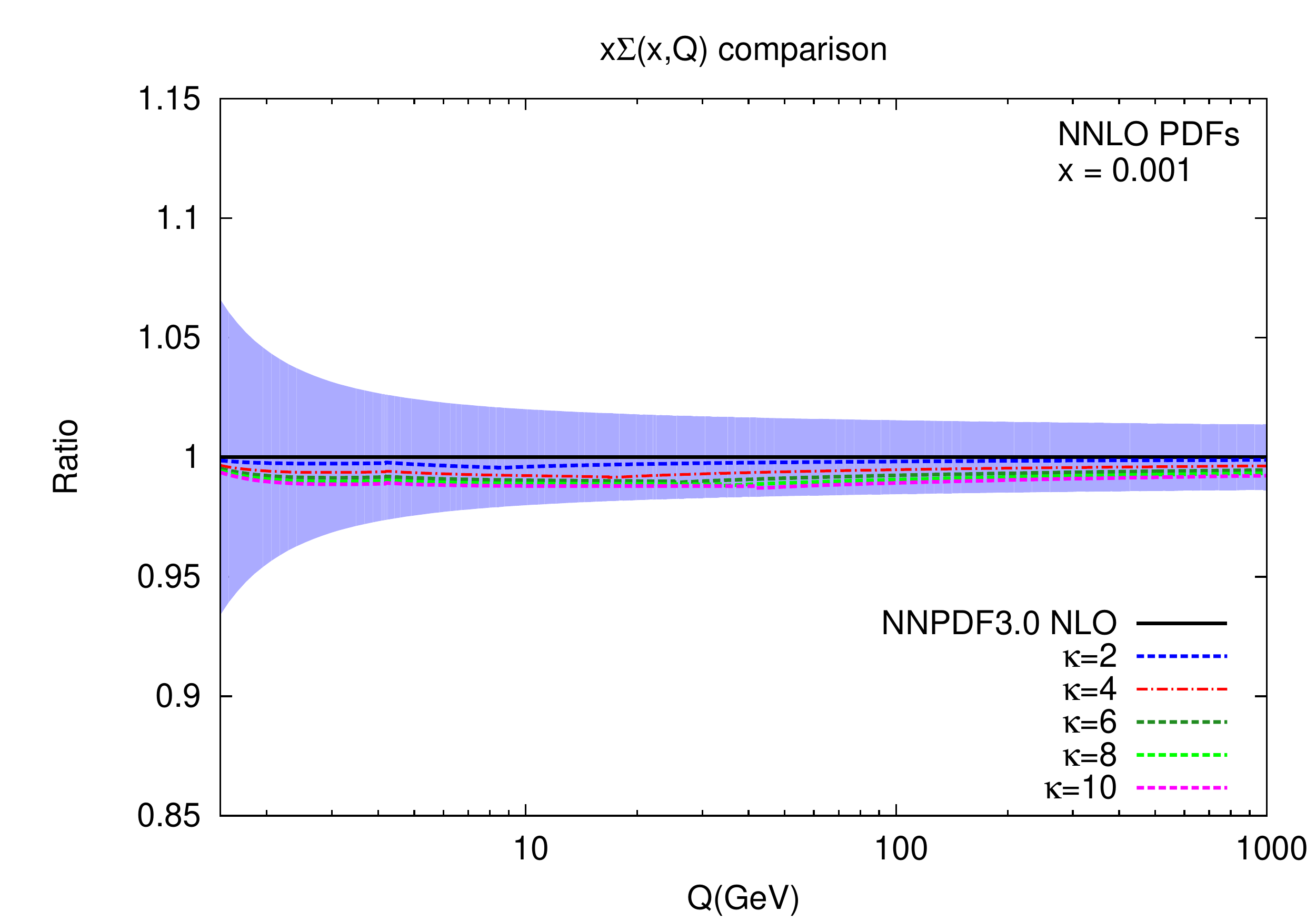}
\includegraphics[width=0.33\textwidth,angle=-90]{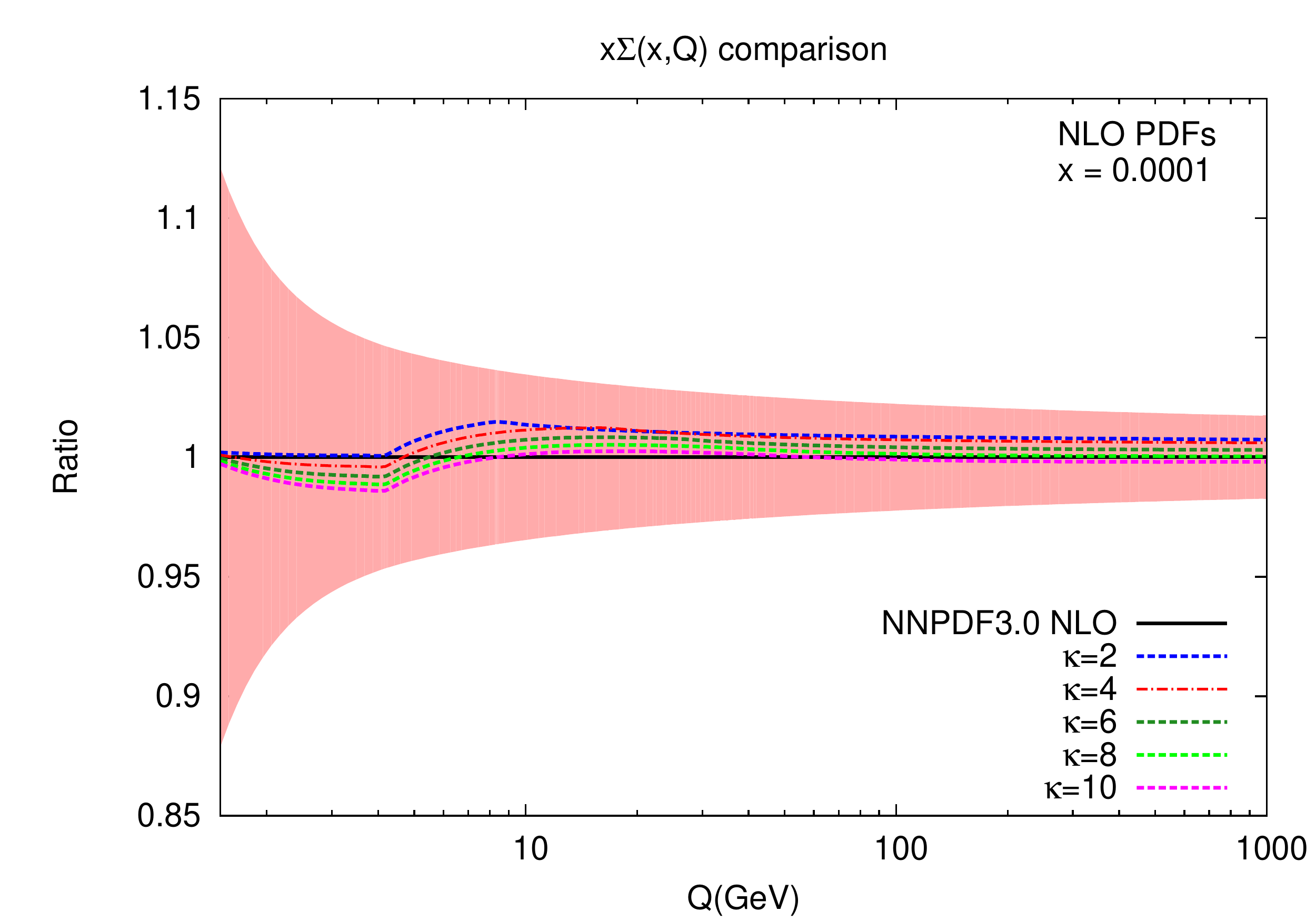}
\includegraphics[width=0.33\textwidth,angle=-90]{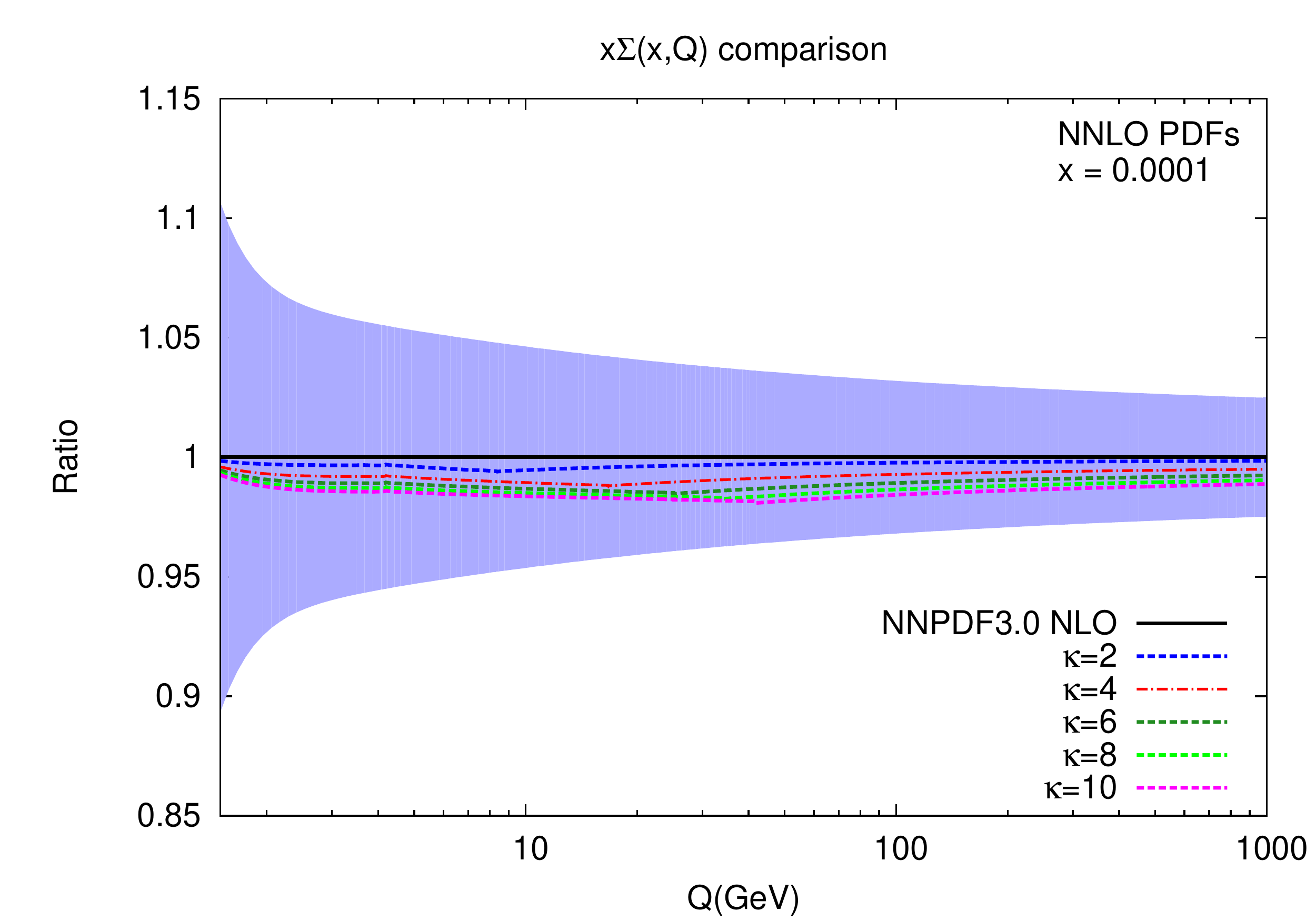}
\caption{As in fig.~\ref{fig:gluonpdf} but for $\Sigma=u+\bar{u}+d+\bar{d}+s+\bar{s}+c+\bar{c}$.}
\label{fig:singlet}
\end{figure}

\end{document}